\documentclass[11pt]{report}

\usepackage{amssymb,amsmath}
\usepackage[dvips]{graphicx}
\usepackage{verbatim}
\usepackage{psfrag}
\usepackage{ifthen}

\newboolean{ucsdformat}
\setboolean{ucsdformat}{false}

\DeclareMathOperator{\poly}{poly}
\DeclareMathOperator{\Tr}{tr}
\DeclareMathOperator{\supp}{supp}
\DeclareMathOperator{\sign}{sign}
\DeclareMathOperator{\vol}{vol}

\newtheorem{thm}{Theorem}[chapter]
\newtheorem{prop}[thm]{Proposition}
\newtheorem{lem}[thm]{Lemma}
\newcommand{\vskipline}{\vskip 11pt}
\newcommand{\tab}{\hskip 0.25in}

\newcommand{\set}[1]{\lbrace #1 \rbrace}
\newcommand{\union}{\cup}
\newcommand{\intersect}{\cap}

\newcommand{\bra}[1]{\langle #1 |}
\newcommand{\ket}[1]{| #1 \rangle}

\newcommand{\expect}[1]{\langle #1 \rangle}
\newcommand{\norm}[1]{\lVert #1 \rVert}
\newcommand{\tensor}{\otimes}
\newcommand{\Tensor}{\bigotimes}

\newcommand{\RR}{\mathbb{R}}
\newcommand{\CC}{\mathbb{C}}
\newcommand{\PP}{\mathcal{P}}
\newcommand{\SSS}{\mathcal{S}}

\newcommand{\be}{\begin{eqnarray}}
\newcommand{\ee}{\end{eqnarray}}
\newcommand{\bea}{\begin{eqnarray}}
\newcommand{\eea}{\end{eqnarray}}
\newcommand{\bma}{\begin{subequations}}
\newcommand{\ema}{\end{subequations}}

\def\qed{\leavevmode\unskip\penalty9999 \hbox{}\nobreak\hfill
     \quad\hbox{\leavevmode  \hbox to.77778em{%
               \hfil\vrule   \vbox to.675em%
               {\hrule width.6em\vfil\hrule}\vrule\hfil}}
     \par\vskip3pt}

\def\N{N}   
\def\m{d}   
\def\n{N}   

\begin{document}

\title{The Complexity of the Consistency and $N$-representability Problems for Quantum States\\
}

\author{Yi-Kai Liu\\
Computer Science and Engineering\\
University of California, San Diego\\
\texttt{y9liu@cs.ucsd.edu}}

%

\date{Aug. 22, 2007\\
(slightly revised version, Dec. 17, 2007)}

\maketitle



\begin{abstract}

QMA (Quantum Merlin-Arthur) is the quantum analogue of the class NP.  There are a few QMA-complete problems, most of which are variants of the ``Local Hamiltonian'' problem introduced by Kitaev.  In this dissertation we show some new QMA-complete problems which are very different from those known previously, and have applications in quantum chemistry.  

The first one is ``Consistency of Local Density Matrices'':  given a collection of density matrices describing different subsets of an $n$-qubit system (where each subset has constant size), decide whether these are consistent with some global state of all $n$ qubits.  This problem was first suggested by Aharonov.  We show that it is QMA-complete, via an oracle reduction from Local Hamiltonian.  Our reduction is based on algorithms for convex optimization with a membership oracle, due to Yudin and Nemirovskii.  

Next we show that two problems from quantum chemistry, ``Fermionic Local Hamiltonian'' and ``$N$-representability,'' are QMA-complete.  These problems involve systems of fermions, rather than qubits; they arise in calculating the ground state energies of molecular systems.  $N$-representability is particularly interesting, as it is a key component in recently developed numerical methods using the contracted Schrodinger equation.  Although these problems have been studied since the 1960's, it is only recently that the theory of quantum computation has provided the right tools to properly characterize their complexity.

Finally, we study some special cases of the Consistency problem, pertaining to 1-dimensional and ``stoquastic'' systems.  We also give an alternative proof of a result due to Jaynes:  whenever local density matrices are consistent, they are consistent with a Gibbs state.
\end{abstract}


\begin{center}
\textbf{Acknowledgements}
\end{center}

\noindent
The path that led to this dissertation was neither easy nor predictable.  I am especially grateful to the following people (listed in alphabetical order) who helped me along the way:  Dorit Aharonov, Andrew Childs, Matthias Christandl, Sanjoy Dasgupta, Russell Impagliazzo, David Meyer, John Preskill and Frank Verstraete.  

I have also benefitted from many conversations with the other graduate students in the theory group, including Chris Calabro, Sashka Davis, Ragesh Jaiswal, Kirill Levchenko, Vadim Lyubashevsky and Nathan Segerlind.  

Thanks to my various friends who do interesting things other than computer science.  And thanks to my parents, for always being there.  

I was supported by a Quantum Computing Graduate Research Fellowship (QuaCGR), provided by the US Army Research Office (ARO) and the Disruptive Technology Office (DTO).  Without their financial support, this work probably would not have taken place.  

Chapter 2 contains some material that was previously published in the paper:  Y.-K. Liu, ``Consistency of Local Density Matrices is QMA-complete,'' \textit{Proc. RANDOM 2006}, pp.438-449, Springer-Verlag (2006).  The dissertation author was the primary investigator and author of this paper.  

Chapter 3 contains some material that was previously published in the paper:  Y.-K. Liu, M. Christandl and F. Verstraete, ``$N$-representability is QMA-complete,'' \textit{Phys. Rev. Lett.} 98, 110503 (2007).  The dissertation author was the primary investigator and author of this paper.  


\chapter{Introduction}

\ifthenelse{\boolean{ucsdformat}}{\thispagestyle{chappage}}{}

\section{Overview}

Beginning in the 1980's, the field of quantum mechanics was reinvigorated by a new idea:  that quantum mechanics has important consequences for machines that store and manipulate information.  In particular, it appeared that quantum computers might be more powerful than classical computers.  This opened up a new direction in computer science, and led to discoveries such as Shor's algorithm for factoring and discrete logarithms \cite{Shor}, Grover's algorithm for black-box search \cite{Grover}, and the first schemes for fault-tolerant quantum computation \cite{Shor2}.  Since then, the field of quantum computation has developed rapidly, and there is considerable interest in building practical quantum computers and finding new quantum algorithms.  (See \cite{NC} for a survey of this area, as it stood in 2000.)  

In this dissertation we study complexity classes based on quantum computation.  On one hand, this is motivated by the possibility that we may eventually succeed in building scalable quantum computers (thus demonstrating that this is a ``reasonable'' model of computation).  But quantum complexity theory is also interesting because it gives new insights into problems that we care about, whether or not we have a quantum computer.  This dissertation focuses on a few such problems, including some ``real-world'' problems from quantum chemistry, whose complexity is best characterized using ideas from quantum (as opposed to classical) computation.  

We study the ``consistency problem for local quantum states,'' which is defined as follows (omitting some details).  Suppose we have a system of $n$ qubits, and we are given a collection of local density matrices $\rho_1,\ldots,\rho_m$, where each $\rho_i$ describes a subset $C_i$ of the qubits.  We assume that $|C_i| \leq k$, for some fixed constant $k$.  Then the problem is to decide whether the $\rho_i$ are ``consistent,'' i.e., whether there exists some global state $\sigma$ (on all $n$ qubits) that matches each of the $\rho_i$ on the subsets $C_i$.  This problem was originally proposed by Dorit Aharonov \cite{A}.  This dissertation presents new results on the computational complexity of the consistency problem, as well as related problems from quantum chemistry and condensed matter physics.

\vskipline

In chapter 2, we show that the consistency problem is QMA-complete, where QMA is the natural generalization of the complexity class NP to the setting of quantum computation.  Before this, there was a canonical QMA-complete problem, the Local Hamiltonian problem, as shown by Kitaev \cite{KSV}.  Local Hamiltonian can be viewed as a generalization of Max-$k$-SAT, or in physical terms, as the problem of estimating the ground state energy of a system of spins with local interactions.  Subsequent work showed that the problem remains QMA-hard for 2-body interactions \cite{KKR}, even when restricted to nearest neighbors on a 2-D square lattice \cite{OT}; the problem is also QMA-hard for nearest neighbors on a 1-D chain where each site is not a qubit, but a qudit of dimension $d \geq 8$ \cite{qma1d-1,qma1d-2}.  However, these were essentially the only known QMA-complete problems (aside from a few problems which are closely related to the definition of QMA).  With the Consistency problem, we give the first real example of a QMA-complete problem that is not a variant of Local Hamiltonian.  In particular, Consistency is best described as a constraint satisfaction problem, while Local Hamiltonian is an optimization problem.  

We give a poly-time oracle reduction from Local Hamiltonian to Consistency, using algorithms for convex optimization with a membership oracle.  This kind of reduction is quite unusual.  After our paper was published, we became aware of work by Gurvits \cite{Gurvits} that used a similar technique to show NP-hardness of the separability problem for quantum states, and also an older paper by Grotschel et al \cite{GLS-1981} that used a weaker technique (convex optimization with a separation oracle) to show NP-hardness of weighted fractional chromatic number; but these seem to be the only previous examples.  Here we develop the technique in greater detail.  The usual approach is to use algorithms such as the shallow-cut ellipsoid method of Yudin and Nemirovskii \cite{YN,GLS}, or the random-walk algorithm of Bertsimas and Vempala \cite{BV,KV}.  We find that much simpler algorithms are sufficient for this application, because we only need to find approximate solutions (with accuracy $\pm 1/\poly(n)$), as opposed to exact solutions (accuracy $\pm 2^{-n}$).  

\vskipline

In chapter 3, we study the $N$-representability problem, which is an analogue of the consistency problem for fermionic systems.  (This chapter is joint work with Matthias Christandl and Frank Verstraete.)  $N$-representability was first introduced by quantum chemists in the 1960's, as a route to computing the ground states of molecular systems \cite{Coulson,Tredgold,Coleman}; beginning in the 1990's, it has received renewed attention, thanks to improved variational methods based on semidefinite programming, and brand new methods such as the contracted Schrodinger equation \cite{book1,book,Mazziotti}.  (Collectively these are known as 2-RDM methods.)

We show that fermionic Local Hamiltonian is QMA-hard, by constructing a mapping from spin systems to fermionic systems.  Then we show that $N$-representability is QMA-hard, using the convex optimization technique from chapter 2.  Ironically, this is the same idea that the quantum chemists use to design algorithms, but restated in a much more general form:  we show that any efficient solution to $N$-representability would imply an efficient algorithm to compute ground state energies, not just for molecules, but for generic local Hamiltonians---and this is QMA-hard.  Finally, we show that fermionic Local Hamiltonian and $N$-representability are in QMA, and hence are QMA-complete.  In addition, we show that a related problem, pure-state $N$-representability, is in the class QMA(2) (see chapter 3 for details).  

Our hardness result implies that 2-RDM methods must break down in the general case.  But there is empirical evidence that 2-RDM methods perform well on instances that arise in quantum chemistry.  It would be wonderful to find some theoretical explanation for this.  Is there some fundamental property of these instances that explains the success of 2-RDM methods?  Also, it is not clear whether 2-RDM methods still give accurate results when scaled up to larger molecules; a theoretical analysis would be helpful in answering this question.

\vskipline

In chapter 4, we study the consistency problem for 1-dimensional and ``stoquastic'' systems.  These are interesting special cases, for which the Local Hamiltonian problem may not be QMA-hard.  (Local Hamiltonian on a 1-D chain of qudits (for $d \geq 8$) is QMA-hard \cite{qma1d-1,qma1d-2}, but this is not known for smaller values of $d$, e.g., qubits.  Also, there is complexity-theoretic evidence that Stoquastic Local Hamiltonian is not QMA-hard, though it is at least MA-hard \cite{stoq-1}.)  We show that 1-D Consistency has the same complexity as 1-D Local Hamiltonian, up to poly-time oracle reductions.  Also, we propose a stoquastic version of the Consistency problem, which appears to be equivalent to Stoquastic Local Hamiltonian, up to poly-time oracle reductions.  These results suggest that, for special classes of systems, Consistency may provide an alternative route to solving Local Hamiltonian.  (This is the approach used in the 2-RDM methods in quantum chemistry.)  

For these special cases, the reduction from Local Hamiltonian to Consistency uses the same ideas as before, but the reverse direction requires a new technique, since we can no longer use the machinery of QMA-hardness.  We give a new reduction from Consistency to Local Hamiltonian that is based on Lagrange duality (combined with convex optimization using a membership oracle).  The duality idea is similar to recent work by Hall \cite{Hall} on the ``subsystem compatibility problem''; this resembles the Consistency problem, except that one is given density matrices describing all proper subsets of the system.  (Thus the description of the problem is exponentially large in the number of qubits, and the problem is poly-time solvable, using an amount of time that is polynomial in the size of the input, but exponential in the number of qubits.)  Previously, duality techniques have also been used in the study of entanglement, e.g., the notion of an ``entanglement witness'' \cite{Horodecki}.

\vskipline

In chapter 5, we show an interesting structural property of consistent quantum states:  if $\rho_1,\ldots,\rho_m$ are consistent with some state $\sigma \succ 0$, then they are also consistent with a Gibbs state $\sigma' = (1/Z) \exp(M_1+\cdots+M_m)$.  This result was previously proved by Jaynes \cite{Jaynes-2} in connection with the maximum-entropy principle; here we give a somewhat different proof, using the partition function.  

\section{Quantum computation}

Consider a quantum mechanical system.  For our purposes, the \textit{state} of the system is described by a unit vector $\ket{\psi}$ in a vector space $\mathbb{C}^d$.  Here we assume that the dimension $d$ is finite, i.e., we do not consider systems with continuous degrees of freedom, arbitrarily many particles, unbounded energy, etc.  We also assume that the state is pure, or deterministic (later we will come back to discuss mixed states).  We remark that the complex phase of the vector $\ket{\psi}$ is unimportant:  for any $\theta \in \RR$, the vectors $\ket{\psi}$ and $e^{i\theta} \ket{\psi}$ have the same physical meaning.  We use ``bracket'' notation:  $\ket{\psi}$ denotes a column vector, while $\bra{\psi}$ denotes its adjoint, or conjugate transpose, which is a row vector.

Various operations on the system are described by linear operators on $\mathbb{C}^d$, that is, complex $d \times d$ matrices.  For an operator $A$, we define $A^\dagger$ to be the adjoint, or conjugate transpose, of $A$.  If the system is closed (it does not interact with an outside environment), then the state evolves via \textit{unitary} operations:  $\ket{\psi}$ evolves to $U \ket{\psi}$, where $U$ is a unitary matrix, that is, $U^\dagger = U^{-1}$.  Note that this operation preserves the length of the vector $\ket{\psi}$.  

For our purposes, a measurement is described by an \textit{observable} $O$, which is a Hermitian matrix, that is, $O^\dagger = O$.  By the spectral theorem, $O$ can be written in the form $O = \sum_i \lambda_i \Pi_i$, where the $\lambda_i$ are distinct real numbers, and the $\Pi_i$ are projectors onto orthogonal subspaces.  Here the $\lambda_i$ represent the possible outcomes of the measurement:  if the system is in state $\ket{\psi}$, then the measurement yields outcome $\lambda_i$ with probability $p_i = \bra{\psi}\Pi_i\ket{\psi}$; following the measurement, the system will be in state $(1/\sqrt{p_i}) \Pi_i\ket{\psi}$.  Thus the expectation value of the measurement is given by $\bra{\psi} O \ket{\psi}$.  

A special kind of measurement is the following:  we have an orthonormal basis $\set{\ket{\varphi_1},\ldots,\ket{\varphi_d}}$, and we let $O = \sum_{i=1}^d i \ket{\varphi_i}\bra{\varphi_i}$.  If the system is in state $\ket{\psi} = \sum_{i=1}^d \alpha_i \ket{\varphi_i}$, then the measurement yields outcome $i$ with probability $|\alpha_i|^2$; following the measurement, the system will be in state $\ket{\varphi_i}$.  This is called a measurement in the basis $\set{\ket{\varphi_1},\ldots,\ket{\varphi_d}}$.  

\vskipline

The basic building block of a quantum computer is the \textit{qubit}.  This is a two-dimensional system, whose state is a unit vector in $\mathbb{C}^2$.  We fix an orthonormal basis for $\mathbb{C}^2$ which consists of two states, $\ket{0}$ and $\ket{1}$; then we can write $\ket{\psi} = \alpha\ket{0} + \beta\ket{1}$.  

We can construct more complex systems by assembling multiple qubits.  To describe this, we need to define the \textit{tensor product}.  Let $A$ and $B$ be vector spaces, of dimension $d_A$ and $d_B$.  For any vectors $a \in A$ and $b \in B$, the tensor product $a \tensor b$ is a vector of dimension $d_A d_B$, where the tensor operation satisfies the following properties:  (1) for any vectors $a \in A$, $b \in B$, and any scalar $s$, we have $s(a \tensor b) = (sa) \tensor b = a \tensor (sb)$; (2) for any vectors $a,a' \in A$, $b \in B$, we have $(a+a') \tensor b = a \tensor b + a' \tensor b$; (3) for any vectors $a \in A$, $b,b' \in B$, we have $a \tensor (b+b') = a \tensor b + a \tensor b'$.  

Furthermore, $A \tensor B$ is the vector space of dimension $d_A d_B$, consisting of all linear combinations of tensor products, that is, all vectors of the form 
\[
\sum_{a \in A,\; b \in B} u_{ab} (a \tensor b).  
\]
There is a natural inner product on this space:  one defines $\langle a \tensor b, a' \tensor b' \rangle = \langle a,a' \rangle \langle b,b' \rangle$, and extends it using linearity to get 
\[
\Bigl\langle \sum_{a,b} u_{ab} (a \tensor b), \;
             \sum_{a',b'} v_{a'b'} (a' \tensor b') \Bigr\rangle 
= \sum_{a,b} \sum_{a',b'} 
  \bar{u}_{ab} v_{a'b'} \langle a,a' \rangle \langle b,b' \rangle.  
\]
Note that, if $\set{a^{(i)}}$ is an orthonormal basis for $A$, and $\set{b^{(j)}}$ is an orthonormal basis for $B$, then $\set{a^{(i)} \tensor b^{(j)}}$ is an orthonormal basis for $A \tensor B$.  Also, given operators $P$ and $Q$ acting on the spaces $A$ and $B$, respectively, one can construct an operator $P \tensor Q$ acting on the space $A \tensor B$, by defining $(P \tensor Q) (a \tensor b) = (Pa) \tensor (Qb)$, and extending it by linearity.  

More concretely, one can write the tensor product $a \tensor b$ by taking the vector $a$ and replacing each scalar entry $a_i$ with a block consisting of the vector $a_i b$ (this is known as the Kronecker product).  For example, $(a_1, a_2)^T \tensor (b_1, b_2)^T = (a_1 b_1, a_1 b_2, a_2 b_1, a_2 b_2)^T$.  One can write the tensor product of two matrices in a similar way.  

If we have two quantum systems, described by state spaces $A$ and $B$, then the combined system is described by the state space $A \tensor B$.  Also, if $P$ is a unitary operation or an observable for the first system, then $P \tensor I$ is the equivalent operation for the combined system; likewise, if $Q$ is an operation for the second system, the $I \tensor Q$ is the equivalent operation for the combined system.  

So, a system of $n$ qubits has a state space $(\mathbb{C}^2)^{\tensor n} = \mathbb{C}^2 \tensor \cdots \tensor \mathbb{C}^2$, that is, a tensor product of $n$ copies of $\mathbb{C}^2$.  This is a vector space of dimension $2^n$, with an orthonormal basis consisting of the vectors $\ket{z} = \ket{z_1} \tensor \cdots \tensor \ket{z_n}$, $z \in \set{0,1}^n$.  We refer to this as the \textit{standard} or \textit{computational} basis.  

\vskipline

We are ready to introduce the \textit{quantum circuit} model of computation.  We define a quantum computer to be a device that can perform the following tasks on $n$ qubits (using resources that scale polynomially in $n$):  (1) prepare qubits in the computational basis states; (2) implement a universal family of quantum gates, which can be applied to any subset of qubits; (3) measure qubits in the computational basis.  Here, a \textit{quantum gate} is simply a unitary operation on a fixed number of qubits (that does not grow with $n$).  We assume a fixed, finite set of gates; circuits on $n$ qubits are built by composing these gates.  We say that a set of gates $S$ is \textit{universal} if, for any unitary operation $U$, one can approximate $U$ with error $\varepsilon$ by using a circuit of size $O(\poly(1/\varepsilon))$ consisting of gates from $S$.  (Note that the $O(\poly(1/\varepsilon))$ contains a hidden constant that depends on $U$.)  This implies that, for any set of gates $S'$, a circuit of size $m$ consisting of gates from $S'$ can be simulated with error $\varepsilon$ by using a circuit of size $O(\poly(m/\varepsilon))$ consisting of gates from $S$.  (Again, the hidden constant depends on $S'$.)  

For example, the following gates are a universal set:  controlled-NOT ($CNOT$), Hadamard ($H$), phase ($S$), $\pi/8$ gate ($T$).  Controlled-NOT is a two-qubit gate, while the others are single-qubit gates.  They are defined as follows:  
\[
CNOT = \begin{pmatrix}
1& 0& 0& 0\\
0& 1& 0& 0\\
0& 0& 0& 1\\
0& 0& 1& 0
\end{pmatrix}, \quad
H = \frac{1}{\sqrt{2}} \begin{pmatrix}
1& 1\\
1& -1
\end{pmatrix}, \quad
S = \begin{pmatrix}
1& 0\\
0& i
\end{pmatrix}, \quad
T = \begin{pmatrix}
1& 0\\
0& e^{\pi i/4}
\end{pmatrix}.  
\]
Furthermore, for any unitary transformation $U$, the number of gates needed to approximate $U$ with error $\varepsilon$ grows like $O(\log^c(1/\varepsilon))$, $c \approx 2$; this is knows as the Solovay-Kitaev theorem.  See \cite{NC} for more details and proofs of these results.

We remark that there are other equivalent models of quantum computation, such as the quantum Turing machine \cite{BV97} (see also \cite{NC} for references to earlier work in this area), and models motivated by possible experimental implementations of quantum computers.  

\section{Quantum Complexity Classes}

We define the class BQP (``bounded-error quantum polynomial time''), by analogy with BPP (``bounded-error probabilistic polynomial time'') \cite{BV97}:  a language $L$ is in BQP if there exists a poly-time quantum algorithm $A$ such that 
\begin{itemize}
\item If $x \in L$, then $A(x)$ accepts with probability $\geq 2/3$.  
\item If $x \notin L$, then $A(x)$ accepts with probability $\leq 1/3$.  
\end{itemize}
To be precise, $A$ is a uniform family of quantum circuits, of polynomial size.  Similarly to BPP, the success probabilities can be amplified via repetition.  

The class QMA, or ``Quantum Merlin-Arthur,'' is defined as follows \cite{Wat00,KSV,AN}:  a language $L$ is in QMA if there exists a poly-time quantum verifier $V$ and a polynomial $p$ such that 
\begin{itemize}
\item If $x \in L$, then there exists a quantum state $\rho$ on $p(|x|)$ qubits such that $V(x,\rho)$ accepts with probability $\geq 2/3$.  
\item If $x \notin L$, then for all quantum states $\rho$ on $p(|x|)$ qubits, $V(x,\rho)$ accepts with probability $\leq 1/3$.  
\end{itemize}
Here, $|x|$ denotes the length of the string $x$.  This is similar to the definition of MA or NP, except that the witness is allowed to be a quantum state, and the verifier is a quantum circuit with bounded error probability.  The success probabilities can be amplified via parallel repetition; see the discussion in \cite{AN}.  

Note that one can easily restate these definitions in terms of promise problems, rather than languages.  

We give a brief summary of the known relationships between BQP, QMA and other complexity classes.  Definitions of the other classes can be found in \cite{Papadimitriou}.  

First, it is not hard to see that BPP $\subseteq$ BQP, BQP $\subseteq$ QMA, and MA $\subseteq$ QMA.  

BQP and QMA are contained in ``counting'' classes such as \#P.  In particular, BQP $\subseteq$ PP \cite{ADH97} (see \cite{DHHMNO} for a simpler proof); a stronger result is BQP $\subseteq$ AWPP \cite{FR98}.  Also, QMA $\subseteq$ PP \cite{KW}; a stronger result is QMA $\subseteq$ A$_0$PP \cite{Vyal}.  However, these upper bounds do not seem to be tight.  PP is quite a  powerful class; note that $\text{P}^\text{PP}$ contains the polynomial hierarchy PH (Toda's theorem).  Also, PP seems to be much more powerful than BQP; note that PP = PostBQP (BQP with postselection) \cite{Aaronson}.  

Much less is known about the relationship between BQP and the polynomial hierarchy PH.  (Recall that PH is the union of the classes NP, coNP, $\Sigma_2^\text{P} = \text{NP}^\text{NP}$, $\Pi_2^\text{P} = (\text{coNP})^\text{NP}$,....)  We do know that, relative to a random oracle, with probability 1, BQP does not contain NP \cite{BBBV97}.  Since BPP and MA are contained in PH, one might expect that BQP would be in PH, but this is not known.  

\section{The Local Hamiltonian Problem}

The Local Hamiltonian problem is defined as follows \cite{KSV,AN}:  
\begin{quote}
Consider a system of $n$ qubits.  We are given a Hamiltonian $H = H_1+\cdots+H_m$, where each $H_i$ acts on a subset of qubits $C_i \subseteq \set{1,\ldots,n}$ (and so has dimension $2^{|C_i|} \times 2^{|C_i|}$).  The $H_i$ are Hermitian matrices, with norm $\norm{H_i} \leq 1$.  Also, each subset $C_i$ has size $|C_i| \leq k$, for some fixed $k$.  

All numbers are specified with $\gamma$ bits of precision.  

In addition, we are given a string ``$1^s$'' (the unary encoding of a natural number $s$), and two real numbers $a$ and $b$, such that $b-a \geq 1/s$.  

The problem is to distinguish between the following two cases:  
\begin{itemize}
\item If $H$ has an eigenvalue that is $\leq a$, output ``YES.''  
\item If all the eigenvalues of $H$ are $\geq b$, output ``NO.''  
\end{itemize}
\end{quote}

Note that one may have multiple terms in the Hamiltonian that act on the same subset; so the subsets $C_i$ might not all be distinct.  

The string ``$1^s$'' is simply a device to ensure that the gap between the ``YES'' and ``NO'' cases is not too small, relative to the ``size'' of the problem.  

Intuitively, we think of $n$ as the ``size'' of the problem, and we are interested in instances where $k$ is a constant, $m \leq \poly(n)$, $\gamma \leq \poly(n)$ and $s \leq \poly(n)$ (so the gap $b-a$ is at least $1/\poly(n)$).  We say an algorithm is efficient if it solves these instances in time $\poly(n)$.  

Formally, an instance of the problem is described by a string of length $\ell = \Theta(4^k m \gamma + s)$.  
We say an algorithm is efficient if it takes time polynomial in $\ell$.  
Although this formal definition looks different from our intuition, it is equivalent, as we will see in the next section.

Finally, note that this is a \textit{promise problem}:  we are promised that the input is either a ``YES'' instance or a ``NO'' instance.  

Special cases of the problem include 2-Local Hamiltonian (where $k=2$), and 2-Local Hamiltonian on a graph $G$ (where $k=2$, and the graph $G'$, consisting of vertices $1,\ldots,n$ and edges $C_1,\ldots,C_m$, is restricted to be a subgraph of $G$).  

Kitaev showed that Local Hamiltonian is in QMA, and the case of $k=5$ is QMA-hard \cite{KSV,AN}.  With greater effort, one can show that 2-Local Hamiltonian is also QMA-hard \cite{KR,KKR}.  These hard instances of Local Hamiltonian do have the property that $m \leq \poly(n)$ and $s \leq \poly(n)$.

(This is a slight abuse of notation, because QMA is a class of languages, whereas Local Hamiltonian is a promise problem.)  

\section{Promise Problems and Polynomial Time}

In the previous section we considered two notions of what it means to solve the Local Hamiltonian problem efficiently.  Assume $k$ is constant, so an instance of the problem is described by a string of length $\ell = \Theta(m\gamma+s)$.  Intuitively, we believe an algorithm is efficient if, on instances where $m \leq \poly(n)$, $\gamma \leq \poly(n)$ and $s \leq \poly(n)$, the algorithm takes time $\poly(n)$.  Formally, we say an algorithm is efficient if, on all instances, it takes time $\poly(\ell)$.  We now show that, under some mild conditions, these two notions are equivalent.  

We say that Local Hamiltonian is \textit{polynomial-time solvable} if:  
\begin{quote}
There exists an algorithm $A$ and a polynomial $t$, such that on all instances, $A$ returns the correct answer in time $t(\ell)$.  
\end{quote}

Let $(S)$ denote the following statement, which corresponds more closely to our intuition:  
\begin{quote}
There exists an algorithm $A$ and a polynomial $t$, and there exist constants $\alpha,\beta>0$, such that for any instance that satisfies $m,\gamma,s \leq \alpha n^\beta$, $A$ returns the correct answer in time $t(n)$.  
\end{quote}
Statement $(S)$ asserts that, for some specific bounds on the size of $m$, $\gamma$ and $s$ as a function of $n$, the algorithm $A$ runs in time $\poly(n)$.  These bounds can be very restrictive, for instance, they may be sublinear in $n$.  Thus $(S)$ appears to be a weaker condition, because it does not say anything about the running time for other values of $m$, $\gamma$ and $s$.  

Obviously, if Local Hamiltonian is poly-time solvable, then $(S)$ holds.  We will show the reverse implication, using a padding argument:  Suppose that $(S)$ holds.  We will construct a modified algorithm $\tilde{A}$ that solves arbitrary instances of Local Hamiltonian.  $\tilde{A}$ that takes an instance $x$, modifies it by adding extra ``dummy'' qubits to the problem, thus increasing $n$ until it satisfies the promises stated in condition $(S)$, and then runs algorithm $A$.  On an input of length $\ell$, algorithm $\tilde{A}$ takes time 
\[
\max \set{t(n), t((m/\alpha)^{1/\beta}), t((\gamma/\alpha)^{1/\beta}), t((s/\alpha)^{1/\beta})} \leq \poly(\ell), 
\]
hence Local Hamiltonian is poly-time solvable.

So statement $(S)$ is equivalent to poly-time solvability.  So we can use either of these notions; it turns out that the latter one is more convenient and less cumbersome.  Similar arguments apply to other promise problems.

\section{Density Matrices}

Consider a system of $n$ qubits.  Up to this point we have dealt with pure states, which are represented by vectors $\ket{\psi}$ in $\mathbb{C}^{2^n}$.  However, one may also encounter \textit{mixed} states, which are ensembles of pure states, where each state $\ket{\psi_i}$ appears with some probability $p_i$.  (For simplicity we assume a discrete ensemble $\set{\ket{\psi_i}}$; continuous ensembles can be treated in a similar way.)  It turns out that a mixed state is represented by a \textit{density matrix}, which is a positive semidefinite matrix on $\mathbb{C}^{2^n}$ with trace 1, defined by 
\[
\rho = \sum_i p_i \ket{\psi_i}\bra{\psi_i}.
\]
In particular, a pure state $\ket{\psi}$ is represented by the density matrix $\ket{\psi}\bra{\psi}$.  Also, if we have an ensemble where each element is a mixed state $\rho_i$, which appears with probability $p_i$, then the ensemble is described by the density matrix $\rho = \sum_i p_i \rho_i$.  

Interestingly, it is possible for two seemingly different ensembles to be represented by the same density matrix.  For instance, an equal mixture of $\ket{0}$ and $\ket{1}$ yields the same density matrix as an equal mixture of $\ket{+} = \frac{1}{\sqrt{2}} (\ket{0}+\ket{1})$ and $\ket{-} = \frac{1}{\sqrt{2}} (\ket{0}-\ket{1})$.  Quantum mechanics asserts that all of the physically accessible information is contained in the density matrix.  So in cases like this, the two ensembles cannot be distinguished by an observer.  

One can reformulate the basic facts of quantum mechanics, using density matrices instead state vectors.  A unitary operation $U$ transforms a density matrix $\rho$ to $U \rho U^\dagger$.  If we measure an observable $O = \sum_i \lambda_i \Pi_i$, we get outcome $\lambda_i$ with probability $p_i = \Tr(\Pi_i \rho)$; following the measurement, the system will be in state $(1/p_i) \Pi_i \rho \Pi_i$.  Thus the expectation value of the measurement is given by $\Tr(O\rho)$.  In particular, if we measure in a complete orthonormal basis $\set{\ket{\varphi_1},\ldots,\ket{\varphi_d}}$, we get outcome $i$ with probability $\bra{\varphi_i} \rho \ket{\varphi_i}$ (these are simply the diagonal elements of $\rho$ in the basis $\set{\ket{\varphi_1},\ldots,\ket{\varphi_d}}$); following the measurement, the system will be in state $\ket{\varphi_i}\bra{\varphi_i}$.  Finally, if two quantum systems $A$ and $B$ are in states $\rho_A$ and $\rho_B$, then the combined system is in state $\rho_A \tensor \rho_B$.  

\vskipline

Density matrices are a convenient tool for describing \textit{subsets} of a quantum system.  Here the situation is more complicated than in the classical world, because of the phenomenon of entanglement.  For example, consider the following two-qubit state, $\ket{\Phi^+} = \frac{1}{\sqrt{2}} (\ket{00}+\ket{11})$.  This is a pure state, and in the classical world, that would imply that the two individual bits were pure (i.e., deterministic), and uncorrelated.  But for this quantum state, even though the overall state is pure, the two individual bits are mixed (they can be either 0 or 1), and correlated (they are always equal).  In fact, for a quantum state, it is possible for a subset of the system to have higher entropy than the system as a whole.  These unusual effects are caused by entanglement; see \cite{NC} for a further discussion of entanglement and its applications to quantum computation.

A subset of a quantum system is described by a \textit{reduced density matrix}.  Say we have two quantum systems, with state spaces $A$ and $B$.  Let $\rho$ be the state of the combined system, i.e., a density matrix $\rho$ on the space $A \tensor B$, where $\rho$ is \textit{not} necessarily of the form $\sigma \tensor \tau$.  Let $\set{\ket{a_1},\ldots,\ket{a_d}}$ be a basis for $A$, and let $\set{\ket{b_1},\ldots,\ket{b_{d'}}}$ be a basis for $B$.  Then $\set{\ket{a_i} \tensor \ket{b_{i'}}}$ is a basis for $A \tensor B$, and we can write $\rho$ in the form 
\[
\begin{split}
\rho &= \sum_{i,i',j,j'} \rho_{i,i',j,j'} (\ket{a_i} \tensor \ket{b_{i'}}) 
        (\bra{a_j} \tensor \bra{b_{j'}}) \\
     &= \sum_{i,i',j,j'} \rho_{i,i',j,j'} (\ket{a_i}\bra{a_j}) \tensor 
        (\ket{b_{i'}}\bra{b_{j'}}).
\end{split}
\]
Then the subset $A$ is described by the reduced density matrix 
\[
\rho^{[A]} = \Tr_B(\rho).
\]
Here we define the \textit{partial trace} over $B$ by 
\[
\begin{split}
\Tr_B(\rho) &= \sum_{i,i',j,j'} \rho_{i,i',j,j'} (\ket{a_i}\bra{a_j}) 
               \Tr(\ket{b_{i'}}\bra{b_{j'}}) \\
            &= \sum_{i,j} \Bigl(\sum_{i'} \rho_{i,i',j,i'}\Bigr) \ket{a_i}\bra{a_j}.
\end{split}
\]
This is also called ``tracing over $B$.''  Intuitively, it is like computing a marginal probability distribution, by summing over all possible values of $B$.  It can be shown that the result does not depend on the choice of basis for $B$.  Furthermore, for any observable $O$ on the subsystem $A$, one can show that measuring $O$ with the reduced state $\rho^{[A]}$ yields the same outcomes as measuring $O \tensor I$ with the original state $\rho$.  

\vskipline

Finally, we introduce the Pauli matrices, which are a useful tool for working with density matrices.  Let $X$, $Y$ and $Z$ denote the Pauli matrices for a single qubit, 
\[
X = \begin{pmatrix} 0 & 1 \\ 1 & 0 \end{pmatrix}, \quad
Y = \begin{pmatrix} 0 & -i \\ i & 0 \end{pmatrix}, \quad
Z = \begin{pmatrix} 1 & 0 \\ 0 & -1 \end{pmatrix}, 
\]
and define $\PP = \set{I,X,Y,Z}$.  We can construct $n$-qubit Pauli matrices by taking tensor products $P = P_1 \tensor \cdots \tensor P_n \in \PP^{\tensor n}$.  

Any $2^n$-dimensional Hermitian matrix can be written as a real linear combination of $n$-qubit Pauli matrices.  Furthermore, the $n$-qubit Pauli matrices are orthogonal with respect to the Hilbert-Schmidt inner product:  $\Tr(P^\dagger Q) = 2^n$ if $P = Q$, and 0 otherwise.  So, if $\sigma$ is an $n$-qubit state, we can write it in the form 
\[
\sigma = \frac{1}{2^n} \sum_{P \in \PP^{\tensor n}} \alpha_P P, 
\]
where the coefficients are uniquely determined by $\alpha_P = \Tr(P\sigma)$; note that these are the expectation values of the Pauli matrices $P$.  This application of the Pauli matrices is closely related to quantum state tomography.  

One can also write a reduced density matrix $\sigma^{[A]}$, where $A \subseteq \set{1,\ldots,n}$, in terms of the Pauli matrices.  We say that a Pauli matrix $P$ is supported on the set $A$ if, for all $i \notin A$, $P_i = I$.  Also, define the restriction of $P$ to $A$, $P|_A = \tensor_{i \in A} P_i$.  

The partial trace acts on $P$ as follows:  $\Tr_{\set{1,\ldots,n}-A} (P) = 2^{n-|A|} P|_A$ if $P$ is supported on $A$, and 0 otherwise.  Thus we have 
\[
\sigma^{[A]} = \Tr_{\set{1,\ldots,n}-A} (\sigma)
 = \frac{1}{2^{|A|}} \sum_{\text{$P$ supported on $A$}} \alpha_P P|_A.  
\]
In other words, the information contained in $\sigma^{[A]}$ is precisely the expectation values of those Pauli matrices $P$ that are supported on $A$.  

\vskipline

We state a few definitions from matrix analysis \cite{Bhatia}.  For a vector $v \in \CC^n$, we define the $\ell_2$ and $\ell_1$ norms, 
\[
\norm{v} = \norm{v}_2 = (\sum_i |v_i|^2)^{1/2}, \qquad
\norm{v}_1 = \sum_i |v_i|.  
\]
For a matrix $A \in \CC^{n \times n}$, we let $A^\dagger$ denote the conjugate transpose, and $|A| = \sqrt{A^\dagger A}$.  We define the sup, $\ell_2$ and $\ell_1$ norms, 
\[
\norm{A} = \sup_{\norm{v}=1} \norm{Av}, \qquad
\norm{A}_2 = \Tr(A^\dagger A) = \sum_{ij} |A_{ij}|^2, \qquad
\norm{A}_1 = \Tr|A|.  
\]

\section{Consistency of Local Density Matrices}

We define the Consistency problem as follows \cite{A}:  
\begin{quote}
Consider a system of $n$ qubits.  We are given a collection of local density matrices $\rho_1,\ldots,\rho_m$, where each $\rho_i$ acts on a subset of qubits $C_i \subseteq \set{1,\ldots,n}$ (and so has dimension $2^{|C_i|} \times 2^{|C_i|}$).  Each subset $C_i$ has size $|C_i| \leq k$, for some constant $k$.  

All numbers are specified with $\gamma$ bits of precision.  

In addition, we are given a string ``$1^s$'' (the unary encoding of a natural number $s$), and a real number $\beta$, such that $\beta \geq 1/s$.  

The problem is to distinguish between the following two cases:  
\begin{itemize}
\item There exists an $n$-qubit state $\sigma$ such that, for all $i$, $\Tr_{\set{1,\ldots,n}-C_i}(\sigma) = \rho_i$.  In this case, output ``YES.''  \footnote{Here the equality holds up to $\gamma$ bits of precision.}
\item For all $n$-qubit states $\sigma$, there exists some $i$ such that $\norm{\Tr_{\set{1,\ldots,n}-C_i}(\sigma) - \rho_i}_1 \geq \beta$.  In this case, output ``NO.''  
\end{itemize}
\end{quote}

Without loss of generality, we can assume that the subsets $C_i$ are all distinct; thus $m \leq \binom{n}{k} \leq n^k$.  As in the Local Hamiltonian problem, the string ``$1^s$'' is simply a device to ensure that the gap between the ``YES'' and ``NO'' cases is not too small, relative to the ``size'' of the problem.  Here, we use the norm $\norm{A}_1 = \Tr|A|$ to measure the distance between $\rho_i$ and the corresponding reduced density matrix of $\sigma$.  When multipled by 1/2, this is the trace distance.  

An instance of this problem is described by a string of length $\ell = \Theta(4^k m \gamma + s)$.  We say that an algorithm is efficient if it takes time $\poly(\ell)$.  The remarks made earlier about polynomial-time solvability of Local Hamiltonian apply to this problem as well.  We will be interested in instances where $k$ is a constant, $m \leq \poly(n)$ (see above), $\gamma \leq \poly(n)$ and $s \leq \poly(n)$; these instances are described by strings of length $\ell \leq \poly(n)$.  

An important special case is where $k=2$.  We can visualize the system as a graph with nodes $1,\ldots,n$ and edges given by the subsets $C_1,\ldots,C_m$.

\chapter{Consistency of Local Density Matrices is QMA-complete}

\ifthenelse{\boolean{ucsdformat}}{\thispagestyle{chappage}}{}

%


\section{Introduction}

Quantum mechanical systems exhibit many unusual phenomena, such as coherent superpositions and nonlocal entanglement.  It is interesting to compare this with the behavior of classical probabilistic systems.  In a classical system, such as a Markov chain or a graphical model, one may have correlations or dependencies among different parts of the system; in particular, local properties can affect the joint probability distribution of the entire system.  Many quantum systems have a similar flavor, though their behavior is more complicated.  In this paper, we investigate one problem of this kind, and its relationship to the complexity class QMA.  

First, consider a classical problem.  Suppose we have random variables $X_1,\ldots,$ $X_n$, with some unknown joint distribution $D$, and we are given marginal distributions $D_1,\ldots,D_m$, where each $D_i$ describes a subset $C_i$ of the variables.  (We assume that the random variables $X_j$ take on values in some fixed finite set, and the subsets $C_i$ have size at most some constant $k$.)  Does there exist a joint distribution $D$ that matches each of the marginals $D_i$ on the subsets $C_i$?  If so, we say that the marginals $D_i$ are ``consistent.''  

Deciding the consistency of marginal distributions is NP-hard, by a straightforward reduction from 3-coloring.  (We are given a graph $G = (V,E)$.  For each vertex $v \in V$, construct a random variable $X_v$ which takes on values in $\set{r,g,b}$.  For each edge $(u,v) \in E$, specify that the marginal distribution of $X_u$ and $X_v$ must be uniform over the set $\set{r,g,b}^2 \setminus \set{rr,gg,bb}$.  These marginals are consistent iff $G$ is 3-colorable.)  

Now consider the generalization of this problem to quantum states.  (This problem was first suggested to me by Dorit Aharonov, in connection with the class QCMA \cite{A}.)  Suppose we have an $n$-qubit system, and we are given local density matrices $\rho_1,\ldots,\rho_m$, where each $\rho_i$ describes a subset $C_i$ of the qubits.  Does there exist a global state $\sigma$ on all $n$ qubits that matches each of the local states $\rho_i$ on the subsets $C_i$?  If so, we say that the local states $\rho_i$ are ``consistent.''  

We will show that this problem is QMA-complete, where QMA is the quantum analogue of NP.  QMA is the class of languages that have poly-time quantum verifiers, where the witness is allowed to be a quantum state.  QMA arises naturally in the study of quantum computation, and it also has a complete problem, Local Hamiltonian, which is a generalization of $k$-SAT \cite{KSV,AN}.  

Our result is interesting, because we only know of a few QMA-complete problems, and most of them look like universal models of quantum computation.  For instance, the fact that Local Hamiltonian is QMA-complete \cite{KSV,AN,KR,KKR,OT} is closely related to the fact that adiabatic quantum computation is equivalent to the standard quantum circuit model \cite{ADKLLR}.  Other QMA-complete problems such as Identity Check involve properties of quantum circuits \cite{JWB}.  The Consistency problem, however, does not seem to embody any particular model of quantum computation; this will become clearer when we present our reduction from Local Hamiltonian.  

Why are there so few QMA-complete problems, when there is such an astonishing variety of NP-complete problems?  The reason seems to be that the techniques used to show NP-hardness, such as mapping reductions using combinatorial gadgets, break down when we apply them to a ``quantum'' problem like Local Hamiltonian.  For instance, to reduce Local Hamiltonian to the Consistency problem, we would try to use local density matrices to ``simulate'' local Hamiltonians.  But we run into problems due to the presence of non-commuting matrices.  (In cases where quantum gadgets do work, such as \cite{KKR,OT}, they are much more subtle than classical gadgets.)  

Instead, our proof that the consistency problem is QMA-hard uses a poly-time oracle reduction from Local Hamiltonian.  The basic idea is that Local Hamiltonian can be expressed as a convex program in polynomially many variables, which can be solved using convex optimization algorithms, given an oracle for the Consistency problem.  In particular, we use a class of convex optimization algorithms \cite{YN,GLS,BV,KV,Vsurvey} that only require a membership oracle, rather than a separation oracle.  We also use a simple representation of the local density matrices in terms of the expectation values of Pauli observables.  

Note that the Consistency problem has a rather different structure from Local Hamiltonian.  For instance, a local density matrix contains complete information about the local state of the system, whereas in many cases a local Hamiltonian only constrains the local state of the system to lie within a certain subspace.  

Finally, we remark that our reduction from Local Hamiltonian to Consistency preserves the ``neighborhood structure'' of the problem, in that the local density matrices act on the same subsets of qubits as the local terms in the Hamiltonian.  So, using the QMA-hardness results for 2-Local Hamiltonian \cite{KKR} and Local Hamiltonian on a 2-D square lattice \cite{OT}, we can immediately get QMA-hardness results for the corresponding special versions of the Consistency problem.  

We also mention some related work.  In \cite{BravyiVyalyi}, one considers the Common Eigen\-space Problem, verifying the consistency of a set of eigenvalue equations $H_i \ket{\psi} = \lambda_i \ket{\psi}$, where the operators $H_i$ commute.  We do something similar, translating each local density matrix into constraints on the expectation values of Pauli matrices, though in our case the Pauli matrices do not commute.  Also, in \cite{Bravyi}, one considers a quantum analogue of 2-SAT, where we seek a state $\ket{\psi}$ whose local density matrices have support on prescribed subspaces.  However, this problem is more closely related to Local Hamiltonian than to Consistency, since the constraints can be written in the form $\Pi_i \ket{\psi} = 0$ where the $\Pi_i$ are local projectors.  

After our result was published, we became aware of some related work by Gurvits \cite{Gurvits}, who used convex optimization with a membership oracle to show NP-hardness of the separability problem for quantum states.  Also, an older paper by Grotschel et al \cite{GLS-1981} used a simpler tool, convex optimization with a separation oracle, to show NP-hardness of weighted fractional chromatic number.  

This chapter is organized as follows.  First, we show that Consistency is in QMA.  Then we develop the technique of convex optimization with a membership oracle.  We go into considerable detail, because we will use this tool in the following chapters as well.  One particular contribution is to give algorithms for ``approximate'' convex optimization, where one is allowed to make additive errors of size $1/\poly(n)$; these algorithms are much simpler than the algorithms of \cite{YN,GLS,BV,KV}.  Finally, we show that Consistency is QMA-hard, by a reduction from Local Hamiltonian.


\section{Consistency is in QMA}
\label{ch2-in-qma}

\begin{thm}\label{thm-QMA}
Consistency is in QMA.  
\end{thm}

\noindent
Proof sketch:  The basic idea is as follows.  Given a witness state $\sigma$, the verifier will pick a subset $C_i$ at random, and perform measurements to compare $\sigma$ (on the subset $C_i$) to $\rho_i$.  There is a complication, however, because the verifier requires many independent copies of the witness $\sigma$, and a dishonest prover might try to cheat by entangling the different copies.  In spite of this, one can show that the verifier is still sound, using a Markov argument.  This argument is due to Aharonov and Regev, who used it to give an alternative definition of QMA, known as QMA+ \cite{AR}.  Using the QMA+ definition, one can easily see that Consistency is in QMA.  For the sake of clarity, however, we will explicitly construct a QMA verifier for Consistency.  

The verifier works as follows:  
\begin{quote}
Set $\varepsilon = (1/2) (\beta/4^k)$ and $r = (16/\varepsilon^2) \ln(8\cdot 4^km/\varepsilon)$.  (These are polynomially related to the length of the input.)

Given a witness $\tau$, which is a quantum state on $rn$ qubits.  (We view this as $r$ registers, each consisting of $n$ qubits.)  

Choose $i \in \set{1,\ldots,m}$ at random.  Choose a Pauli matrix $Q \in \PP^{\tensor |C_i|}$ (acting on the subset $C_i$) at random.  

Perform the following measurements on $\tau$:  for $j = 1,\ldots,r$, measure the observable $Q$ on the $j$'th register, and let $X_j \in \set{1,-1}$ denote the result.\footnote{One can measure $Q$ using the following procedure:  introduce an ancilla qubit in the state $\ket{0}$, apply a Hadamard gate on the ancilla, apply $Q$ controlled by the ancilla, apply another Hadamard gate on the ancilla, and then measure the ancilla in the 0/1 basis.  The ``0'' and ``1'' measurement outcomes correspond to the $+1$ and $-1$ eigenvalues of $Q$.}  

Compute $Y = (1/r) \sum_{j=1}^r X_j$.  If $|Y - \Tr(Q\rho_i)| \leq \varepsilon$, then output ``YES''; otherwise, output ``NO.''  
\end{quote}

Suppose we have a ``YES'' instance of Consistency, i.e., there exists an $n$-qubit state $\sigma$ such that, for all $i$, $\Tr_{\set{1,\ldots,n}-C_i} (\sigma) = \rho_i$.  Then the correct witness is $\tau = \sigma^{\tensor r}$.  For all choices of $i$ and $Q$, the random variables $X_1,\ldots,X_r$ are i.i.d., with expectation value $E(X_j) = \Tr((Q \tensor I) \sigma) = \Tr(Q \rho_i)$.  We use the Chernoff bound.  The following lemma can be derived from \cite{Motwani-Raghavan}, and gives a simple but not especially tight bound.  
\begin{lem}
Let $X_1,\ldots,X_n$ be independent, 0-1-valued random variables, with $E(X_i) = p_i$, $0 < p_i < 1$.  Let $X = \sum_{i=1}^n X_i$, and let $\mu = E(X) = \sum_{i=1}^n p_i$.  Then, for all $\delta \leq 1$, 
\[
\Pr\Bigl[ \frac{X}{n} < \frac{\mu}{n} - \delta \Bigr] \leq e^{-\delta^2 n/4}, 
\]
\[
\Pr\Bigl[ \frac{X}{n} > \frac{\mu}{n} + \delta \Bigr] \leq e^{-\delta^2 n/4}.  
\]
\end{lem}
Hence 
\[
\Pr[|Y-\Tr(Q\rho_i)| > \varepsilon] \leq 2e^{-\varepsilon^2 r/16}.  
\]
So the verifier rejects with probability $\leq 2e^{-\varepsilon^2 r/16} = (1/4) (\varepsilon/4^km)$.  

Now suppose we have a ``NO'' instance of Consistency, i.e., for all $n$-qubit states $\sigma$, there exists some $i$ such that $\norm{\Tr_{\set{1,\ldots,n}-C_i} (\sigma) - \rho_i}_1 \geq \beta$.  We claim that, for any witness state $\tau$, the verifier rejects.  

Let $\tau^{(j)}$ denote the reduced state for the $j$'th register, and define $\tau^* = (1/r)$ $\sum_{j=1}^r \tau^{(j)}$.  The significance of this state comes from the following two observations:  
\[
E(X_j) = \Tr((Q \tensor I) \tau^{(j)}), 
\]
\[
E(Y) = (1/r) \sum_{j=1}^r E(X_j) = \Tr((Q \tensor I) \tau^*).  
\]

We know there exists some $i$ such that $\norm{\Tr_{\set{1,\ldots,n}-C_i} (\tau^*) - \rho_i}_1 \geq \beta$.  We can write 
\[
\Tr_{\set{1,\ldots,n}-C_i}(\tau^*) - \rho_i 
 = \frac{1}{2^{|C_i|}} \sum_{Q \in \PP^{\tensor |C_i|}} 
   \Bigl( \Tr((Q \tensor I) \tau^*) - \Tr(Q \rho_i) \Bigr) Q.  
\]
By the triangle inequality, 
\[
\norm{\Tr_{\set{1,\ldots,n}-C_i}(\tau^*) - \rho_i}_1 
 \leq \sum_{Q \in \PP^{\tensor |C_i|}} 
 \Bigl| \Tr((Q \tensor I) \tau^*) - \Tr(Q \rho_i) \Bigr|, 
\]
hence there exists some $Q$ such that $|\Tr((Q \tensor I) \tau^*) - \Tr(Q \rho_i)| \geq \beta / 4^{|C_i|}$.  

So, with probability $\geq 1/4^k m$, we will choose some $i$ and $Q$ such that $|E(Y) - \Tr(Q \rho_i)| \geq \beta / 4^k = 2\varepsilon$.  We now use a Markov argument to lower-bound the probability that the verifier rejects.  First, consider the case where $E(Y) \leq \Tr(Q \rho_i) - 2\varepsilon$.  The verifier will accept only if $Y \geq E(Y) + \varepsilon$.  Define $Z = Y+1 \geq 0$.  By Markov's inequality, 
\[
\Pr[Z \geq E(Z) + \varepsilon] 
 \leq \frac{E(Z)}{E(Z)+\varepsilon} = 1 - \frac{\varepsilon}{E(Z)+\varepsilon} 
 \leq 1 - \varepsilon/2.  
\]
Hence, the verifier rejects with probability $\geq (1/2) (\varepsilon/4^km)$.  

Now consider the case where $E(Y) \geq \Tr(Q \rho_i) + 2\varepsilon$.  The verifier will accept only if $Y \leq E(Y) - \varepsilon$.  Define $Z = -Y+1 \geq 0$.  By Markov's inequality, 
\[
\Pr[Z \geq E(Z) + \varepsilon] 
 \leq \frac{E(Z)}{E(Z)+\varepsilon} = 1 - \frac{\varepsilon}{E(Z)+\varepsilon} 
 \leq 1 - \varepsilon/2.  
\]
Hence, the verifier rejects with probability $\geq (1/2) (\varepsilon/4^km)$.  

The gap between the probability that the verifier rejects on a ``NO'' instance and the probability that the verifier rejects on a ``YES'' instance is $\geq (1/4) (\varepsilon/4^km)$.  This gap is inverse polynomial in the size of the input, and it can be amplified via parallel repetition.  $\square$

\section{Convex Optimization using a Membership Oracle}

Convex optimization is the problem of minimizing a convex function $f$ subject to convex contraints, i.e., let $K$ be the set of feasible solutions (which is convex), and find some $x \in K$ that minimizes $f(x)$.  Convex optimization includes linear programming and semidefinite programming as special cases, and has numerous applications in operations research, statistics and other areas \cite{BV}.  Many algorithms are known for convex optimization.  On one hand there are general methods such as the ellipsoid algorithm, which solve general convex programs and are theoretically (if not practically) efficient.  There are also interior-point methods, which typically work on special classes of convex programs (e.g., linear or semidefinite programs), and are efficient in practice.  

We will be concerned with convex programs of the following form:  
\begin{verse}
Let $K \subseteq \RR^n$ be a convex set specified by a membership oracle, i.e., given a point $x$, the oracle tells us whether or not $x$ is in $K$.\\
Assume that $K$ contains a ball of radius $r$ around a known point $p$, and $K$ is contained within a ball of radius $R$ around the origin.\\
Let $f:\: \RR^n \rightarrow \RR$ be a linear function, which is efficiently computable.\\
Find some $x \in K$ that minimizes $f(x)$.
\end{verse}
These programs are quite challenging to solve, because we do not have an explicit description of the convex constraints; we only have an oracle that tells us whether or not a proposed solution is feasible.  Moreover, when the solution is not feasible, the oracle does not give us any additional information (such as a violated constraint or a separating hyperplane) that could help us fix the solution.  (However, we at least have a starting point $p$ which is feasible.)  

Remarkably, there are algorithms that solve these convex programs in polynomial time.  The first such algorithm was the shallow-cut ellipsoid method, due to Yudin and Nemirovskii \cite{YN,GLS}; recently a different algorithm based on random walks in convex bodies was devised by Bertsimas and Vempala \cite{BV,KV}.  These algorithms even give ``exact'' solutions, in the following sense:  if the membership oracle can resolve the boundary of the set $K$ with precision $\pm\delta$ (for any $\delta$) while taking time $\poly(n,\log(1/\delta))$, then the algorithm can find the optimal solution with precision $\pm\varepsilon$ (for any $\varepsilon$) while taking time $\poly(n,\log(R/r),\log(1/\varepsilon))$.  

Our problem is a little different, however.  We are given a weaker membership oracle, that runs in time $\poly(n,(1/\delta))$.  But our goal is also more modest:  we desire an algorithm that finds the optimal solution in time $\poly(n,(R/r),(1/\varepsilon))$.  We refer to this as ``approximate'' convex optimization.  (Intuitively, in the ``approximate'' setting, we are promised that $\delta$ and $\varepsilon$ are at least $1/\poly(n)$, and $R/r$ is at most $\poly(n)$.  (For more discussion of what it means to solve a gap promise problem in polynomial time, see chapter 1.)  Note the contrast with the ``exact'' setting, where $\delta$ and $\varepsilon$ may be exponentially small, and $R/r$ may be exponentially large.)  

In addition, we care about some other aspects of the algorithm.  We will eventually use this to give a reduction from Local Hamiltonian to Consistency; hence the running time is less important (so long as it is polynomial), but we are interested in the relationship between $\delta$ and $\varepsilon$, i.e., for a given value of $\varepsilon$, how small does $\delta$ have to be.  

It turns out that the ``exact'' algorithms mentioned earlier can be adapted to the ``approximate'' setting.  But in fact there are much simpler algorithms in the ``approximate'' setting, for which the relationship between $\delta$ and $\varepsilon$ is just as good, though the running time is larger.  In this section we will describe one such algorithm in detail, and then sketch some of the other more sophisticated methods.  

\vskipline

Now we will define the problem more precisely.  We take a similar approach to \cite{GLS}, though there are some differences which we will discuss presently.  First, some notation:  let $S(p,r)$ denote the closed ball of radius $r$ around the point $p$, 
\[
S(p,r) = \set{x \in \RR^n \;|\; \norm{x-p} \leq r}.  
\]
Also, for any set $K$, we define the ball of radius $\varepsilon$ around $K$, 
\[
S(K,\varepsilon) = \set{x \in \RR^n \;|\; 
\text{there exists $y \in K$ s.t. $\norm{x-y} \leq \varepsilon$}}, 
\]
and we define the interior of $K$ with radius $\varepsilon$, 
\[
S(K,-\varepsilon) = \set{x \in \RR^n \;|\; S(x,\varepsilon) \subseteq K}.
\]

Let $K$ be a closed convex set in $\RR^n$, and suppose we are given a point $p \in \RR^n$, and inner and outer radii $r,R \in \RR$, such that $S(p,r) \subseteq K \subseteq S(0,R)$.  (This implies that $K$ is bounded and full-dimensional.)  We want to show a \textit{reduction} from the problem of optimizing a linear function over $K$, to the problem of deciding membership in $K$.  

In the following sections, we will represent real numbers with $\Gamma$ bits of precision; arithmetic operations will then take time $\poly(\Gamma)$.  (Usually, we will have $\Gamma = \poly(n)$.)

We define the weak optimization problem $WOPT_\varepsilon$ as follows:  (The adjective ``weak'' refers to the fact that we allow additive errors of size $\varepsilon$.)  
\begin{verse}
Given $c \in \RR^n$, $\norm{c} = 1$, $\gamma \in \RR$, and $\varepsilon \in \RR$, $\varepsilon > 0$.\\
If there exists a vector $y \in S(K,-\varepsilon)$ with $c \cdot y \geq \gamma + \varepsilon$, then answer ``YES.''\\
If for all $x \in S(K,\varepsilon)$, $c \cdot x \leq \gamma - \varepsilon$, then answer ``NO.''\\
\end{verse}
We have formulated this as a decision problem, rather than a search problem, because this suffices for our application.  (This is different from the convention used in \cite{GLS}, where $WOPT$ refers to the search problem, and $WVAL$ is the decision problem.  However, the same reductions hold true for both $WVAL$ and $WOPT$.)

We define the weak membership problem $WMEM_\delta$ as follows:
\begin{verse}
Given $y \in \RR^n$, and $\delta \in \RR$, $\delta > 0$.\\
If $y \in S(K,-\delta)$, then answer ``YES.''\\
If $y \notin S(K,\delta)$, then answer ``NO.''\\
\end{verse}

We also define the weak separation problem $WSEP_\delta$ as follows:
\begin{verse}
Given $y \in \RR^n$, and $\delta \in \RR$, $\delta > 0$.\\
If $y \in S(K,-\delta)$, then answer ``YES.''\\
If $y \notin S(K,\delta)$, then return a vector $c \in \RR^n$, $\norm{c} = 1$, such that for every $x \in S(K,-\delta)$, $c \cdot x \leq c \cdot y + \delta$.\\
\end{verse}
This problem is similar to the membership problem, except that when $y$ lies outside of $K$, one is asked to find a hyperplane that separates $y$ from $K$.  This problem serves as an intermediate step in the reduction from $WOPT$ to $WMEM$.  

Finally, we define special versions of these problems that capture the notion of ``approximate'' convex optimization.  We define $WOPT_{\text{1/poly}}$ in the same way as $WOPT_\varepsilon$, except that the input now includes a unary string ``$1^s$,'' such that $\varepsilon \geq 1/s$.  Intuitively, this amounts to a promise that $\varepsilon$ is at least inverse-polynomial in the length of the input.  In a similar way, we define $WMEM_{\text{1/poly}}$ and $WSEP_{\text{1/poly}}$.  Also, when we deal with these problems, we will often assume that $R/r \leq \poly(n)$.  

There are a few differences between our definitions and the ones in \cite{GLS}.  We construct gap promise problems, where the input is promised to fall under one of two (disjoint) cases, and the algorithm must answer ``YES'' or ``NO'' accordingly.  \cite{GLS} uses a different style, where the algorithm must assert either ``$A$ is true'' or ``$B$ is true''; on every input, at least one of them is true, however it is also possible for both $A$ and $B$ to hold simultaneously.  In fact this formulation is equivalent to a promise problem, where ``$A$ and not $B$'' and ``$B$ and not $A$'' are the two disjoint cases, which the algorithm must distinguish.  

Also, unlike here, \cite{GLS} does not make any assumptions about how many bits of precision are used to specify the input; they show that the running time is polynomial in the length of the input, which is not necessarily polynomial in $n$.  Our setting, where the input has $\poly(n)$ bits of precision and the running time is $\poly(n)$, can be viewed as a special case.


\vskipline

Our main result is the following:  

\begin{thm}\label{ch2-thm-opt-mem}
Let $K$ be any closed convex set in $\RR^n$, such that $S(p,r) \subseteq K \subseteq S(0,R)$, as defined above.  Suppose $R/r \leq \poly(n)$.  Then there is a poly-time oracle reduction from $WOPT_{\text{1/poly}}$ to $WMEM_{\text{1/poly}}$.  
\end{thm}

We will prove this theorem in the following sections.  (We will also give more detailed bounds on the various parameters.)  The techniques used for ``exact'' convex optimization \cite{GLS} can be adapted to our ``approximate'' setting.  However, one can give other, simpler reductions in the ``approximate'' case---in particular, one can do away with the ellipsoid method entirely.  The general picture is as follows:

In the ``exact'' setting \cite{GLS}, one can reduce $WOPT$ to $WSEP$ using the central-cut ellipsoid method.  It is not known whether one can reduce $WSEP$ to $WMEM$, but one can reduce $WOPT$ to $WMEM$ via the shallow-cut ellipsoid method.

In the ``approximate'' setting, one can give similar reductions.  This is because the above algorithms have the property that, when $R/r$ is at most polynomial, $\varepsilon$ and $\delta$ are polynomially related.  Alternatively, one can reduce $WOPT_{\text{1/poly}}$ to $WSEP_{\text{1/poly}}$ using a simple perceptron-like algorithm.  Furthermore, one can reduce $WSEP_{\text{1/poly}}$ to $WMEM_{\text{1/poly}}$, using a clever non-ellipsoidal algorithm (this was actually a preprocessing step in the shallow-cut ellipsoid method).  Combining these steps gives a simpler reduction from $WOPT_{\text{1/poly}}$ to $WMEM_{\text{1/poly}}$, for which the relationship between $\varepsilon$ and $\delta$ is just as good, but the running time is larger.  

Finally, there are the algorithms based on random walks \cite{BV,KV}.  These are notable for a couple of reasons.  First, they can solve convex programs where the objective function $f$ is not linear.  Roughly speaking, one needs a membership oracle for the set $K$, and a separation oracle for the level sets of $f$ (which one could obtain by computing the gradient of $f$).  We will not need this extra degree of generality here.  

Second, these algorithms have a simple error-tolerance property, which is quite different from the ellipsoid method.  The intuition is as follows.  These algorithms work by performing a random walk inside the set $K$, which converges to the uniform distribution.  The membership oracle makes mistakes near the boundary of $K$.  If this ``boundary layer'' is sufficiently thin, then its volume will be small compared to the total volume of $K$, and so with significant probability, the random walk will never visit that part of the set.  

These random-walk algorithms might in some cases achieve a better relationship between $\varepsilon$ and $\delta$, compared to the shallow-cut ellipsoid method.  It would be interesting to carry out this analysis in detail.


\subsection{A Simple Reduction}

We will give a simple reduction from $WOPT$ to $WMEM$ in the approximate setting.  

Note:  All calculations are done with $\poly(n)$ bits of precision.  However, in order to give a more streamlined exposition, in this section we assume that all arithmetic operations yield exact results.  Later, in section \ref{ch2-rounding}, we will analyze the effect of round-off errors.  

We present the reduction in several steps.  First, consider a variant of the weak membership problem with 1-sided error (call it $WMEM^1_\delta$):  
\begin{verse}
Given $y \in \RR^n$, and $\delta \in \RR$, $\delta > 0$, all specified with $\poly(n)$ bits of precision.\\
Distinguish between the following two cases:\\
If $y \in K$, then answer ``YES.''\\
If $y \notin S(K,\delta)$, then answer ``NO.''\\
\end{verse}

\begin{lem}\label{ch2-asr-mem1-mem}
(This is Lemma 4.3.3 in \cite{GLS}.)  
There exists an algorithm $A$ and a polynomial $t$, 
such that for any convex set $K$ with parameters $(n,R,r,p)$ as defined above, 
and for any $\delta>0$, there exists $\delta' \geq r\delta/4R$, 
such that $A((n,R,r,p),\ldots)$ is an oracle reduction from $WMEM^1_\delta$ to $WMEM_{\delta'}$, which runs in time $t(n,\log(R))$.  
\end{lem}

\noindent
Proof:  The algorithm $A$ is as follows:  
\begin{verse}
Given $(n,R,r,p)$ as defined above, $y \in \RR^n$, $\delta>0$.\\
If $\norm{y-p} \geq 2R$, then answer ``NO.''\\
Run the $WMEM_{\delta'}$ oracle on the point $y' = (1-\delta/4R)y + (\delta/4R)p$,\\
and return the answer given by the oracle.\\
\end{verse}
The analysis is straightforward; see \cite{GLS} for details.  $\square$

\vskipline

Next, consider a variant of the weak separation problem with parameter $\beta$ (call this $WSEP^\beta_\delta$):  
\begin{verse}
Given a point $y \in \RR^n$, $0<\delta<1$, and $0<\beta<1$, specified with $\poly(n)$ bits of precision.\\
If $y \in S(K,-\delta)$, answer ``YES.''\\
If $y \notin S(K,\delta)$, return a vector $c \in \RR^n$, $\norm{c} = 1$, such that for every $x \in K$, $c \cdot x \leq c \cdot y + \delta + \beta \norm{x-y}$.\\
\end{verse}
Intuitively, we now have a weaker form of separation:  instead of a separating hyperplane, we have a cone with slope $\beta$.  Points $x \in K$ that are far away from $y$ can violate the inequality $c \cdot x \leq c \cdot y + \delta$ by an amount proportional to $\norm{x-y}$.  

\begin{lem}\label{ch2-asr-sepb-mem1}
(This is Lemma 4.3.4 in \cite{GLS}.)  
There exists an algorithm $A$ and a polynomial $t$, 
such that for any convex set $K$ with parameters $(n,R,r,p)$ as defined above, 
and for any $0<\delta<1$ and $0<\beta<1$, there exists $\varepsilon \geq \beta^2r^2\delta / 128n^5R^2$, 
such that $A((n,R,r,p),\ldots)$ is an oracle reduction from $WSEP^\beta_\delta$ to $WMEM^1_\varepsilon$, which runs in time $t(n,(1/\beta),\log(R/r),\log(1/\delta))$.  
\end{lem}

\noindent
Proof:  The algorithm is as follows:  
\begin{verse}
Given $(n,R,r,p)$ as defined above, $y \in \RR^n$, $0<\delta<1$, $0<\beta<1$.\\
Run the $WMEM^1_\varepsilon$ oracle on the point $y$.  If the oracle answers ``YES,'' then return ``YES.''\\
Define $\delta_1 = \frac{r}{R+r} \delta$, $r_1 = \frac{r}{4nR} \delta_1$, $\varepsilon = \varepsilon_1 = \frac{\beta^2}{16n^4} r_1$, and $\alpha = \arctan(\beta/4n^2)$.\\
Do binary search to find two points $v$ and $v'$ on the line segment connecting $y$ and $p$, such that $v$ is closer to $y$, $v'$ is closer to $p$, the $WMEM^1_\varepsilon$ oracle answers ``NO'' at $v$ and ``YES'' at $v'$, and $\norm{v-v'} \leq \delta_1/(2n)$.  Then define $v'' = \frac{1}{r+\varepsilon_1} ((r-r_1)v' + (r_1+\varepsilon_1)p)$.  Translate the coordinate system so that $v'' = 0$.\\
Repeat the following procedure:\\
\tab Let $H$ be the $(n-1)$-dimensional hyperplane perpendicular to $v$ and containing the point $(\cos^2 \alpha)v$.  Let $v_1,\ldots,v_n$ be the vertices of a regular simplex in $H$, centered at $(\cos^2 \alpha)v$, such that for all $i=1,\ldots,n$, the angle between $v_i$ and $v$ equals $\alpha$.  (Note that $\norm{v_i} = (\cos \alpha) \norm{v}$.)\\
\tab Run the $WMEM^1_\varepsilon$ oracle at each of the points $v_1,\ldots,v_m$.  If the oracle returns ``NO'' on some of the $v_i$, then choose one such $v_i$, set $v := v_i$ (replacing the previous value of $v$), and go back to the beginning of the loop.\\
\tab If the oracle returns ``YES'' on all of the $v_i$, then break out of the loop, and return the vector $c = v/\norm{v}$.\\
\end{verse}
The analysis of this algorithm is rather intricate.  We will sketch the general ideas; details can be found in \cite{GLS}.  

First, we run the $WMEM^1_\varepsilon$ oracle on the point $y$.  If this is a ``YES'' instance of the problem, then we are done.  If this is a ``NO'' instance of the problem, then we proceed to the remainder of the algorithm; furthermore, we can conclude that $y \notin K$.  

Note that the angle $\alpha$ is defined by a right triangle with side lengths $\sqrt{r_1}$ and $\sqrt{\varepsilon_1}$:  
\begin{center}
\psfrag{alpha}{$\alpha$}
\psfrag{a}{$\sqrt{r_1}$}
\psfrag{b}{$\sqrt{\varepsilon_1}$}
\psfrag{c}{$\sqrt{r_1+\varepsilon_1}$}
\includegraphics{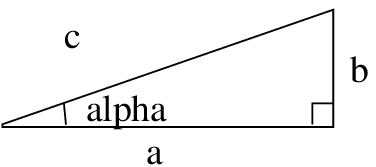}
\end{center}

The binary search produces two points $v$ and $v'$ such that $\norm{v-v'} \leq \delta_1/(2n)$, $v \notin K$ and $v' \in S(K,\varepsilon_1)$.  We construct a point $v''$ that satisfies $S(v'',r_1) \subseteq K$ (this can be seen by a duality argument\footnote{Take any $d \in \RR^n$ and $\gamma \in \RR$ such that $\norm{d} = 1$ and all $x \in K$ satisfy $d \cdot x \leq \gamma$.  Then observe that $d \cdot v' \leq \gamma + \varepsilon_1$ and $d \cdot p \leq \gamma - r$.  This implies $d \cdot v'' \leq \gamma - r_1$.}), and $\norm{v-v''} < \delta_1/n$.  When we translate the coordinates so that $v'' = 0$, we get that $S(0,r_1) \subseteq K$ and $\norm{v} < \delta_1/n$.  

Next we do an iterative procedure that continues until it finds a simplex $v_1,\ldots,$ $v_n \in S(K,\varepsilon_1)$, where the simplex was constructed from a vector $v \notin K$.  Let $p$ denote the number of iterations; we can upper-bound it as follows.  Note that with every iteration, $\norm{v}$ decreases by a factor of $(\cos \alpha)$.  Initially, $\norm{v} < \delta_1$, and the loop must terminate as soon as $\norm{v} < r_1$, since $S(0,r_1) \subseteq K$.  Then $p$ must satisfy the inequality $(\cos\alpha)^p \, \delta_1 > r_1$.  This implies 
\[
p < \frac{\log(r_1/\delta_1)}{\log(\cos\alpha)}
  = \frac{\log(\delta_1/r_1)}{\log(1/\cos\alpha)}.
\]
Observe that $\log(\delta_1/r_1) = \log(4nR/r)$, and 
\[
\begin{split}
\log(1/\cos\alpha)
 &= -\tfrac{1}{2} \log(1-\sin^2\alpha)\\
 &\geq \tfrac{1}{2} \, (\sin^2\alpha) \quad \text{[since $\log(1+x) \leq x$ for all $x$]}\\
 &= \frac{1}{2} \, \frac{\beta^2}{\beta^2+16n^4} \quad \text{[by the definition of $\alpha$]}\\
 &\geq \frac{1}{2} \, \frac{\beta^2}{17n^4}.  
\end{split}
\]
Hence the number of iterations $p$ is at most $\poly(n, \log(R/r), (1/\beta))$.  

We claim that $c = v/\norm{v}$ has the desired property, namely that for all $x \in K$, 
\begin{equation}
\label{wsepb-eqn1}
c \cdot x \leq c \cdot y + \delta + \beta \norm{x-y}.  
\end{equation}
Consider the following simpler statement, that for all $x \in K$, 
\begin{equation}
\label{wsepb-eqn2}
c \cdot x \leq \beta \norm{x} + \delta_1.  
\end{equation}

First, we show that (\ref{wsepb-eqn2}) implies (\ref{wsepb-eqn1}).  Given some $x \in K$, consider the point 
\[
x' = \frac{r}{R+r} (x-y) = \frac{r}{R+r} x + \frac{R}{R+r} \frac{r}{R} (-y).
\]
We claim that $x' \in K$.  Geometrically, the picture is as follows:  
\begin{center}
\psfrag{vpp}{$v'' = 0$}
\psfrag{y}{$y$}
\psfrag{x}{$x$}
\psfrag{xmy}{$x-y$}
\psfrag{xp}{$x'$}
\psfrag{rRmy}{$(r/R)(-y)$}
\psfrag{p}{$p$}
\includegraphics{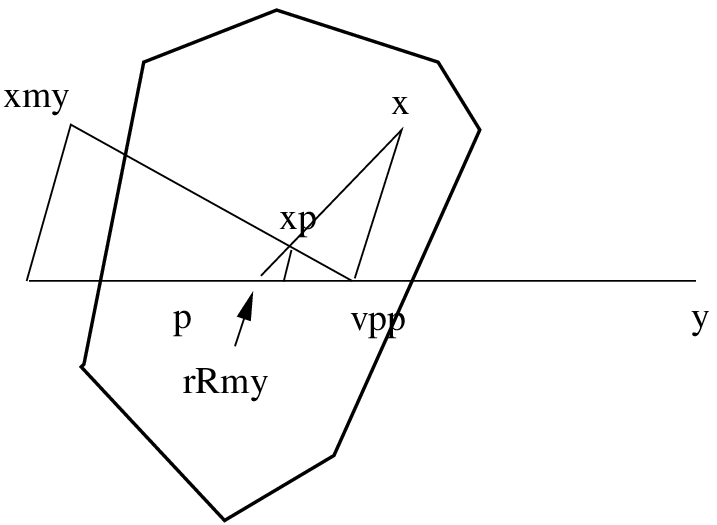}
\end{center}
The vector $x'$ is proportional to $x-y$, and is a convex combination of $x$ and $(r/R)(-y)$.  $(r/R)(-y)$ lies along the line $py$.  In our picture, $y$ is on the right of $v'' = 0$, and $p$ is on the left.  So $(r/R)(-y)$ is on the left of $v'' = 0$.  Also, without loss of generality, $\norm{y} \leq R$, so $(r/R)(-y)$ is on the right of $p - r (y/\norm{y})$.  Hence, by convexity, $(r/R)(-y) \in K$, and this implies $x' \in K$.  

Now substitute $x'$ into (\ref{wsepb-eqn2}); this yields (\ref{wsepb-eqn1}), as desired.  

Next, we will show that (\ref{wsepb-eqn2}) holds, i.e., that for all $x \in K$, 
\[
c \cdot x \leq \beta \norm{x} + \delta_1.  
\]
Define $v'_i = \frac{r_1}{\varepsilon_1+r_1} v_i$.  Observe that $v'_1,\ldots,v'_n \in K$ (this follows from the fact that $S(0,r_1) \subseteq K$ and a duality argument).  Also note that $\frac{r_1}{\varepsilon_1+r_1} = \cos^2 \alpha$.  Define $w = (1/n) \sum_{i=1}^n v'_i$, and note that $w = \gamma v$ where we define $\gamma = \cos^4 \alpha$.  

We write $x$ in the form $x = \lambda v + u$, where $u \cdot v = 0$.  Notice that $c \cdot x = \frac{v}{\norm{v}} \cdot x = \lambda \norm{v}$; also recall that $\norm{v} \leq \delta_1/n$.  If $\lambda \leq 1$, then the claim follows easily.  However, if $\lambda > 1$, we need a more clever argument.

If $\lambda > 1$, then the geometric picture is as follows:  
\begin{center}
\psfrag{K}{$K$}
\psfrag{vpp}{$v''=0$}
\psfrag{vpi}{$v'_i$}
\psfrag{w}{$w$}
\psfrag{v}{$v$}
\psfrag{lambdav}{$\lambda v$}
\psfrag{x}{$x$}
\psfrag{z}{$z$}
\includegraphics{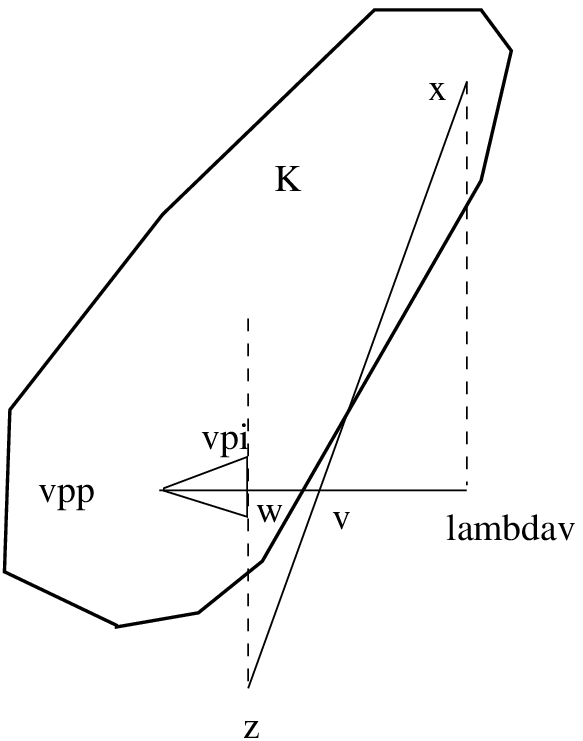}
\end{center}
We draw a line through $x$ and $v$.  This line intersects the hyperplane $w + v^\perp$ at some point; call this point $z$.  We will make the following argument.  Since $x \in K$ and $v \notin K$, we know that $z \notin K$.  Thus, within the hyperplane $w + v^\perp$, $z$ cannot lie within the simplex generated by $v'_1,\ldots,v'_n$.  Thus $z$ must be far from $w$, so $x$ must be far from $\lambda v$, which implies that $\norm{x}$ is large and $\lambda$ is relatively small.  

This can be made precise as follows (see \cite{GLS} for the step-by-step details).  We can rewrite $x = \lambda v + u$ in the form 
\[
\gamma v + \frac{\gamma-1}{\lambda-1} u = \frac{\gamma-1}{\lambda-1} x + \frac{\lambda-\gamma}{\lambda-1} v.  
\]
Then we set $z$ equal to either side of this equation.  Using the above geometric argument, we deduce a lower bound on $\norm{z-w}$, 
\[
\norm{z-w} \geq (1/n) \norm{v'_1-w} = (1/n) (\tan \alpha) \norm{w}.
\]
From the definition of $z$, we have that $u = \frac{\lambda-1}{\gamma-1} (z-\gamma v)$.  Also recall that $w = \gamma v$.  Hence 
\[
\norm{u} \geq \frac{|\lambda-1|}{|\gamma-1|} (1/n) (\tan \alpha) \gamma \norm{v}.  
\]
After some manipulation, this yields the bound 
\[
\norm{u} \geq (\lambda-1) \norm{v} \frac{n}{\beta}, 
\]
which implies 
\[
\lambda-1 \leq \frac{\norm{u}}{\norm{v}} \frac{\beta}{n}
 \leq \frac{\norm{x}}{\norm{v}} \frac{\beta}{n}.  
\]
Substitute this into $c \cdot x = \frac{v}{\norm{v}} \cdot x = \lambda \norm{v}$ and the claim follows.  

Note:  the reader may have noticed that we proved an inequality that is stronger than (\ref{wsepb-eqn2}), by a factor of $1/n$ on the right hand side.  This has to do with a slight difference between our algorithm and the one in \cite{GLS}.  Our algorithm outputs $c = v/\norm{v}$, whereas the algorithm in \cite{GLS} outputs $c = v/\norm{v}_\infty$.  By not fully normalizing $c$, they avoid some potential problems with numerical precision; however, this is only a concern when one is doing ``exact'' convex optimization.  $\square$

\vskipline

In fact, with a slight modification, the above algorithm solves the $WSEP$ problem in the approximate setting (but not in the exact setting, because the running time is polynomial in $1/\beta$, not $\log(1/\beta)$).  Thus, by combining the previous two lemmas, we can get a reduction from $WSEP$ to $WMEM$ in the approximate setting.

\begin{lem}\label{ch2-asr-sep-mem}
There exists an algorithm $B$ and a polynomial $t$, 
such that for any convex set $K$ with parameters $(n,R,r,p)$ as defined above, 
and for any $0<\varepsilon<1$, there exists $\delta \geq r^3\varepsilon^3/16384n^5R^5$, 
such that $B((n,R,r,p),\ldots)$ is an oracle reduction from $WSEP_\varepsilon$ to $WMEM_\delta$, which runs in time $t(n,\log(1/r),(R/\varepsilon))$.  
\end{lem}

\noindent
Proof:  Using the previous two lemmas, we can give a reduction from $WSEP^\beta_{\varepsilon/2}$ to $WMEM_\delta$.  Then set $\beta = \varepsilon/(4R)$.  Modify the algorithm so that it first checks if $\norm{y} > R$, and if so, returns $c := y/\norm{y}$.  This algorithm correctly solves the $WSEP_\varepsilon$ problem:  If $y \in S(K,-\varepsilon/2)$, it answers ``YES.''  If $y \notin S(K,\varepsilon/2)$ and $\norm{y} > R$, then $c = y/\norm{y}$ defines a separating hyperplane (since for all $x \in K$, $\norm{x} \leq R$).  If $y \notin S(K,\varepsilon/2)$ and $\norm{y} \leq R$, then we have that for every $x \in K$, $c \cdot x \leq c \cdot y + \varepsilon/2 + 2R \beta \leq c \cdot y + \varepsilon$.  $\square$

\vskipline

Next, one can use a simple perceptron-like algorithm to reduce from $WOPT$ to $WSEP$ in the approximate setting.

\begin{lem}\label{ch2-asr-opt-sep}
There exists an algorithm $C$ and a polynomial $t$, 
such that for any convex set $K$ with parameters $(n,R,r,p)$ as defined above, 
and for any $\varepsilon > 0$, there exists $\delta \geq \varepsilon/3$, 
such that $C((n,R,r,p),\ldots)$ is an oracle reduction from $WOPT_\varepsilon$ to $WSEP_\delta$, which runs in time $t(n,R,(1/\varepsilon))$.  
\end{lem}

\noindent
Proof:  Assume we have an oracle for $WSEP_\delta$; we will specify $\delta$ later in the proof.  We wish to construct an algorithm $C$ that solves $WOPT_\varepsilon$.  Let $c$, $\gamma$ and $\varepsilon$ be given.  

Define the set 
\[
K'(c,\gamma) = K \cap \set{x \in \RR^n \;|\; c \cdot x \geq \gamma}.  
\]
Clearly $K'(c,\gamma)$ has outer radius $R$.  We have to distinguish between the following two cases:  (1) If there exists a vector $y \in S(K,-\varepsilon)$ with $c \cdot y \geq \gamma + \varepsilon$, then $K'(c,\gamma)$ contains a ball of radius $\varepsilon$.  (2) If for all $x \in S(K,\varepsilon)$, $c \cdot x \leq \gamma - \varepsilon$, then $K'(c,\gamma)$ is empty.

We can construct a $WSEP_\delta$ oracle for $K'(c,\gamma)$ as follows:
\begin{verse}
Given $y \in \RR^n$.\\
Run the $WSEP_\delta$ oracle for $K$ on input $y$.\\
If the oracle returns a separating hyperplane $s$, then return $s$.\\
Else, if $c \cdot y < \gamma$, then return $-c$.\\
Else, return ``YES.''\\
\end{verse}

Now we construct the following algorithm $C$ that solves $WOPT_\varepsilon$.  (This is essentially the same as the classical perceptron algorithm.)
\begin{verse}
Given $c \in \RR^n$, $\gamma \in \RR$ and $\varepsilon > 0$.\\
Initialize $z = (0,...,0) \in \RR^n$.\\
Repeat the following at most $R^2/(\varepsilon-2\delta)^2$ times.\\
\tab Run the $WSEP_\delta$ oracle for $K'(c,\gamma)$ on input $z$.\\
\tab If the oracle returns ``YES,'' then return ``YES.''\\
\tab Else, the oracle returns a separating hyperplane $s$.  Set $z = z - (\varepsilon-2\delta) s$.\\
If the oracle never returned ``YES,'' then return ``NO.''\\
\end{verse}

Also, we set $\delta = \varepsilon/3$.  It is straightforward to see that this algorithm runs in time $\poly(n,R,(1/\varepsilon))$.  It remains to show that the algorithm correctly solves the $WOPT_\varepsilon$ problem.  

First, consider case (2):  for all $x \in S(K,\varepsilon)$, $c \cdot x \leq \gamma - \varepsilon$.  Then the $WSEP_\delta$ oracle for $K'(c,\gamma)$ will never answer ``YES''; if it did, that would imply $y \in S(K'(c,\gamma),\delta)$, and thus, $y \in S(K,\delta)$ and $c \cdot y \geq \gamma - \delta$, a contradiction.  Therefore, the algorithm returns ``NO.''  

Now consider case (1):  there exists a vector $y \in S(K,-\varepsilon)$ with $c \cdot y \geq \gamma + \varepsilon$.  Thus $K'(c,\gamma)$ contains a ball of radius $\varepsilon$ centered around $y$.  Let $z_t$ denote the value of $z$ after the $t$'th iteration of the algorithm.  Consider what happens on the $(t+1)$'st iteration.  If the $WSEP_\delta$ oracle for $K'(c,\gamma)$ returns ``YES,'' then the algorithm returns ``YES,'' as desired.  Otherwise, the oracle returns a vector $s$ such that for every $x \in S(K'(c,\gamma),-\delta)$, $s \cdot x \leq s \cdot z_t + \delta$.  If we consider the case of $x = y + (\varepsilon-\delta)s$, we see that $s \cdot y + \varepsilon - \delta \leq s \cdot z_t + \delta$.  In other words, 
\[
s \cdot (y-z_t) \leq -\varepsilon + 2\delta.  
\]
This implies that $z_{t+1}$ will be closer to $y$ than $z_t$ was.  In particular, 
\begin{align*}
\norm{z_{t+1}-y}^2
 &= \norm{z_t-y}^2 - 2 (z_t-y) \cdot (\varepsilon-2\delta) s + \norm{(\varepsilon-2\delta)s}^2\\
 &\leq \norm{z_t-y}^2 - 2 (\varepsilon-2\delta)^2 + (\varepsilon-2\delta)^2\\
 &= \norm{z_t-y}^2 - (\varepsilon-2\delta)^2.  
\end{align*}
We know that our starting point $z_0$ was not too far from $y$, specifically, $\norm{z_0-y}^2 \leq R^2$.  Thus, after at most $R^2/(\varepsilon-2\delta)^2$ iterations, the algorithm will find the point $y$ and return ``YES.''

Thus, the algorithm correctly solves the $WOPT_\varepsilon$ problem.  $\square$

\vskipline

Combining all of these steps, we get a reduction from $WOPT$ to $WMEM$.  

\begin{prop}\label{ch2-asr-opt-mem}
There exists an algorithm $A$ and a polynomial $t$, 
such that for any convex set $K$ with parameters $(n,R,r,p)$ as defined above, 
and for any $0<\varepsilon<1$, there exists $\delta \geq r^3\varepsilon^3 / 442368n^5R^5$, 
such that $A((n,R,r,p),\ldots)$ is an oracle reduction from $WOPT_\varepsilon$ to $WMEM_\delta$, which runs in time $t(n,R,(1/\varepsilon),\log(1/r))$.
\end{prop}

\noindent
Proof:  This follows from Lemmas \ref{ch2-asr-opt-sep} and \ref{ch2-asr-sep-mem}.  $\square$

\vskipline

This directly implies Theorem \ref{ch2-thm-opt-mem}.  

A few remarks about the precision requirement, i.e., the dependence of $\delta$ on $\varepsilon$.  First, the constant factor of 442368 can be substantially improved by doing a more careful analysis.  But it is less clear whether one can improve on the overall form of the expression $r^3\varepsilon^3 / n^5R^5$.  Note that this expression comes mostly from the reduction from $WSEP^\beta$ to $WMEM$.  This step also appears in the more sophisticated reductions based on the ellipsoid method; so the precision requirement for those reductions is comparable.  On the other hand, it may be possible to improve on the precision requirement by using algorithms based on random walks instead; this would give a randomized (rather than deterministic) reduction.  

\subsection{Round-off Errors}
\label{ch2-rounding}

We now consider the effect of round-off errors in the algorithms described above.  We claim that if we do all calculations with $\poly(n)$ bits of precision, then the errors are negligible.  Since we are doing ``approximate'' convex optimization, rather than ``exact,'' our situation is much less delicate than the one in \cite{GLS}.  

\vskipline

First, some general remarks:  We represent numbers using $\poly(n)$ bits of precision.  For simplicity, we use fixed-point notation, where the position of the decimal point is fixed.  This is less powerful than floating-point notation, but it suffices for our needs.  See \cite{Knuth-vol2} for a detailed discussion of how to implement the basic arithmetic operations.  

If the algorithm returns some number $r'$, and the true answer is $r$, we want to bound the \textit{absolute error}, i.e., we want to show that $|r-r'| \leq \varepsilon$.  (Alternatively, one could bound the relative error, i.e., $|r-r'| \leq \varepsilon |r|$.  But this is less useful for our purposes.)  

Errors come from various sources.  When we round a number to $\poly(n)$ bits of precision, the absolute error increases by $2^{-\poly(n)}$, which is not too serious; the real concern is that subsequent arithmetic operations can amplify the error.  

The absolute error behaves well under addition and subtraction, but can blow up after multiplication by a very large number or division by a very small number.  In particular, if $|r-r'| \leq \varepsilon$ and $|s-s'| \leq \delta$, then we have the following bounds:  
\[
|(r+s) - (r'+s')| \leq \varepsilon + \delta, 
\]
\[
|(r-s) - (r'-s')| \leq \varepsilon + \delta, 
\]
\[
|rs - r's'| = |r(s-s') + (r-r')s'| \leq |r|\delta + \varepsilon|s| + \varepsilon\delta, 
\]
\[
\Bigl| \frac{r}{s} - \frac{r'}{s'} \Bigr| = \Bigl| \frac{r(s'-s)+(r-r')s}{ss'} \Bigr|
 \leq \frac{|r|\delta+\varepsilon|s|}{|ss'|}
 = \Bigl( \Bigl|\frac{r}{s}\Bigr| \delta + \varepsilon \Bigr) \Bigl|\frac{1}{s'}\Bigr|.  
\]

In addition to the usual arithmetic operations, we will occasionally need to calculate the square root.  This can be done using Newton's method, or just binary search.  (Given a number $r \geq 0$, we want to find some $t \geq 0$ such that $t^2-r = 0$.)  The behavior of the absolute error depends on the magnitude of $r$---it can blow up when $r$ is very small.  

In particular, suppose $r \geq 0$, $r' \geq 0$, $|r-r'| \leq \varepsilon$.  In the case where $r \leq r'$, we have that $\sqrt{r'} \leq \sqrt{r} + \frac{r'-r}{2\sqrt{r}}$ (this follows from the concavity of the square root function, and taking the first derivative at the point $r$).  Thus $\sqrt{r'}-\sqrt{r} \leq \frac{\varepsilon}{2\sqrt{r}}$.  A similar argument applies in the case where $r \geq r'$.  So we have the general bound 
\[
|\sqrt{r'}-\sqrt{r}| \leq \frac{\varepsilon}{2\sqrt{\min(r,r')}}.  
\]

\vskipline

Now we consider the algorithms described in the previous section.  

\vskipline

In lemma \ref{ch2-asr-mem1-mem}, the reduction from $WMEM^1$ to $WMEM$ is quite straightforward.  We are multiplying and dividing numbers whose magnitude is order $R$, so we need order $\log(R)$ bits of precision.  

\vskipline

In lemma \ref{ch2-asr-sepb-mem1}, the reduction from $WSEP^\beta$ to $WMEM^1$ is much more complicated, because of the iterative procedure where, on every round, one constructs a simplex $v_1,\ldots,v_n$ centered around a given vector $v$.  First, let us describe one procedure for constructing the simplex.  
\begin{verse}
Take the standard basis vectors $e_1,\ldots,e_n \in \RR^n$, where $e_i = (0,\ldots,0,1,0,\ldots,$ $0)$, with a 1 in the $i$'th coordinate.  These vectors define a regular simplex in the $(n-1)$-dimensional hyperplane $\set{x \in \RR^n \;|\; u \cdot x = 1}$, where $u = (1,1,\ldots,1)$.\\
Define $\hat{u} = u/\norm{u}$ and $\hat{v} = v/\norm{v}$, and apply a rotation $Q$ that maps $\hat{u}$ to $\hat{v}$.  $Q$ is given by the formula $Q = A+I-P$, where $A$ is the desired rotation within $\text{span}(\hat{u},\hat{v})$, and $P$ is a projector onto $\text{span}(\hat{u},\hat{v})$.  We construct $A$ and $P$ as follows.  Define $w = \hat{v} - (\hat{v} \cdot \hat{u}) \hat{u}$, and $\hat{w} = w/\norm{w}$.  Then $\hat{u}$ and $\hat{w}$ form an orthonormal basis for $\text{span}(\hat{u},\hat{v})$, and we can write $\hat{v} = \alpha\hat{u} + \beta\hat{w}$, or equivalently, $\hat{w} = (1/\beta)(\hat{v}-\alpha\hat{u})$.  (Note:  in this paragraph only, $\alpha$ and $\beta$ have a completely different meaning from the $\alpha$ and $\beta$ used elsewhere in the algorithm.)  We define 
\[
\begin{split}
A &= \hat{v}\hat{u}^T + (-\beta\hat{u}+\alpha\hat{w}) \hat{w}^T\\
  &= \hat{v}\hat{u}^T + (-\beta\hat{u}+(\alpha/\beta)(\hat{v}-\alpha\hat{u})) \hat{w}^T\\
  &= \hat{v}\hat{u}^T + (1/\beta) (-\hat{u} + \alpha\hat{v}) \hat{w}^T\\
  &= \hat{v}\hat{u}^T + (1/\beta^2) (-\hat{u} + \alpha\hat{v}) (\hat{v}-\alpha\hat{u})^T.  
\end{split}
\]
And we define 
\[
\begin{split}
P &= \hat{u}\hat{u}^T + \hat{w}\hat{w}^T\\
  &= \hat{u}\hat{u}^T + (1/\beta^2) (\hat{v} - \alpha\hat{u}) (\hat{v} - \alpha\hat{u})^T.  
\end{split}
\]\\
Finally, we scale the simplex so it has the correct shape.  Currently, the center of the simplex lies at distance $1/\sqrt{n}$ from the origin, and the vertices are at distance $\sqrt{1-(1/n)}$ from the center.  We want these distances to be $(\cos^2\alpha) \norm{v}$ and $(\sin\alpha\cos\alpha) \norm{v}$, respectively.  To accomplish this, we apply the transformation 
\[
T = \sqrt{n} (\cos^2\alpha) \norm{v} \hat{v}\hat{v}^T
    + (1-(1/n))^{-1/2} (\sin\alpha\cos\alpha) \norm{v} (I-\hat{v}\hat{v}^T), 
\]
where $\sin\alpha$ and $\cos\alpha$ are obtained from the formulas 
\[
\sin\alpha = \frac{1}{\sqrt{1+16n^4/\beta^2}}, \quad
\cos\alpha = \frac{1}{\sqrt{1+\beta^2/16n^4}}.  
\]\\
\end{verse}

There are a few places where trouble could occur.  First, if the vector $v$ is small, then $\norm{v}$ may have a large error.  However, we know that the algorithm must stop iterating when $\norm{v} < r_1$, so $v$ cannot be too small.  

The second difficulty occurs when we construct the rotation $Q$.  If $\hat{u}$ and $\hat{v}$ are close together, then the vector $w$ will be small, so $\hat{w}$ may have a large error; furthermore, $\beta$ will be small, so expressions containing a $(1/\beta)$ factor may have a large error.  However, in this case one can avoid the problem by simply skipping this step, and not applying any rotation $Q$.  Note that the vertices of the simplex, $e_1,\ldots,e_n$, are at distance 1 from the origin.  If $\norm{\hat{u}-\hat{v}} \leq \varepsilon$, then the correct rotation would move each point $e_i$ by a distance of at most $\varepsilon$; so omitting the rotation only increases the error by $\varepsilon$.  

Finally, there is the question of how these errors in constructing the simplex affect the correctness and running time of the iterative procedure in lemma \ref{ch2-asr-sepb-mem1}.  To maintain correctness, we must ensure that, even with the errors, each vertex $v_i$ satisfies the following two properties:  $v_i \cdot \hat{v} \geq (\cos^2\alpha) \norm{v}$, and the angle between $v_i$ and $\hat{v}$ is at least $\alpha$.  We can accomplish this by slightly adjusting each point $v_i$ in such a way that the simplex moves away from the origin and expands outward.  (One can imagine many ways to do this adjustment; the details are not important.)  

This, of course, hurts the running time---because of the adjustments, the vectors $v_i$ may not shrink as quickly, so the algorithm may need to perform more iterations.  However, if we use polynomially many bits of precision, then the adjustments will be sufficiently small, so that the vectors $v_i$ will shrink quickly and the algorithm will need at most polynomially many iterations.  In particular, if an adjustment moves a point $v_i$ by a distance at most $\eta$, then we have that 
\[
\norm{v_i} \leq (\cos\alpha) \norm{v} + \eta \leq (\cos\alpha + \tfrac{\eta}{r_1}) \norm{v}.  
\]
We can bound the number of iterations $p$, using the same argument as before:  
\[
p < \frac{\log(\delta_1/r_1)}{\log(1/(\cos\alpha + \tfrac{\eta}{r_1}))}, 
\]
and one can show that 
\[
\log(1/(\cos\alpha + \tfrac{\eta}{r_1})) \geq \frac{1}{2} \Bigl(\frac{\beta^2}{17n^4} - \frac{3\eta}{r_1}\Bigr).  
\]
Using polynomially many bits of precision, we can easily ensure that $\eta/r_1 \leq (1/6)$ $(\beta^2/17n^4)$; then $\log(1/(\cos\alpha + \tfrac{\eta}{r_1})) \geq (1/4) (\beta^2/17n^4)$, which means that the algorithm will need at most polynomially many iterations.  

\vskipline

Finally, consider the reduction from $WOPT$ to $WSEP$ in lemma \ref{ch2-asr-opt-sep}.  This algorithm also involves an iterative procedure, but it is quite straightforward; note that after each iteration, the vector $z$ is updated by an addition operation, so the errors accumulate gradually without blowing up.  

\subsection{The Ellipsoid Method}

In place of the perceptron algorithm, one can use the standard (central-cut) ellipsoid method \cite{GLS} to reduce $WOPT$ to $WSEP$.  This gives a faster running time which is logarithmic in $R/r$ and $1/\varepsilon$, while the precision requirement is comparable to what we had before.  

\begin{prop}\label{ch2-ccem-opt-sep}
There exists an algorithm $A$, and there exist polynomials $q$ and $t$, 
such that for any convex set $K$ with parameters $(n,R,r,p)$ as defined above, 
and for any $\varepsilon>0$, there exists $\delta \geq 1/q(n,(R/r),(1/\varepsilon))$, 
such that $A((n,R,r,p),\ldots)$ is an oracle reduction from $WOPT_\varepsilon$ to $WSEP_\delta$, which runs in time $t(n,\log(R/r),\log(1/\varepsilon))$.
\end{prop}

The analysis of this algorithm is similar to \cite{GLS}; the main difference is that, since we are doing ``approximate'' convex optimization, we need to pay more attention to the precision required for the $WSEP$ oracle.  In particular, we need $\delta$ to be polynomial, not exponential, in $\varepsilon$.  

First, some notation.  Let $E(A,a)$ denote an ellipsoid, 
\[
E(A,a) = \set{x \in \RR^n \;|\; (x-a)^T A^{-1} (x-a) \leq 1}, 
\]
where $A$ is a positive definite $n \times n$ matrix and $a \in \RR^n$.  Note that $E(A,a) = \sqrt{A}B + a$, where $B$ is the closed ball of radius 1 around the origin.  Also, let $\lambda_{\max}(A)$ and $\lambda_{\min}(A)$ denote the largest and smallest eigenvalues of $A$.  Note that $\norm{A} = \lambda_{\max}(A)$ and $\norm{A^{-1}} = 1/\lambda_{\min}(A)$.  

To solve the $WOPT$ problem, we have to decide whether the set 
\[
K'(c,\gamma) = K \cap \set{x \in \RR^n \;|\; c \cdot x \geq \gamma} 
\]
contains a ball of radius $\varepsilon$ or is empty.  We have access to a $WSEP$ oracle for $K'(c,\gamma)$.  At every iteration $k$, the algorithm computes an ellipsoid $E(A_k,a_k)$ that contains $K'(c,\gamma)$.  The required precision $\delta$ for the $WSEP$ oracle scales roughly like $\sqrt{\lambda_{\min}(A)}$.  The key observation is that if $\lambda_{\min}(A) < \varepsilon^2$, then the ellipsoid $E(A_k,a_k)$ cannot contain $K'(c,\gamma)$ unless $K'(c,\gamma)$ is empty; and if this happens, the algorithm can stop and answer ``NO.''  Thus, the required precision $\delta$ scales roughly like $\varepsilon$.  

\vskipline

A more powerful idea is contained in the shallow-cut ellipsoid method \cite{YN,GLS}.  This gives a reduction from $WOPT$ to $WMEM$, with a faster running time which is logarithmic in $R/r$ and $1/\varepsilon$, and roughly the same precision requirement as before.  

\begin{prop}\label{ch2-scem-opt-mem}
There exists an algorithm $A$, and there exist polynomials $q$ and $t$, 
such that for any convex set $K$ with parameters $(n,R,r,p)$ as defined above, 
and for any $\varepsilon>0$, there exists $\delta \geq 1/q(n,(R/r),(1/\varepsilon))$, 
such that $A((n,R,r,p),\ldots)$ is an oracle reduction from $WOPT_\varepsilon$ to $WMEM_\delta$, which runs in time $t(n,\log(R/r),\log(1/\varepsilon))$.
\end{prop}

Again, this follows from the analysis of the algorithm in \cite{GLS}; the only new ingredient is the claim that $\delta$ is polynomial in $\varepsilon$.  

A key idea is the notion of a \textit{shallow separation oracle} for a convex set $K$.  This oracle solves the following problem:
\begin{verse}
Given a positive definite matrix $A \in \RR^{n \times n}$ and a vector $a \in \RR^n$.\\
Find a vector $c \in \RR^n$, $\norm{c} = 1$, such that for all $x \in K$, 
\[
c \cdot x \leq c \cdot a + \tfrac{1}{n+1} \sqrt{c^T A c}.
\]\\
Or output ``NO'' if no such vector $c$ exists.\\
\end{verse}
This has a simple geometric interpretation.  Consider the ellipsoid $E(A,a)$.  Now, given a vector $c$, $\norm{c} = 1$, find the point where the ray $\set{a + \lambda c \;|\; \lambda \geq 0}$ intersects the boundary of the ellipsoid $E((n+1)^{-2}A,a)$.  Then construct a hyperplane orthogonal to $c$ that contains this point.  If $K$ lies entirely behind this hyperplane, then we say that $c$ is a ``shallow cut.''  


The shallow separation oracle has two important properties:  it can be constructed from a $WSEP^\beta$ oracle with $\beta = 1/(n+2)$, and it is powerful enough to support a special version of the ellipsoid method.  

To solve the $WOPT$ problem, we proceed as follows.  We construct a shallow separation oracle for the set $K'(c,\gamma)$.  The shallow-cut ellipsoid method works by computing a series of ellipsoids $E(A_k,a_k)$ that contain $K'(c,\gamma)$.  When it queries the shallow separation oracle on an ellipsoid $E(A_k,a_k)$, the precision required for the $WSEP^\beta$ oracle is roughly $\sqrt{\lambda_{\min}(A_k)}$.  Now we make the same observation as before:  if $\lambda_{\min}(A_k) < \varepsilon^2$, then $K'(c,\gamma)$ must be empty, and we can stop the algorithm and answer ``NO.''  Thus the precision required for the $WSEP^\beta$ oracle is roughly $\varepsilon$.  

\subsection{Algorithms using Random Walks}


As an alternative to the shallow-cut ellipsoid method, one can also use some recently developed algorithms which are based on random walks in convex bodies \cite{BV,Vsurvey}.  These algorithms actually solve a slightly more general class of convex programs, where the objective function $f$ need not be linear; when $f$ is linear, one can use a slightly faster algorithm based on simulated annealing \cite{KV}.  

These algorithms are not necessarily faster or more accurate than the shallow-cut ellipsoid method, but they have other intriguing features.  The points where the algorithm queries the membership oracle are chosen randomly from some set $P$ (which changes over successive iterations of the algorithm).  Thus we get a randomized oracle reduction from $WOPT$ to $WMEM$, rather than a deterministic oracle reduction.  Also, there is a simple reason why the randomized algorithm can tolerate imprecision in the membership oracle:  most of the points that it queries will not lie close to the boundary of the set $P$.  (In contrast, a deterministic algorithm must do some work to correct for possible errors, as in Lemma \ref{ch2-asr-sep-mem}.)

The analysis given by Bertsimas and Vempala \cite{BV} assumes a real-valued model of computation, and does not account for the precision of the membership oracle.  However, this can be done using techniques due to Lovasz and Simonovits \cite{LS93}.  Here we sketch the idea.  It would be interesting to prove a tight bound on the precision requirement, and see how it compares with the precision requirement of the ellipsoid method.  






The Bertsimas-Vempala algorithm is built around a subroutine that solves the feasibility problem (the $WOPT$ problem).  The basic idea is as follows:  
\begin{verse}
Given $c \in \RR^n$, $\gamma \in \RR$, $\varepsilon > 0$.\\
Let $P$ be the set $K$.\\
Randomly sample some points from $P$, and compute an approximate centroid of $P$; call this point $z$.\\
If $c \cdot z \geq \gamma$, stop and output ``YES.''  Otherwise, use the vector $c$ to cut out a portion of the set $P$.\footnote{Specifically, we can deduce a hyperplane that separates $z$ from the set $\set{x \:|\: c \cdot x \geq \gamma}$.  Then we take the intersection of $P$ with the half-space that does not contain $z$.}\\
Repeat the procedure starting from line 3.  If $P$ gets too small, stop and output ``NO.''
\end{verse}

The critical step is to sample random points from the set $P$.  (Note that $P$ is convex, and we have a membership oracle for $P$.)  One way is to do a random walk known as the ``ball walk'':  
\begin{quote}
Pick a point $y$ uniformly at random in the ball of radius $\delta$ centered at the current position $x$.  If $y \in P$, then move to $y$, otherwise stay at $x$.  Repeat.  
\end{quote}

The points where the membership oracle makes mistakes all lie close to the boundary of $P$; call this the ``boundary layer'' $P_b$.  Intuitively, if the boundary layer is thin, it should not have much effect on the random walk.  Using an argument by Lov\'asz and Simonovits \cite{LS93}, one can prove (omitting some details):  
\begin{lem}\label{lemma-LS}
For any polynomial $t$, there exists a polynomial $q$ such that, if we run the ball walk for at most $t(n)$ steps, and $\vol(P_b)/\vol(P) \leq 1/q(n)$, then with probability $2/3$ we will never enter the region $P_b$.  
\end{lem}
So, if we can show that the boundary layer is small compared to the total volume of $P$, then our algorithm will work fine.  (As long as the random walk does not enter the boundary layer, the algorithm will perform exactly as if it had access to a perfect membership oracle.)  



Define the set 
\[
K'(c,\gamma) = K \cap \set{x \in \RR^n \;|\; c \cdot x \geq \gamma}.  
\]
The algorithm has to distinguish between the following two cases:  (1) If there exists a vector $y \in S(K,-\varepsilon)$ with $c \cdot y \geq \gamma + \varepsilon$, then $K'(c,\gamma)$ contains a ball of radius $\varepsilon$.  (2) If for all $x \in S(K,\varepsilon)$, $c \cdot x \leq \gamma - \varepsilon$, then $K'(c,\gamma)$ is empty.

In case (1), the set $P$ always contains a ball of radius $\varepsilon$.  Let $p$ be the polynomial such that after at most $p(n)$ steps we will find a solution with the desired precision $\varepsilon$ (assuming a perfect membership oracle).  Let $q$ be the polynomial given by Lemma \ref{lemma-LS}.  Now set the precision of the membership oracle to be $\delta = \varepsilon/(2nq(n))$.  We will show that the boundary layer $P_b$ is small compared to the total volume of $P$.  Define $P^+$ to be the set $P$ expanded by an amount $\delta$, that is, $P^+ = P + \delta B$, where $B$ is the unit ball.  We have that 
\[
P^+ \subseteq P + (\delta/\varepsilon) P = (1 + \delta/\varepsilon) P, 
\]
where the equality holds because $P$ is convex.  This implies that 
\[
\vol(P^+)
 \leq (1 + \delta/\varepsilon)^n \vol(P)
 \leq e^{1/(2q(n))} \vol(P)
 \leq (1 + 1/q(n)) \vol(P).  
\]
So we can conclude that $\vol(P_b) \leq \vol(P^+) - \vol(P) \leq (1/q(n)) \vol(P)$.  Therefore, by Lemma \ref{lemma-LS}, the algorithm will work correctly in this case.  

In case (2), it is easy to see that, as long as the precision of the membership oracle satisfies $\delta \leq \varepsilon$, the oracle will never answer ``YES,'' and so the algorithm will output ``NO.''




\section{Consistency is QMA-hard}

\begin{thm}\label{thm-QMA-hard}
Consistency is QMA-hard, via a poly-time oracle reduction from Local Hamiltonian.  Furthermore, the reduction uses the same value of $k$ for both problems, so we get that Consistency with $k=2$ is QMA-hard.  The reduction yields an instance of Consistency with $\beta \geq \Omega((b-a)^3 / 4^{11k} m^{14})$.  
\end{thm}
We will prove this theorem in the following sections.  First we describe the basic idea of the reduction, which uses convex optimization with a membership oracle; we also discuss some of the technical complications that arise.  Next, we show how to write our convex program in a particular form that is needed for the reduction.  Finally, we deal with the issue of numerical precision, and prove the theorem.


\subsection{The Basic Idea}

We want to solve the Local Hamiltonian problem, i.e., to estimate the smallest eigenvalue of a local Hamiltonian $H = H_1+\cdots+H_m$, where $H_i$ acts on the subset $C_i$.  To this end, we consider the following convex program:  
\begin{verse}
Let $\rho$ be any $2^n \times 2^n$ complex matrix.\\
Find some $\rho$ that minimizes $\Tr(H\rho)$,\\
such that $\rho \succeq 0$ and $\Tr(\rho) = 1$.
\end{verse}
It is easy to see that $H$ has an eigenvalue $\leq \gamma$ if and only if the convex program has optimal value $\Tr(H\rho) \leq \gamma$.  (Note that, although the convex program allows mixed states $\rho$, the optimal solution $\rho$ can always be chosen to be a pure state.)  Unfortunately, this convex program has $4^n$ variables, so solving it requires exponential time.  

We now construct another convex program, which is equivalent to the previous one, but has only a polynomial number of variables:  
\begin{verse}
Let $\rho_1,\ldots,\rho_m$ be complex matrices, where $\rho_i$ has size $2^{|C_i|} \times 2^{|C_i|}$.\\
(We interpret each $\rho_i$ as the reduced density matrix for the subset $C_i$.)\\
Find some $\rho_1,\ldots,\rho_m$ that minimize $\Tr(H_1\rho_1) + \cdots + \Tr(H_m\rho_m)$,\\
such that each $\rho_i$ satisfies $\rho_i \succeq 0$ and $\Tr(\rho_i) = 1$,\\
and $\rho_1,\ldots,\rho_m$ are consistent.
\end{verse}
Note that consistency implies that $\rho_i \succeq 0$ and $\Tr(\rho_i) = 1$, so these constraints are redundant.  One can easily check that the set of feasible solutions is indeed convex:  if $(\rho_1,\ldots,\rho_m)$ are consistent, and $(\rho'_1,\ldots,\rho'_m)$ are consistent, then any convex combination $(\rho''_1,\ldots,\rho''_m)$, where $\rho''_i = q\rho_i+(1-q)\rho'_i$ ($0 \leq q \leq 1$), is also consistent.  

The optimal value of this convex program is equal to the optimal value of the previous convex program; this is because, if $\rho_1,\ldots,\rho_m$ are consistent with some $n$-qubit state $\sigma$, then $\Tr(H\sigma) = \Tr(H_1\rho_1) + \cdots + \Tr(H_m\rho_m)$.  Also, the number of variables is $\sum_{i=1}^m 4^{|C_i|} \leq 4^k m$, which is polynomial in the length of the input.  

This convex program has a ``consistency'' constraint, which we do not know how to evaluate.  But if we have an oracle for the Consistency problem, then we can solve this convex program in polynomial time, using the techniques from the previous section.  Let $K$ be the set of feasible solutions, 
\[
K = \set{(\rho_1,\ldots,\rho_m) \text{ which are consistent}}.  
\]
Local Hamiltonian is equivalent to the $WOPT$ problem, and Consistency is equivalent to the $WMEM$ problem.  So we can apply Theorem \ref{ch2-thm-opt-mem} which shows a poly-time oracle reduction from $WOPT$ to $WMEM$.  

We have to deal with a couple of technical issues.  First, in order for the reduction to work, the set $K$ must contain a ball of radius $r$, and be contained within a ball of radius $R$, where $R/r$ is at most polynomially large.  In particular, $K$ cannot lie in a lower-dimensional subspace.  This requires us to represent each element $(\rho_1,\ldots,\rho_m) \in K$ in a way that has the right number of ``degrees of freedom.''  

We could represent $(\rho_1,\ldots,\rho_m)$ by writing down the matrix entries for the $\rho_i$, to form a vector in $\CC^d$, $d = \sum_{i=1}^m 4^{|C_i|}$.  But this won't work, because the $\rho_i$ must satisfy some algebraic constraints, in order to be consistent:  each $\rho_i$ must be Hermitian, $(\rho_i)^\dagger = \rho_i$, and $\rho_i$ and $\rho_j$ must agree on their intersection $C_i \intersect C_j$, that is, $\Tr_{C_i - (C_i \intersect C_j)}(\rho_i) = \Tr_{C_j - (C_i \intersect C_j)}(\rho_j)$.  These constraints imply that the set $K$ actually lies in a lower-dimensional subspace of $\CC^d$.  In the next section, we will show how to represent $(\rho_1,\ldots,\rho_m)$ in a way that satisfies these constraints automatically.

The other issue concerns numerical precision.  Local Hamiltonian and Consistency are equivalent to the $WOPT$ and $WMEM$ problems with inverse-polynomial precision.  For our reduction, we will bound the amount of precision required of the Consistency oracle, in terms of the precision desired for the Local Hamiltonian problem.


\subsection{How to represent $(\rho_1,\ldots,\rho_m)$}

We will represent each element of $K$ using the expectation values of the ``local'' Pauli matrices on the subsets $C_1,\ldots,C_m$.  These local Pauli matrices form a basis for the space of all local Hamiltonians (acting on the subsets $C_i$).  For an $n$-qubit state $\sigma$, knowing the expectation values of these Pauli matrices is equivalent to knowing the projection of $\sigma$ onto this subspace; and this is equivalent to knowing the local density matrices of $\sigma$.  

First, some notation.  Let $P$ be an $n$-qubit Pauli matrix, $P = \Tensor_{i=1}^n P_i$.  Define the ``support'' of $P$ be the set of qubits on which $P$ acts nontrivially; that is, $\supp(P) = \set{i \:|\: P_i \neq I}$.  Also, for any subset of qubits $C$, define the ``restriction'' of $P$ to $C$, $P|_C = \Tensor_{i \in C} P_i$.  

Define $\SSS_i$ to be the set of Pauli matrices supported on $C_i$, excluding the identity matrix because its expectation value is always 1:  
\[
\SSS_i = \set{P \in \PP^{\tensor n} \:|\: \supp(P) \subseteq C_i} - \set{I}.  
\]
Let $\SSS = \bigcup_{i=1}^m \SSS_i$; this is the set of all ``local'' Pauli matrices.  Let $d = |\SSS|$, and note that $d \leq 4^k m - 1$, which is polynomial in the length of the input.  

For each local Pauli matrix $P \in \SSS$, let $\alpha_P$ be the corresponding expectation value; and let $(\alpha_P)_{P \in \SSS}$ denote the collection of these $\alpha_P$.  We define the set $K' \subseteq \RR^d$, 
\[
K' = \set{(\alpha_P)_{P \in \SSS} \text{ which are consistent}}, 
\]
where we say the $\alpha_P$ are ``consistent'' if there exists an $n$-qubit state $\sigma$ such that for all $P \in \SSS$, $\alpha_P = \Tr(P\sigma)$.  Clearly the set $K'$ is convex.  

So we can restate our convex program using the expectation values $\alpha_P$ ($P \in \SSS$), rather than the density matrices $(\rho_1,\ldots,\rho_m)$:  
\begin{verse}
Let $\alpha_P$ (for $P \in \SSS$) be real numbers.\\
Find some $\alpha_P$ that minimize 
\[
\sum_{i=1}^m \frac{1}{2^{|C_i|}} 
\Bigl( \Tr(H_i) + \sum_{P \in \SSS_i} \alpha_P \Tr(H_i (P|_{C_i})) \Bigr), 
\]\\
such that $(\alpha_P)_{P \in \SSS} \in K'$ (i.e., the $\alpha_P$ are consistent).
\end{verse}
This is justified by the following two lemmas:

\begin{lem}\label{lemma-K-represent}
There is a linear bijection between $K$ and $K'$.  
\end{lem}

\noindent
Proof:  Given some $(\rho_1,\ldots,\rho_m) \in K$, we can construct $(\alpha_P)_{P \in \SSS} \in K'$ as follows:  
\begin{quote}
For each $P \in \SSS$:  We know that $P \in \SSS_i$ for some $i$.  So we can write $P$ in the form $P = (P|_{C_i}) \tensor I$.  Then we set $\alpha_P = \Tr((P|_{C_i}) \rho_i)$.  
\end{quote}
If the $\rho_i$ are consistent with some $n$-qubit state $\sigma$, then the $\alpha_P$ are also consistent with $\sigma$.  To see this, write $\alpha_P = \Tr((P|_{C_i}) \rho_i) = \Tr(P\sigma)$.  (Note that in the case where $\supp(P) \subseteq C_i \intersect C_j$, it makes no difference whether we pick $i$ or $j$ in the above procedure, because $\rho_i$ and $\rho_j$ yield the same reduced density matrix on $C_i \intersect C_j$.)  

Going in the opposite direction, given some $(\alpha_P)_{P \in \SSS} \in K'$, we can construct $(\rho_1,\ldots,$ $\rho_m) \in K$ as follows:  
\begin{quote}
For each $i = 1,\ldots,m$:  We construct $\rho_i$ by using the $\alpha_P$ for all $P \in \SSS_i$.  Note that we can write $P$ in the form $P = (P|_{C_i}) \tensor I$.  We set 
\[
\rho_i = \frac{1}{2^{|C_i|}} \Bigl( I + \sum_{P \in \SSS_i} \alpha_P (P|_{C_i}) \Bigr).  
\]
\end{quote}
If the $\alpha_P$ are consistent with some $n$-qubit state $\sigma$, then the $\rho_i$ are also consistent with $\sigma$.  To see this, write $\sigma$ in terms of the $\alpha_P$, where we now include the expectation values $\alpha_P = \Tr(P\sigma)$ for all $P \in \PP^{\tensor n}$, 
\[
\sigma = \frac{1}{2^n} \sum_{P \in \PP^{\tensor n}} \alpha_P P.  
\]
Note that when we trace out the qubits not in $C_i$, we get that $\Tr_{\set{1,\ldots,n}-C_i}(P)$ equals $2^{n-|C_i|} (P|_{C_i})$ if $\supp(P) \subseteq C_i$, and 0 otherwise.  Thus we have 
\[
\Tr_{\set{1,\ldots,n}-C_i}(\sigma)
 = \frac{1}{2^{|C_i|}} \sum_{P \::\: \supp(P) \subseteq C_i} \alpha_P (P|_{C_i})
 = \rho_i.  
\]

Finally, observe that these maps (between $K$ and $K'$) are linear, and they are inverses of each other.  $\square$

\begin{lem}\label{lemma-convex-prog}
The optimal value of this convex program is equal to the smallest eigenvalue of the local Hamiltonian $H = H_1+\cdots+H_m$.  
\end{lem}

\noindent
Proof:  This follows from the remarks in the previous section, and Lemma \ref{lemma-K-represent}.  In particular, we have that 
\[
\begin{split}
\sum_{i=1}^m \Tr(H_i \rho_i) 
 &= \sum_{i=1}^m \Tr\Bigl( H_i \frac{1}{2^{|C_i|}} 
    \Bigl( I + \sum_{P \in \SSS_i} \alpha_P (P|_{C_i}) \Bigr) \Bigr) \\
 &= \sum_{i=1}^m \frac{1}{2^{|C_i|}} 
    \Bigl( \Tr(H_i) + \sum_{P \in \SSS_i} \alpha_P \Tr(H_i (P|_{C_i})) \Bigr).  
\end{split}
\]
(Note that we view $H_i$ as an operator acting on the subset of qubits $C_i$ only, not the entire system.  So $\Tr(H_i)$ is a trace over $2^{|C_i|}$ dimensions.)  $\square$

\vskipline

Next, we prove some bounds on the geometry of the set $K' \subseteq \RR^d$.  

\begin{lem}\label{lemma-R}
$K'$ is contained in a ball of radius $R = \sqrt{d}$ centered at the origin.  
\end{lem}

\noindent
Proof:  Suppose $(\alpha_P)_{P \in \SSS} \in K'$, and say it is consistent with some state $\sigma$.  Since $\alpha_P = \Tr(P\sigma)$, it follows that $-1 \leq \alpha_P \leq 1$, which implies the result.  $\square$

\begin{lem}\label{lemma-r}
The ball of radius $r = 1/\sqrt{d}$ around the origin is contained in $K'$.  
\end{lem}

\noindent
Proof:  Let $(\alpha_P)_{P \in \SSS}$ be any vector in $\RR^d$ of length at most $1/\sqrt{d}$.  By the Cauchy-Schwartz inequality, $\sum_{P \in \SSS} |\alpha_P| \leq 1$; let $p = \sum_{P \in \SSS} |\alpha_P|$.  Now define $\sigma = (1/2^n) (I + \sum_{P \in \SSS} \alpha_P P)$.  This is a legal density matrix, because it can be written as 
\[
\begin{split}
\sigma
 &= \frac{1}{2^n} \Bigl( (1-p)I + \sum_{P \in \SSS} (|\alpha_P|I + \alpha_P P) \Bigr)\\
 &= (1-p) \frac{I}{2^n} + \sum_{P \in \SSS} |\alpha_P| \frac{I+\sign(\alpha_P)P}{2^n}, 
\end{split}
\]
which is (with probability $1-p$) the fully mixed state, and (with probability $|\alpha_P|$, for $P \in \SSS$) the mixture of all eigenstates of $P$ with eigenvalue $\sign(\alpha_P)$.  Furthermore, the $\alpha_P$ are consistent with $\sigma$; thus we conclude that $(\alpha_P)_{P \in \SSS} \in K'$.  $\square$


\subsection{Numerical Precision}

In this section we deal with the issue of numerical precision.  We give reductions from Local Hamiltonian to $WOPT$, from $WOPT$ to $WMEM$ (using the general tools of section 2.3), and finally from $WMEM$ to Consistency.  

(Note added later:  one can simplify these proofs by using a slightly different reduction, from $WOPT^*$ to $WMEM^*$, which is described in section 3.6.)

\begin{lem}
\label{ch2-localham-wopt}
There is a poly-time mapping reduction from Local Hamiltonian to \linebreak $WOPT_{\text{1/poly}}$ (on the set $K'$).  This reduction yields an instance of $WOPT$ with $\varepsilon \geq \Omega((b-a) / (2^k m^{3/2}))$.  
\end{lem}

\noindent
Proof:  We have an $n$-qubit system, and subsets $C_1,\ldots,C_m \subseteq \set{1,\ldots,n}$, where $|C_i| \leq k$.  Accordingly we define $\SSS$ to be the set of local Pauli matrices, and let $d = |\SSS|$.  We let $\alpha = (\alpha_P)_{P \in \SSS}$ denote a vector of expectation values of local Pauli matrices.  Then $K' = \set{\alpha \in \RR^d \;|\; \alpha \text{ is consistent with some \textit{n}-qubit state } \sigma}$.  Note that $d \leq 4^k m - 1$ is polynomial in the length of the input to the Local Hamiltonian problem.  

We are given a local Hamiltonian $H = \sum_{i=1}^m H_i$, two numbers $a,b \in \RR$, and a unary string ``$1^s$,'' such that $b-a \geq 1/s$.  (Note that $\norm{H} \leq \sum_{i=1}^m \norm{H_i} \leq m$, so we can assume $|a|,|b| \leq m$.)  If $H$ has an eigenvalue $\leq a$, we should answer ``YES''; if all eigenvalues of $H$ are $\geq b$, we should answer ``NO.''  

We will reduce this to an instance of $WOPT_{\text{1/poly}}$.  In this problem, one is given $c \in \RR^d$, $\norm{c} = 1$, $\gamma \in \RR$, $\varepsilon \in \RR$, and a unary string ``$1^t$,'' such that $\varepsilon \geq 1/t$.  If there exists some $y \in S(K',-\varepsilon)$ such that $c \cdot y \geq \gamma + \varepsilon$, then we should answer ``YES''; if for all $x \in S(K',\varepsilon)$, $c \cdot x \leq \gamma - \varepsilon$, then we should answer ``NO.''  

As shown in the previous section, the smallest eigenvalue of $H$ is equal to the optimal value $f(\alpha)$ for the following convex program:  find some $\alpha \in K'$ that minimizes the function 
\[
f(\alpha) = \sum_{i=1}^m \frac{1}{2^{|C_i|}} 
\Bigl( \Tr(H_i) + \sum_{P \in \SSS_i} \alpha_P \Tr(H_i (P|_{C_i})) \Bigr).  
\]

We can write $f(\alpha)$ using simpler notation.  For each $i = 1,\ldots,m$, define a vector $\eta_i = (\eta_{i,P})_{P \in \SSS}$, where $\eta_{i,P} = 2^{-|C_i|} \Tr(H_i(P|_{C_i}))$ if $P$ is supported on $C_i$, and $\eta_{i,P} = 0$ otherwise.  For each $i = 1,\ldots,m$, also define a scalar $\nu_i = 2^{-|C_i|} \Tr(H_i)$.  Then we can write 
\[
f(\alpha) = \sum_{i=1}^m (\nu_i + \alpha \cdot \eta_i).  
\]
Define $\eta = \sum_{i=1}^m \eta_i$ and $\nu = \sum_{i=1}^m \nu_i$.  Then we can write 
\[
f(\alpha) = \nu + \alpha \cdot \eta.
\]

In addition, we can bound the size of $\eta$ and $\nu$ as follows.  Observe that $H_i$ can be written in terms of $\eta_i$ and $\nu_i$,  
\[
H_i = \nu_i I + \sum_{P \in \SSS_i} \eta_{i,P} (P|_{C_i}).  
\]
Therefore 
\[
\norm{H_i}_2^2 = \Tr(H_i^2)
 = 2^{|C_i|} (\nu_i^2 + \sum_{P \in \SSS_i} \eta_{i,P}^2)
 = 2^{|C_i|} (\nu_i^2 + \norm{\eta_i}^2).  
\]
Also, note that $\norm{H_i}_2^2 \leq 2^{|C_i|} \norm{H_i}^2$.  So we conclude that $|\nu_i| \leq \norm{H_i} = 1$ and $\norm{\eta_i} \leq \norm{H_i} = 1$.  Hence, $|\nu| \leq m$ and $\norm{\eta} \leq m$.  

Now we construct an instance of $WOPT_{\text{1/poly}}$ as follows.  Let $c = -\eta / \norm{\eta}$.  We will specify $\gamma$ and $\varepsilon$ later in the proof.  

Consider what happens on a ``YES'' instance of Local Hamiltonian.  There exists some $\alpha^* \in K'$ such that $\eta \cdot \alpha^* \leq a - \nu$.  Furthermore, we claim that there exists a point $\alpha$ in the interior of $K'$ such that $\eta \cdot \alpha$ is not much larger than $a - \nu$.  To see this, let $\sigma^*$ be the $n$-qubit density matrix corresponding to $\alpha^*$.  Now consider the density matrix 
\[
(1-q) \sigma^* + q (I/2^n) + \sum_{P \in \SSS} u_P (P/2^n).  
\]
This is a legal density matrix (positive semidefinite with trace 1) provided that $0 \leq q \leq 1$ and $\sum_{P \in \SSS} |u_P| \leq q$.  When we write down the expectation values of the local Pauli matrices $P \in \SSS$, this density matrix corresponds to the point $(1-q) \alpha^* + u$.  This point is in $K'$ provided that $0 \leq q \leq 1$ and $\norm{u}_1 \leq q$.  Note that $\norm{u}_1 \leq \sqrt{d} \norm{u}$.  We conclude that a ball of radius $q/\sqrt{d}$ around the point $(1-q) \alpha^*$ is contained in $K'$.  In other words, 
\[
(1-q) \alpha^* \in S(K',-q/\sqrt{d}).  
\]
Also, note that 
\[
\eta \cdot ((1-q) \alpha^*) \leq (1-q) (a-\nu) \leq a-\nu + 2qm.  
\]
Now let $q = \varepsilon \sqrt{d}$ (assuming $\varepsilon \leq 1/\sqrt{d}$).  We have shown that there exists some $\alpha \in S(K',-\varepsilon)$, such that $\eta \cdot \alpha \leq a - \nu + 2\varepsilon\sqrt{d}m$.  This implies 
\[
-c \cdot \alpha \leq \frac{1}{\norm{\eta}} (a - \nu + 2\varepsilon\sqrt{d}m).  
\]
We will choose $\gamma$ and $\varepsilon$ so that the right side of this inequality equals $-\gamma - \varepsilon$.  Then this is a ``YES'' instance of $WOPT_{\text{1/poly}}$.  

On the other hand, suppose we have ``NO'' instance of Local Hamiltonian, so that for all $\alpha \in K'$, $\eta \cdot \alpha \geq b - \nu$.  Furthermore, for all $\alpha$ close to $K'$, $\eta \cdot \alpha$ is not much smaller than $b - \nu$.  In particular, using the fact that $\norm{\eta} \leq m$, we get that for any $\alpha \in S(K',\varepsilon)$, 
\[
\eta \cdot \alpha \geq b - \nu - \varepsilon m.  
\]
This implies 
\[
-c \cdot \alpha \geq \frac{1}{\norm{\eta}} (b - \nu - \varepsilon m).  
\]
We will choose $\gamma$ and $\varepsilon$ so that the right side of this inequality equals $-\gamma + \varepsilon$.  Then this is a ``NO'' instance of $WOPT_{\text{1/poly}}$.  

Now we choose $\gamma$ and $\varepsilon$.  We set $\gamma$ according to 
\[
-\gamma = \frac{1}{\norm{\eta}} (a - \nu + 2\varepsilon\sqrt{d}m) + \varepsilon
 = \frac{1}{\norm{\eta}} (b - \nu - \varepsilon m) - \varepsilon.  
\]
In order for this to work, $\varepsilon$ must satisfy the equation 
\[
2\varepsilon = \frac{1}{\norm{\eta}} (b - a - \varepsilon m - 2\varepsilon\sqrt{d}m), 
\]
which has a solution 
\[
\varepsilon = \frac{b-a}{2\norm{\eta} + m + 2\sqrt{d}m}
 \geq \frac{b-a}{(2\sqrt{d}+3)m}
 \geq \Omega((b-a) / (2^k m^{3/2})).
\]
(Note that $\varepsilon$ is inverse-polynomial in the length of the input.)  This concludes the proof.  $\square$

\begin{lem}
\label{ch2-wmem-consistency}
There is a poly-time mapping reduction from $WMEM_{\text{1/poly}}$ (on the set $K'$) to Consistency.  This reduction yields an instance of Consistency with $\beta \geq \delta / (2^k \sqrt{m})$.  
\end{lem}

\noindent
Proof:  We have an $n$-qubit system, and subsets $C_1,\ldots,C_m \subseteq \set{1,\ldots,n}$, where $|C_i| \leq k$.  Accordingly we define $\SSS$ to be the set of local Pauli matrices, and let $d = |\SSS|$.  We let $\alpha = (\alpha_P)_{P \in \SSS}$ denote a vector of expectation values of local Pauli matrices.  Then $K' = \set{\alpha \in \RR^d \;|\; \alpha \text{ is consistent with some \textit{n}-qubit state } \sigma}$.  

We will eventually use this lemma as the final step in a reduction from Local Hamiltonian.  Note that $d \leq 4^k m - 1$ is polynomial in the length of the input to the Local Hamiltonian problem.

The $WMEM_{\text{1/poly}}$ problem is as follows.  We are given $\alpha \in \RR^d$, $\delta \in \RR$, and a unary string ``$1^s$,'' where $\delta \geq 1/s$.  If $\alpha \in S(K',-\delta)$, we should answer ``YES.''  If $\alpha \notin S(K',\delta)$, we should answer ``NO.''  

We reduce this to the following instance of the Consistency problem.  We construct the local density matrices $\rho_1,\ldots,\rho_m$ from the expectation values $\alpha_P$ ($P \in \SSS$), as described in Lemma \ref{lemma-K-represent}.  We set $\beta = \delta/\sqrt{d}$.  Note that $\beta \geq \delta / (2^k \sqrt{m})$ is inverse-polynomial in the length of the input to $WMEM$, and it is also inverse-polynomial in the length of the input to Local Hamiltonian.  

Clearly, a ``YES'' instance of $WMEM_{\text{1/poly}}$ maps to a ``YES'' instance of Consistency.  Now suppose we have a ``NO'' instance of $WMEM_{\text{1/poly}}$.  Then for all $n$-qubit states $\sigma$, 
\[
\bigl( \sum_{P \in \SSS} (\Tr(P\sigma)-\alpha_P)^2 \bigr)^{1/2} \geq \delta.  
\]
Thus there is some $P \in \SSS$ such that $|\Tr(P\sigma)-\alpha_P| \geq \delta/\sqrt{d}$.  We know that $P$ is supported on some subset $C_i$, so we can write $P = \tilde{P} \tensor I$ where $\tilde{P}$ acts on $C_i$.  Note that $\alpha_P = \Tr(\tilde{P} \rho_i)$.  Also, let $\tilde{\sigma} = \Tr_{\set{1,\ldots,n}-C_i}(\sigma)$.  Then we have 
\[
\bigl| \Tr(\tilde{P} \tilde{\sigma})
 - \Tr(\tilde{P} \rho_i) \bigr| \geq \delta/\sqrt{d}.  
\]

We will use $\tilde{P}$ to construct a measurement (POVM) that distinguishes between $\tilde{\sigma}$ and $\rho_i$.  Since the eigenvalues of $\tilde{P}$ are all $\pm 1$, we can write $\tilde{P} = \Pi_1-\Pi_2$, where $\Pi_1$ and $\Pi_2$ are projectors on orthogonal subspaces, and $\Pi_1+\Pi_2 = I$.  Thus $\set{\Pi_1,\Pi_2}$ is a POVM.  For the state $\tilde{\sigma}$, let $s_j$ be the probability of measuring $j$ (for $j = 1,2$); and for the state $\rho_i$, let $r_j$ be the probability of measuring $j$ (for $j = 1,2$).  

Then we have 
\[
\bigl| \Tr(\tilde{P} \tilde{\sigma})
 - \Tr(\tilde{P} \rho_i) \bigr|
 = |(s_1-s_2) - (r_1-r_2)|
 = 2|s_1-r_1|.  
\]
Observe that the $\ell_1$ distance between $s$ and $r$ is $\norm{s-r}_1 = |s_1-r_1| + |s_2-r_2| = 2|s_1-r_1|$.  Also, this is a lower bound for the $L_1$ (matrix) distance between $\tilde{\sigma}$ and $\rho_i$.  So we have 
\[
\norm{\tilde{\sigma} - \rho_i}_1
 \geq \norm{s-r}_1 \geq \delta/\sqrt{d} = \beta.  
\]
Thus we have a ``NO'' instance of Consistency.  $\square$

\vskipline

We are now ready to prove that Consistency is QMA-hard.  

\vskipline

\noindent
Proof of Theorem \ref{thm-QMA-hard}:  Use the previous two lemmas, and the reduction from $WOPT_\varepsilon$ to $WMEM_\delta$ in Theorem \ref{ch2-thm-opt-mem}.  Note that by Proposition \ref{ch2-asr-opt-mem}, and the properties of the set $K'$, the reduction from $WOPT_\varepsilon$ to $WMEM_\delta$ has the following precision requirement:  
\[
\delta \geq \Omega\Bigl(\frac{r^3\varepsilon^3}{d^5R^5}\Bigr)
 \geq \Omega\Bigl(\frac{\varepsilon^3}{d^9}\Bigr)
 \geq \Omega\Bigl(\frac{\varepsilon^3}{4^{9k}m^9}\Bigr).  
\]
$\square$

\section{Discussion}

Consistency of local density matrices is an interesting problem that gives some new insight into the class QMA.  The reduction from Local Hamiltonian is nontrivial, and in that sense, Consistency seems to be an easier problem to deal with.  One direction for future work is to try to find additional QMA-complete problems by giving reductions from Consistency (rather than from Local Hamiltonian).  

Another question is whether Consistency remains QMA-hard under mapping reductions.  We mention that we can build zero-knowledge proof systems for Consistency \cite{Liu-ZK}, using techniques developed by Watrous \cite{Wat}.  If we could show that Consistency is QMA-hard under mapping reductions, then we could get zero-knowledge proof systems for any language in QMA.  

\vskipline
\vskipline

\noindent
\textit{Acknowledgements:}  Thanks to Dorit Aharonov for suggesting this problem and pointing out an error in a previous version of the paper; thanks also to Russell Impagliazzo and the anonymous reviewers for their helpful comments.  Supported by an ARO/NSA Quantum Computing Graduate Research Fellowship.  

\vskipline

\noindent
A preliminary version of this paper appeared as \cite{Liu-consistency-qma}:  Y.-K. Liu, ``Consistency of Local Density Matrices is QMA-complete,'' \textit{Proc. RANDOM 2006}, pp.438-449, Springer-Verlag (2006).  That version is copyright Springer-Verlag Berlin Heidelberg.  The present chapter is substantially expanded and revised.  Its use is permitted under the copyright agreement.  
\chapter{$\N$-representability is QMA-complete}

\ifthenelse{\boolean{ucsdformat}}{\thispagestyle{chappage}}{}

(This chapter is joint work with Matthias Christandl and Frank Verstraete.)


\section{Introduction}

The central theoretical problem in the field of many-body strongly
correlated quantum systems is to find efficient ways of simulating
Schr\"odinger's equations. The main difficulty is the fact that the
dimension of the Hilbert space describing a system of $\N$ quantum
particles scales exponentially in $\N$. This makes a direct
numerical simulation intractable: every time an extra particle is
added to the system, the computational resources would have to be
doubled.

The situation is not hopeless, however, as in principle it could be that all physical wavefunctions, i.e., the
ones that are realized in nature, have very special properties and can be parameterized in an efficient way. The
idea would then be to propose a variational class of wavefunctions that capture the physics of the systems of
interest, and then do an optimization over this restricted class. This approach has proven to be very successful,
as witnessed by mean field theory and renormalization group methods. However, it is still an open problem to find
an efficient variational class to describe complex wavefunctions such as those arising in quantum chemistry.

One of the basic problems in quantum chemistry is to find the ground
state of a Hamiltonian describing the many-body system of an atom or
molecule. Here one is mainly interested in the behavior of the electrons; 
the nuclei are assumed to be fixed, possibly in some non-equilibrium geometry. 
These Hamiltonians are very
ungeneric, because they contain at most 2-body interactions. This
implies that the number of free parameters in such Hamiltonians
scales at most quadratically in the number of particles or modes,
and hence the ground states of all such systems form a
small-dimensional manifold.

For a Hamiltonian with only
2-body interactions, the energy corresponding to a wavefunction is
completely determined by its 2-body correlation functions, and as a
consequence the ground state will be the one with extremal 2-body
reduced density operators. This fact was realized a long time ago,
and led Coulson \cite{Coulson,Tredgold} to propose the following problem:
given a set of $\N$ quantum particles, can we characterize the
allowed sets of 2-body correlations or density operators between all
pairs of particles?

If the particles under consideration are fermions, as is the case in quantum chemistry, 
this has been called the \textit{$\N$-representability problem}~\cite{Coleman}.
Here, we consider the reduced density operators acting on pairs of
fermions, and we want to decide whether they are consistent with
some global state over $\N$ fermions.  
An efficient solution to the $\N$-representability problem would be a huge breakthrough, as it would (for example)
allow us to calculate the binding energies of all molecules.  Therefore, a very large effort has been devoted to
solving this problem \cite{book1,book,Mazziotti}.

Here we will give strong evidence that the $\N$-representability problem is intractable, as it is QMA-complete and
hence NP-hard. By ``intractable,'' we mean that, for large $N$, solving the problem in the worst case requires a
number of operations that grows exponentially in $N$. The complexity class QMA (Quantum Merlin-Arthur) is the
natural generalization of the class NP (nondeterministic polynomial time) to the setting of quantum
computing. 
Colloquially, a problem is in QMA if there exists an efficient quantum algorithm that, when given a possible
solution to the problem, can verify whether it is correct; here the ``solution'' may be a quantum state on
polynomially many qubits. A problem is QMA-hard if it is at least as hard as any other problem in QMA; that is,
given an efficient algorithm for this problem, one could solve every other problem in QMA efficiently. We say that
a problem is QMA-complete if it is in QMA and it is also QMA-hard.

In a seminal work, Kitaev \cite{KSV} 
proved that the Local Hamiltonian problem --- determining the ground state energy of a spin Hamiltonian that is a
sum of 5-body terms (on $n$ qubits), with accuracy $\pm \varepsilon$ where $\varepsilon$ is inverse polynomial in
$n$---is QMA-complete. In fact, it was later shown that this problem remains QMA-complete when restricted to
2-body interactions \cite{KKR}, and even in the case of geometrically local interactions \cite{OT}. 
In this paper, we extend these results to fermionic systems, and show that Fermionic 2-Local Hamiltonian is QMA-complete. 

Another problem is to decide whether a given set of local
density operators is \textit{consistent}, i.e., whether they can be realized as the reduced
density operators of the same global state. In a certain sense, this is the dual of the 
Local Hamiltonian problem (see chapter 4 of this dissertation). The consistency problem has been studied 
for spin systems, and it was recently shown to be QMA-complete (see chapter 2 of this dissertation) 
\cite{Liu-consistency-qma}. In the present paper, we will prove that $\N$-representability, which is the
fermionic version of the consistency problem, is also QMA-complete.

\section{Fermions} 


We review some basic facts about fermions; see \cite{Ostlund-Szabo} for more on this, and other topics in quantum chemistry.  Consider a system of $\n$ particles, where each particle has $\m$ energy levels, and the particles obey Fermi statistics.  (For instance, we might have $\n$ electrons, and we fix a basis set consisting of $\m$ single-electron orbitals.)  We assume $\m \geq \n$.  Since the particles are fermions, we only allow $\n$-particle states that are antisymmetric under exchanges of pairs of particles.  This implies that no two particles can occupy the same state (the Pauli exclusion principle); hence the assumption that $\m \geq \n$.  Also, we assume that the interactions in the system do not create or destroy particles, so we are interested in states with exactly $\n$ particles.  

We will now construct a basis for the space of $\n$-particle fermionic states.  Let $\ket{\varphi_1}, \ldots, \ket{\varphi_\m}$ be an orthonormal basis for a single particle.  Fix an ordering of the particles, from 1 to $\n$.  For any indices $i_1,\ldots,i_\n \in \set{1,\ldots,\m}$, we can construct an $\n$-particle fermionic state using a ``Slater determinant'':  
\[
\ket{\varphi_{i_1} \ldots \varphi_{i_\n}} 
:= \frac{1}{\sqrt{\n!}} \det \Bigl[ \varphi^{(a)}_{i_b} \Bigr]_{a,b=1}^\n 
 = \frac{1}{\sqrt{\n!}} \sum_{\pi \in S_\n} (-1)^{\text{sign}(\pi)} 
\Tensor_{a=1}^\n \ket{\varphi_{i_{\pi(a)}}}. 
\]
Here we construct a matrix whose $(a,b)$'th entry is $\varphi^{(a)}_{i_b}$, which means that the $a$'th particle is in state $\ket{\varphi_{i_b}}$; then we take its ``determinant'' and get a superposition of tensor product states.  Note that the determinant is nonzero if and only if the $i_1,\ldots,i_\n$ are distinct, i.e., no two particles can be in the same state.  Also, changing the order of the $i_1,\ldots,i_\n$ only affects the sign of the determinant.  We adopt the convention that the $i_1,\ldots,i_\n$ always appear in increasing order.  For any $I \subseteq \set{1,\ldots,\m}$, $|I| = \n$, we define 
\[
\ket{\varphi_I} = \ket{\varphi_{i_1} \ldots \varphi_{i_\n}}, 
\]
where $I = \set{i_1, \ldots, i_\n}$, and $i_1 < \cdots < i_\n$.  There are $\binom{\m}{\n}$ states of this form, and they form an orthonormal basis for the space of all $\n$-particle fermionic states.  

Let $\sigma$ be a density matrix describing an $\n$-particle state; then the 2-particle reduced density matrix (2-RDM) is given by 
\[
\rho^{[2]} = \Tr_{3,\ldots,\n}(\sigma).  
\]
This is a matrix of dimension $\binom{\m}{2} \times \binom{\m}{2}$.  Since the $\n$-particle state is antisymmetric, the 2-RDM is the same for every pair of particles.  Also, note that it is not necessary to know anything about the single-particle states $\ket{\varphi_1}, \ldots, \ket{\varphi_\m}$, besides the fact that they are orthonormal.  The partial trace and the question of $N$-representability do not depend on the choice of basis.  

We are also interested in fermionic local Hamiltonians, that is, Hamiltonians on $\n$ fermions, that consist of the same 2-particle interaction acting on every pair of particles:  
\[
H = \sum_{\substack{i,j = 1,\ldots,\n \\ i \neq j}} A^{(ij)}.  
\]
Here, $A$ is a matrix of dimension $\binom{\m}{2} \times \binom{\m}{2}$.  

\subsection{Second-Quantized Operators}
\label{ch3-2ndquant}

``Second quantization'' provides a nice way to describe fermionic systems.  The basic idea is that, rather than dealing with the individual particles, one should pay attention to which of the states $\ket{\varphi_1}, \ldots, \ket{\varphi_\m}$ are occupied.  This gives a unified way of describing states with different numbers of particles.  It is particularly helpful in dealing with the 2-RDM, because it avoids the messy step of tracing out the other $\n-2$ particles.  This clarifies the relationship between the $\n$-particle state and the 2-RDM.  (Note:  the formalism of second quantization is used in our proofs, but is not needed in the statement of our results.)

Let $V_\n$ denote the space of $\n$-particle fermionic states.  We will consider the space of all fermionic states, where the number of particles varies from 0 to $\m$; this is given by 
\[
V = \bigoplus_{\n=0}^\m V_\n.  
\]
(This is known as Fock space.)  Note that states with different numbers of particles lie in orthogonal subspaces.  Generally, we will only be interested in states with a fixed number of particles $\n$, so the state is described by a density matrix whose support lies in the subspace $V_\n$.  However, we will find it useful to define operators (e.g., observables and Hamiltonians) that act on the whole space $V$.  In particular, we will do this with local observables and local Hamiltonians---here, the operator acts identically on all pairs of particles, so its meaning is independent of the total number of particles $\n$.  

Annihilation and creation operators are the basic tools for working in Fock space.  For every $i \in \set{1,\ldots,\m}$, we define the annihilation and creation operators, $a_i$ and $a_i^\dagger$, by describing how they act on the Slater basis states $\ket{\varphi_I}$:  
\[
\begin{split}
a_i \ket{\varphi_I}
 &= 0 \text{ if $i \notin I$}\\
 &= (-1)^{f(I,i)} \ket{\varphi_{I \setminus \set{i}}} \text{ if $i \in I$}
\end{split}
\]
\[
\begin{split}
a_i^\dagger \ket{\varphi_I}
 &= 0 \text{ if $i \in I$}\\
 &= (-1)^{f(I,i)} \ket{\varphi_{I \union \set{i}}} \text{ if $i \notin I$}, 
\end{split}
\]
where $f(I,i) = |\set{j \in I \;|\; j<i}|$.  Intuitively, $a_i$ annihilates a particle in state $\ket{\varphi_i}$, or returns 0 if no such particle exists, while $a_i^\dagger$ creates a particle in state $\ket{\varphi_i}$, or returns 0 if such a particle already exists.  (Thus, given an $\n$-particle state, $a_i$ returns an $(\n-1)$-particle state, while $a_i^\dagger$ returns an $(\n+1)$-particle state.)  The particle is annihilated or created in the first (far left) column of the Slater determinant; moving it to its proper position, among the elements of $I$ in ascending order, produces the $(-1)^{f(I,i)}$ phase factor.  (Note that $a_i^\dagger$ is indeed the adjoint of $a_i$.)  

Note that an $\n$-particle Slater basis state $\ket{\varphi_I}$, where $I = \set{i_1,\ldots,i_\n}$, $i_1 < \cdots < i_\n$, can be written in the form 
\[
\ket{\varphi_I} = a_{i_1}^\dagger \cdots a_{i_\n}^\dagger \ket{\Omega}, 
\]
where $\ket{\Omega}$ is the state with zero fermions, i.e., the vacuum state.  Also, any $\n$-particle state $\ket{\psi}$ can be written in the form 
\[
\ket{\psi}
 = \sum_{{\tiny\begin{array}{l} j_1,\ldots,j_\m \in \set{0,1}\\ j_1+\cdots+j_\m=\n \end{array}}}
 c_{j_1,\ldots,j_\m} (a_1^\dagger)^{j_1}\cdots(a_\m^\dagger)^{j_\m} \ket{\Omega}.  
\]
Also note that $a_i$ and $a_i^\dagger$ satisfy the following anticommutation rules:  
\begin{align*}
a_i^\dagger a_j^\dagger &= -a_j^\dagger a_i^\dagger\\
a_i a_j &= -a_j a_i\\
a_i a_j^\dagger &= \delta_{ij} -a_j^\dagger a_i.  
\end{align*}

Second quantization gives a convenient expression for the 2-RDM.  First, if $\ket{\psi}$ is an $\n$-fermion state, then a straightforward calculation shows that 
\[
\Bigl( \bra{\varphi_i} \tensor I^{\tensor (\n-1)} \Bigr) \ket{\psi}
 = \frac{1}{\sqrt{\n}} a_i \ket{\psi}.  
\]
That is, taking the inner product with $\ket{\varphi_i}$ on the first particle is equivalent to applying the annihilation operator $a_i$.  Similarly, when we act on the first and second particles, we get that 
\begin{equation}
\label{eqn-2nd-quant-identity}
\Bigl( \bra{\varphi_i} \tensor \bra{\varphi_j}
       \tensor I^{\tensor (\n-2)} \Bigr) \ket{\psi}
 = \frac{1}{\sqrt{\n(\n-1)}} a_j a_i \ket{\psi}.  
\end{equation}

Now suppose $\rho^{[2]} = \Tr_{3,\ldots,\n} \ket{\psi} \bra{\psi}$ is the 2-RDM corresponding to $\ket{\psi}$.  Then the matrix elements of $\rho^{[2]}$ are given by 
\[
\begin{split}
\rho^{[2]}_{ijkl}
&= \Bigl( \bra{\varphi_i} \tensor \bra{\varphi_j} \Bigr) \rho^{[2]}
   \Bigl( \ket{\varphi_k} \tensor \ket{\varphi_l} \Bigr)\\
&= \Tr \Bigl(
   \Bigl( \bra{\varphi_i} \tensor \bra{\varphi_j}
          \tensor I^{\tensor (\n-2)} \Bigr)
   \ket{\psi} \bra{\psi}
   \Bigl( \ket{\varphi_k} \tensor \ket{\varphi_l}
          \tensor I^{\tensor (\n-2)} \Bigr)
   \Bigr)\\
&= \frac{1}{\n(\n-1)} \Tr \bigl(
   \bigl( a_k^\dagger a_l^\dagger a_j a_i \bigr)
   \ket{\psi} \bra{\psi} \bigr).  
\end{split}
\]
That is, the matrix elements of $\rho^{[2]}$ are equal to the expectation values of products of annihilation and creation operators.  This extends to the general case, where the $\n$-particle state is described by a density matrix $\sigma$, the corresponding 2-RDM is $\rho^{[2]} = \Tr_{3,\ldots,\n} (\sigma)$, and we have that 
\begin{equation}
\label{ch3-2rdm-2ndquant}
\rho^{[2]}_{ijkl}
 = \frac{1}{\n(\n-1)} \Tr \bigl(
   \bigl( a_k^\dagger a_l^\dagger a_j a_i \bigr)
   \sigma \bigr).  
\end{equation}
Note that $\rho^{[2]}_{ijkl} = 0$ if $i=j$ or $k=l$; this is consistent with the fact that no two fermions can occupy the same state.  

Second quantization also gives a convenient expression for a fermionic local Hamiltonian $H = \sum_{i \neq j} A^{(ij)}$.  First, we write down the matrix elements of $A$:  
\[
A_{ijkl} = \Bigl( \bra{\varphi_i} \tensor \bra{\varphi_j} \Bigr) A
           \Bigl( \ket{\varphi_k} \tensor \ket{\varphi_l} \Bigr).  
\]
Observe that, for any $\n$-fermion state $\ket{\psi}$, 
\[
\begin{split}
\bra{\psi} H \ket{\psi}
 &= \n(\n-1) \bra{\psi} A^{(12)} \ket{\psi}\\
 &= \n(\n-1) \bra{\psi}
    \sum_{ijkl} A_{ijkl} \Bigl(
    \Bigl( \ket{\varphi_i} \tensor \ket{\varphi_j} \Bigr)
    \Bigl( \bra{\varphi_k} \tensor \bra{\varphi_l} \Bigr)
    \tensor I^{\tensor (\n-2)} \Bigr)
    \ket{\psi}\\
 &= \bra{\psi} \sum_{ijkl} A_{ijkl}
    \bigl( a_i^\dagger a_j^\dagger a_l a_k \bigr) \ket{\psi}, 
\end{split}
\]
where we used the antisymmetry of the state $\ket{\psi}$, and equation (\ref{eqn-2nd-quant-identity}).  Thus we can write $H$ in the following form:  
\[
H = \sum_{ijkl} A_{ijkl} \bigl( a_i^\dagger a_j^\dagger a_l a_k \bigr).  
\]
Note that those matrix elements $A_{ijkl}$ with $i=j$ or $k=l$ do not contribute to the sum; this is because we only consider the action of $A$ on fermionic states.  

\subsection{Two-Particle Observables}
\label{ch3-observables}

We construct a complete set of 2-particle observables.  First, define $a_I = a_{i_2} a_{i_1}$, for all pairs of modes $I = \set{i_1,i_2}$, $i_1<i_2$.  Also fix an ordering on the pairs $I$.  Let $L$ denote the last pair in the ordering (so $I \prec L$, for all $I \neq L$).  We now define the following observables: 
\begin{align}
X_{IJ} &= a_I^\dagger a_J + a_J^\dagger a_I, \text{ for all $I \prec J$}, \\
Y_{IJ} &= -i a_I^\dagger a_J +i a_J^\dagger a_I, \text{ for all $I \prec J$}, \\
Z_I &= a_I^\dagger a_I, \text{ for all $I$}. 
\end{align}
These operators are Hermitian, with eigenvalues in the interval $[-1,1]$.  Let $\SSS$ be the set of all these observables, except for $Z_L$.  Note that $|\SSS| < d^4$.  

Taking real linear combinations, the operators $S \in \SSS$ form a basis for the space of all 2-local fermionic Hamiltonians, i.e., any 2-local fermionic Hamiltonian can be written in the form 
\[
H = \gamma_0 I + \sum_{S \in \SSS} \gamma_S S, \quad \gamma_0, \gamma_S \in \RR.  
\]

Note that these observables can act on states with arbitrary numbers of particles.  In particular, they can act on an $\n$-particle state $\sigma$, or on the corresponding 2-RDM $\rho = \Tr_{3,\ldots,\n} (\sigma)$.  The expectation values are the same up to a normalization factor:  
\[
\Tr(S\rho) = \frac{1}{\n(\n-1)} \Tr(S\sigma), \quad S \in \SSS.  
\]

\vskipline

The observables $S \in \SSS$ are especially useful for working with 2-particle states.  In particular, the expectation values of $S$ contain complete information about the state.  To see this, let us restrict $S$ to act \textit{only} on the space of 2-particle states.  Then each annihilation operator $a_I$ ``picks out'' a single Slater basis state $\ket{\varphi_I}$, and so the operators $S$ can be written in the following simple way: 
\begin{align*}
Z_I &= \ket{\varphi_I}\bra{\varphi_I} \\
X_{IJ} &= \ket{\varphi_I}\bra{\varphi_J} + \ket{\varphi_J}\bra{\varphi_I} \\
Y_{IJ} &= -i\ket{\varphi_I}\bra{\varphi_J} + i\ket{\varphi_J}\bra{\varphi_I}.  
\end{align*}
Note that $Z_I$ is a projector onto the state $\ket{\varphi_I}$, while $X_{IJ}$ is a rank-2 operator with eigenvalues $\pm 1$ and eigenvectors $\frac{1}{\sqrt{2}} (\ket{\varphi_I} \pm \ket{\varphi_J})$, and $Y_{IJ}$ is a rank-2 operator with eigenvalues $\pm 1$ and eigenvectors $\frac{1}{\sqrt{2}} (\ket{\varphi_I} \pm i\ket{\varphi_J})$.  These operators have the following orthogonality properties:  
\begin{center}
\begin{tabular}{|l|l|l|}
\hline
$A$ & $B$ & $\Tr(AB)$ \\
\hline
$Z_I$ & $Z_{I'}$ & 1 if $I=I'$, 0 otherwise \\
$Z_I$ & $X_{I'J'}$ & 0 \\
$Z_I$ & $Y_{I'J'}$ & 0 \\
\hline
$X_{IJ}$ & $X_{I'J'}$ & 2 if $I=I'$ and $J=J'$, 0 otherwise \\
$X_{IJ}$ & $Y_{I'J'}$ & 0 \\
\hline
$Y_{IJ}$ & $Y_{I'J'}$ & 2 if $I=I'$ and $J=J'$, 0 otherwise \\
\hline
\end{tabular}
\end{center}
(Some of these identities also hold when we consider $\n$-particle states.  However, $Z_I$ and $Z_{I'}$ are not orthogonal when we view them as operators acting on $\n$-particle states.)  

From these orthogonality properties, it follows that any 2-particle state $\rho$ can be written in the form 
\[
\rho = Z_L + \sum_{I \prec L} \alpha_{(Z_I)} (Z_I-Z_L) 
 + \tfrac{1}{2} \sum_{I \prec J} \alpha_{(X_{IJ})} X_{IJ} 
 + \tfrac{1}{2} \sum_{I \prec J} \alpha_{(Y_{IJ})} Y_{IJ}, 
\]
where 
\begin{align*}
\alpha_{(Z_I)} &= \Tr(Z_I \rho), \text{ for all $I \prec L$}, \\
\alpha_{(X_{IJ})} &= \Tr(X_{IJ} \rho), \text{ for all $I \prec J$}, \\
\alpha_{(Y_{IJ})} &= \Tr(Y_{IJ} \rho), \text{ for all $I \prec J$}.  
\end{align*}
(The coefficient in front of $Z_L$ is fixed due to the fact that $\rho$ has trace 1.)  Note that the $\alpha_S$ are simply the expectation values of the observables $S$, that is, $\alpha_S = \Tr(S\rho)$, for all $S \in \SSS$.  

One application of this is to distinguish between two different 2-particle states, $\rho$ and $\rho'$.  We claim that the $\ell_1$ distance $\norm{\rho-\rho'}_1$, and the difference in expectation values $|\Tr(S\rho)-\Tr(S\rho')|$, are related up to a polynomial factor.  More precisely, we show the following:  
\begin{lem}
\label{ch3-lem-observables}
There exists some $S \in \SSS$ such that $|\Tr(S\rho)-\Tr(S\rho')| \geq \norm{\rho-\rho'}_1 / 2d^4$.  Also, for all $S \in \SSS$, $|\Tr(S\rho)-\Tr(S\rho')| \leq \norm{\rho-\rho'}_1$.  
\end{lem}

\noindent
Proof:  For the first claim, we let $\alpha_S = \Tr(S\rho)$ and $\alpha'_S = \Tr(S\rho')$, and we write 
\[
\rho-\rho' = \sum_{I \prec L} (\alpha_{(Z_I)}-\alpha'_{(Z_I)}) (Z_I-Z_L) 
 + \tfrac{1}{2} \sum_{I \prec J} (\alpha_{(X_{IJ})}-\alpha'_{(X_{IJ})}) X_{IJ} 
 + \tfrac{1}{2} \sum_{I \prec J} (\alpha_{(Y_{IJ})}-\alpha'_{(Y_{IJ})}) Y_{IJ}.  
\]
By the triangle inequality, and using the fact that $\norm{Z_I-Z_L}_1$, $\norm{X_{IJ}}_1$, $\norm{Y_{IJ}}_1 \leq 2$ when we view these as operators on 2-particle states, we get 
\[
\norm{\rho-\rho'}_1 
 \leq 2 \sum_{I \prec L} |\alpha_{(Z_I)}-\alpha'_{(Z_I)}| 
 + \sum_{I \prec J} |\alpha_{(X_{IJ})}-\alpha'_{(X_{IJ})}| 
 + \sum_{I \prec J} |\alpha_{(Y_{IJ})}-\alpha'_{(Y_{IJ})}|, 
\]
so there must be some $S \in \SSS$ such that 
\[
|\alpha_S-\alpha'_S| \geq \frac{\norm{\rho-\rho'}_1}{2|\SSS|} \geq \frac{\norm{\rho-\rho'}_1}{2d^4}.  
\]

Now we show the second claim.  For any $S \in \SSS$, let $p$ be the distribution of the outcomes when one measures $S$ on the state $\rho$, and let $p'$ be the distribution of the outcomes when one measures $S$ on the state $\rho'$.  Then, using the fact that the measurement outcomes are in the range $[-1,1]$, we have that 
\[
|\Tr(S\rho)-\Tr(S\rho')| \leq \norm{p-p'}_1 \leq \norm{\rho-\rho'}_1.  
\]
$\square$

\section{The $N$-representability and Fermionic Local Hamiltonian problems} 

We have a system of $\n$ electrons, and a basis set consisting of $\m$ single-electron orbitals.  (The nuclei are assumed to be fixed, possibly in some non-equilibrium geometry.)  For our purposes, $\n$ is the parameter that describes the size of the system.  $\m$ is typically much larger than $\n$, and the space of $\n$-electron states has dimension $\binom{\m}{\n}$; if $\m \geq c\n$ for some constant $c > 1$, then this grows exponentially in $\n$.  However, in practice $\m$ cannot be chosen too large, because the 2-RDM, and the 2-electron interaction in the Hamiltonian, are described by matrices of dimension $\binom{\m}{2}$.  We will be mainly interested in cases where $\n \leq \m \leq \poly(\n)$.  We would like to solve $\n$-representability, or find ground state energies, with additive error $\pm 1/\poly(\n)$.  

Formally, we define the $N$-representability problem as follows:  
\begin{quote}
Consider a system of $\n$ fermions, where each particle has $\m$ energy levels.  We are given a 2-particle density matrix $\rho$, of size $\binom{\m}{2} \times \binom{\m}{2}$.  In addition, we are given a string ``$1^s$'' (the unary encoding of a natural number $s$), and a real number $\beta \geq 1/s$.  

All numbers are specified with $\poly(\n,s)$ bits of precision.  

The problem is to distinguish between the following two cases:  
\begin{itemize}
\item There exists an $\n$-fermion state $\sigma$ such that $\Tr_{3,\ldots,\n}(\sigma) = \rho$.  In this case, answer ``YES.''  
\item For all $\n$-fermion states $\sigma$, $\lVert \Tr_{3,\ldots,\n}(\sigma) - \rho \rVert_1 \geq \beta$.  In this case, answer ``NO.''  
\end{itemize}
If neither of these cases applies, then one may answer either ``YES'' or ``NO.''  

(Note that we use the $\ell_1$ matrix norm, $\lVert A \rVert_1 = \Tr |A|$, to measure the distance between $\sigma$ and $\rho$.)  
\end{quote}

An instance of this problem is described by a string of length $\ell = \Theta(\m^2 \poly(\n,s)$ $+ s)$, and we say an algorithm solves the problem efficiently if it takes time polynomial in $\ell$.  We claim that this formal definition is equivalent to our intuitive notion of what it means to solve the problem.  Intuitively, an algorithm solves the problem efficiently if, on instances where $\n \leq \m \leq \poly(\n)$ and $\beta \geq 1/\poly(\n)$, the algorithm runs in time $\poly(\n)$.  

Clearly, the formal definition implies the intuitive one, since on instances where $\n \leq \m \leq \poly(\n)$ and $\beta \geq 1/\poly(\n)$, the length of the input is $\leq \poly(\n)$.  

To show that the intuitive definition implies the formal one, we use a padding argument.  Suppose the intuitive definition holds.  Then, given an arbitrary instance of the problem, one can solve it in time polynomial in the length of the input, as follows.  One modifies the problem to have $q$ extra modes (energy levels) and $q$ extra particles, and one modifies the 2-fermion state $\rho$ to enforce the constraint that these $q$ extra modes are always occupied.  Also, one decreases the error parameter $\beta$ by a factor of $(\m+q)^2$.  This produces a new instance of the problem, which is equivalent to the old instance.  In this way we can increase $\n$ and $\m$ so that the promises $\n \leq \m \leq \poly(\n)$ and $\beta \geq 1/\poly(\n)$ are satisfied, but $\n$ is still at most polynomially large compared to the length of the input.  Then the problem can be solved in time $\poly(\n)$, which is polynomial in the length of the input.  

\vskipline

We also define the Fermionic Local Hamiltonian problem, as follows:  
\begin{quote}
Consider a system of $\n$ fermions, where each particle has $\m$ energy levels.  We are given a 2-particle Hamiltonian $A$, which is a $\binom{\m}{2} \times \binom{\m}{2}$ Hermitian matrix with $\norm{A} \leq 1$.  In addition, we are given a string ``$1^s$'' (the unary encoding of a natural number $s$), and two real numbers $a$ and $b$, such that $b-a \geq 1/s$.  

All numbers are specified with $\poly(\n,s)$ bits of precision.  

Define the $\n$-particle Hamiltonian to be $H = \sum_{i \neq j} A^{(ij)}$, restricted to the subspace of $\n$-fermion states.  The problem is to distinguish between the following two cases:  
\begin{itemize}
\item If $H$ has an eigenvalue that is $\leq a$, answer ``YES.''  
\item If all the eigenvalues of $H$ are $\geq b$, answer ``NO.''  
\end{itemize}
If neither of these cases applies, then one may answer either ``YES'' or ``NO.''  
\end{quote}

Again, an instance of this problem is described by a string of length $\ell = \Theta(\m^2$ $\poly(\n,s) + s)$, and we say an algorithm solves the problem efficiently if it takes time polynomial in $\ell$.  This formal definition is equivalent to our intuitive notion of what it means to solve the problem (using a padding argument, as above).  Intuitively, an algorithm solves the problem efficiently if, on instances where $\n \leq \m \leq \poly(\n)$ and $\beta \geq 1/\poly(\n)$, the algorithm runs in time $\poly(\n)$.  

\section{Our Results}

First, we show that any 2-local Hamiltonian of spins can be simulated using a 2-local Hamiltonian of fermions with $\m=2\n$, and hence Fermionic Local Hamiltonian is QMA-hard.  Then, using techniques of convex programming, we show that an efficient algorithm for $\N$-representability would allow us to estimate the ground state energies of 2-local Hamiltonians of fermions; thus, $\N$-representability is QMA-hard.

One might expect that Fermionic Local Hamiltonian would be QMA-hard, but it is somewhat surprising to find that $N$-representability, which was believed to be tractable, is also QMA-hard.  In fact, $N$-representability is QMA-hard for precisely the same reasons that first attracted the interest of the quantum chemists:  convex optimization.  Previous work tried to formulate explicit ``$N$-representability conditions'' that could be used in variational calculations.  In this paper we use a more general framework, convex optimization with a membership oracle (see chapter 2) \cite{YN,GLS}, to show that \textit{any} efficient solution to $N$-representability is impossible unless QMA is tractable.

Second, we show that the above two problems are in QMA.  The natural ``witness'' for these problems is a fermionic state; using the Jordan-Wigner transform, this state can be represented using qubits, in such a way that its local properties can be efficiently verified by a quantum computer.  This is similar to the techniques used to simulate fermionic systems on a quantum computer \cite{Ortiz-et-al-00,Bravyi-Kitaev-00,Abrams-Lloyd-97}.  

\section{Fermionic Local Hamiltonian is QMA-hard}

\begin{thm}
There is a poly-time mapping reduction from 2-Local Hamiltonian to Fermionic 2-Local Hamiltonian.
\end{thm}

\noindent
Proof:  We show how 
to map a 2-local Hamiltonian, $H_{\text{qubit}}$, defined on a system of $\n$ qubits, to a 2-local Hamiltonian
on fermions, $H_{\text{fermi}}$, with $\m = 2\n$ modes, such that the ground state energy remains the same. (This is
the opposite of what has been done in \cite{VC}.) 

We represent each qubit $i$ as a single fermion that can be in two different modes $a_i,b_i$; so each $\n$-qubit basis state corresponds to the following $\n$-fermion state:
\begin{equation}
\label{ch3-qf-state}
\ket{z_1} \otimes \cdots \otimes \ket{z_\n} \mapsto
   (a_1^\dagger)^{1-z_1} (b_1^\dagger)^{z_1} \cdots
   (a_\n^\dagger)^{1-z_\n} (b_\n^\dagger)^{z_\n} \ket{\Omega}.
\end{equation}
The fermionic Hamiltonian, $H_{\text{fermi}}$, consists of two parts:  $H_A$, which ``simulates'' $H_{\text{qubit}}$ on the fermionic states shown above; and $H_B$, which enforces the constraint that there is exactly one fermion at each site $i$.  

First we construct $H_A$.  A Pauli matrix acting on qubit $i$ corresponds to a bilinear function of the creation and annihilation operators:
\begin{equation}
\label{ch3-qf-pauli}
\sigma^x_i \mapsto a_i^\dagger b_i + b_i^\dagger a_i; \qquad
\sigma^y_i \mapsto i( b_i^\dagger a_i - a_i^\dagger b_i ); \qquad
\sigma^z_i \mapsto 1 - 2 b_i^\dagger b_i.
\end{equation}
(Note:  when we write $\sigma^x_i$, we mean an operator on all $\n$ qubits, which is a tensor product of $\sigma^x$ on qubit $i$, and the identity matrix on the other $\n-1$ qubits.)  The above operators commute with $a_j^\dagger$ and $b_j^\dagger$, for all $j \neq i$; hence they act correctly on the states in (\ref{ch3-qf-state}).  
We also consider products of two Pauli matrices acting on qubits $i$ and $j$, e.g., $\sigma^x_i \sigma^z_j$.  This corresponds to a product of two fermionic operators, e.g., $(a_i^\dagger b_i + b_i^\dagger a_i) (1 - 2 b_j^\dagger b_j)$.  (Note that $\sigma^x_i \sigma^z_j$ is equal to the tensor product of $\sigma^x$ on qubit $i$, $\sigma^z$ on qubit $j$, and the identity matrix on the other $\n-2$ qubits.)  

$H_{\text{qubit}}$ can be written as a linear combination of terms of the form $\sigma^u_i$ and $\sigma^u_i \sigma^v_j$, where $u,v \in \set{x,y,z}$ and $i,j \in \set{1,\ldots,\n}$.  We then construct $H_A$ by substituting the corresponding fermionic operators.  

Next we construct $H_B$.  We want to guarantee that, for each $i$, exactly one of the modes $a_i$ and $b_i$ is occupied.  This can be achieved by setting $H_B = \sum_{i=1}^\n \Pi_i$, where 
\begin{equation}
\label{ch3-qf-pi}
\Pi_i = 1 + (2a_i^\dagger a_i-1) (2b_i^\dagger b_i-1).
\end{equation}
To see why this works, note that $\Pi_i$ is diagonal in the basis consisting of the states 
\begin{equation*}
(a_1^\dagger)^{s_1} (b_1^\dagger)^{t_1} \cdots
(a_\n^\dagger)^{s_\n} (b_\n^\dagger)^{t_\n} \ket{\Omega}, 
\end{equation*}
and has eigenvalue 2 if $s_i = t_i$, and eigenvalue 0 if $s_i \neq t_i$.  

In addition, we claim that all of the $\Pi_i$ are biquadratic and commute with all of the operators introduced in (\ref{ch3-qf-pauli}).  (To see this, consider how the operators in (\ref{ch3-qf-pauli}) act on the eigenstates of $\Pi_i$.  Observe that each operator in (\ref{ch3-qf-pauli}) maps a 0-eigenstate to a 0-eigenstate, and maps a 2-eigenstate to a 2-eigenstate.)  

The full Hamiltonian $H_{\text{fermi}}$ is given by 
\[
H_{\text{fermi}} = H_A + \beta H_B, 
\]
where $\beta$ is a real number which we will choose later.  We claim that $H_{\text{fermi}}$ has the same ground state energy as $H_{\text{qubit}}$.  We know $H_A$ and $H_B$ commute, so $H_{\text{fermi}}$ is block-diagonal with respect to the eigenspaces of $H_B$.  Note that the eigenvalues of $H_B$ are $0, 2, 4, \ldots, 2\n$.  Now set $\beta$ equal to a constant times the norm of $H_A$.  This guarantees that the ground state of $H_{\text{fermi}}$ will lie in the 0-eigenspace of $H_B$, so it will have exactly one fermion per site.  Thus the ground state of $H_{\text{fermi}}$ corresponds to the ground state of $H_{\text{qubit}}$, and they have the same energy.  

Finally, note that $\norm{H_{\text{fermi}}} \leq O(\n^2 \norm{H_{\text{qubit}}})$.  (To see this, note that $\norm{H_{\text{fermi}}} \leq O(\norm{H_A})$.  We constructed $H_A$ from $H_{\text{qubit}}$ by writing $H_{\text{qubit}}$ as a linear combination of Pauli matrices; there were $O(\n^2)$ terms in the sum, each having norm $O(\norm{H_{\text{qubit}}})$; hence $\norm{H_A} \leq O(\n^2 \norm{H_{\text{qubit}}})$.)  

Also, note that $H_{\text{fermi}}$ only contains terms with at most 2 annihilation and 2 creation operators.  Thus it is a 2-local fermionic Hamiltonian.  $\square$

\vskipline

Since 2-Local Hamiltonian is QMA-hard \cite{KKR}, this implies that Fermionic 2-Local Hamiltonian is QMA-hard.  

We remark that this mapping from qubits to fermions may have other applications.  For instance, one can show that adiabatic quantum computation on fermionic systems is universal.\footnote{Thanks to Stephen Jordan for pointing this out.}  One direction is already known:  one can use a quantum circuit to simulate the time evolution of a local Hamiltonian of fermions \cite{Ortiz-et-al-00, Bravyi-Kitaev-00, Abrams-Lloyd-97}.  We can show the reverse direction as follows:  to simulate a quantum circuit, first construct an adiabatic local Hamiltonian on qubits \cite{ADKLLR}, then use the above mapping to translate it into an adiabatic local Hamiltonian on fermions.  We claim that this mapping preserves the gap between the two lowest energy levels.  To see this, observe that the energy spectrum of $H_{\text{fermi}}$ contains an exact copy of the spectrum of $H_{\text{qubit}}$ (in the 0-eigenspace of $H_B$), along with other higher energy levels (in the other eigenspaces of $H_B$).  Thus the low-lying energy levels of $H_{\text{fermi}}$ and $H_{\text{qubit}}$ are identical.  

\section{$N$-representability is QMA-hard}

\subsection{Convex Optimization with a Membership Oracle}

First we review the basic result of chapter 2:  given a membership oracle for a closed convex set $K \subseteq \RR^n$, one can solve the optimization problem over $K$ in polynomial time.  This holds provided that $K$ contains a ball of radius $r$ centered at a known point $p$, and $K$ is contained in a ball of radius $R$ centered at the origin, such that $R/r \leq \poly(n)$.  Furthermore, the precision required for the membership oracle depends polynomially on the precision desired for the solution of the optimization problem.  Formally, we say that $WOPT_\varepsilon$ poly-time reduces to $WMEM_\delta$, for some $\delta \geq \poly(\varepsilon, (r/R), (1/n))$; this is Proposition \ref{ch2-asr-opt-mem}.  

We rephrase this result slightly, so it will be more convenient to use later.  First, we define a variant of the weak optimization problem, $WOPT^*_\varepsilon$, as follows:  
\begin{verse}
Given $c \in \RR^n$, $\norm{c} = 1$, $\gamma \in \RR$, and $\varepsilon \in \RR$, $\varepsilon > 0$, all specified with $\poly(n)$ bits of precision.\\
If there exists a vector $y \in K$ with $c \cdot y \geq \gamma + \varepsilon$, then answer ``YES.''\\
If for all $x \in K$, $c \cdot x \leq \gamma - \varepsilon$, then answer ``NO.''
\end{verse}
(This problem differs from $WOPT_\varepsilon$ in that $y$ does not have to be deep inside $K$, and we no longer consider $x$ that are slightly outside of $K$.)  We also define a variant of the weak membership problem, $WMEM^*_\delta$, as follows:  
\begin{verse}
Given $y \in \RR^n$, and $\delta \in \RR$, $\delta > 0$, all specified with $\poly(n)$ bits of precision.\\
If $y \in K$, then answer ``YES.''\\
If $y \notin S(K,\delta)$, then answer ``NO.''
\end{verse}
(This problem differs from $WMEM_\delta$ in that $y$ does not have to be deep inside $K$.)  


We show the following result:  
\begin{thm}
\label{ch3-thm-opt-mem}
Let $K$ be any closed convex set in $\RR^n$, such that $S(p,r) \subseteq K \subseteq S(0,R)$, as defined above.  Then there is an oracle reduction from $WOPT^*_\varepsilon$ to $WMEM^*_\delta$, for some $\delta \geq \poly(\varepsilon, (r/R), (1/n))$, which runs in time $\poly(n, (R/r), (1/\varepsilon))$.  
\end{thm}

\noindent
Proof:  First we show a mapping reduction from $WOPT^*_\varepsilon$ to $WOPT_{(\varepsilon r/4R)}$.  The reduction is trivial---we only change the value of the parameter $\varepsilon$.  Suppose we have a ``YES'' instance of $WOPT^*$, i.e., there exists $x \in K$ such that $c \cdot x \geq \gamma + \varepsilon$.  Define $x' = (1-\delta) x + \delta p$, for some $\delta$ to be chosen later.  Then $S(x', \delta r) \subseteq K$, and $c \cdot x' = (1-\delta) c \cdot x + \delta c \cdot p \geq \gamma + \varepsilon - 2\delta R$.  Now set $\delta = \varepsilon/4R$.  Then $S(x', (\varepsilon r/4R)) \subseteq K$, and $c \cdot x' \geq \gamma + \varepsilon/2$, so this is a ``YES'' instance of $WOPT$.  

Now suppose we have a ``NO'' instance of $WOPT^*$, i.e., for all $x \in K$, $c \cdot x \leq \gamma - \varepsilon$.  This implies that for any $x' \in S(K,\varepsilon/2)$, $c \cdot x' \leq \gamma - \varepsilon/2$, so this is a ``NO'' instance of $WOPT$.  

Next, we use Proposition \ref{ch2-asr-opt-mem} to get a reduction from $WOPT$ to $WMEM$.  Finally, $WMEM$ trivially reduces to $WMEM^*$.  $\square$

\subsection{$N$-representability is QMA-hard}

\begin{thm}
There is a poly-time oracle reduction from Fermionic 2-Local Hamiltonian to $N$-representability.  
\end{thm}

\noindent
Proof:  Let us now assume that we have an efficient algorithm for $\N$-representability.  We claim that this allows us to efficiently determine the ground state energy of any 2-local Hamiltonian on fermions, $H_{\text{fermi}}$.  The basic idea is to find the ground state of $H_{\text{fermi}}$ using convex programming.  However, instead of the full $\n$-particle density matrix, we will just find the 2-particle reduced density matrix, subject to the $\n$-representability constraint.  The resulting convex program has polynomially many variables, and by assumption we have an algorithm that can test whether the $\n$-representability constraint is satisfied.  Thus this program can be solved, using convex optimization with a membership oracle.  

We now describe the details.  First, note that the interesting behavior in $H_{\text{fermi}}$ occurs in the subspace of states with exactly $\N$ particles.  (We are assuming that $H_{\text{fermi}}$ comes from the reduction given in the previous section.)  Restricting ourselves to this subspace, we have the identity $a_i^\dagger a_j = \frac{1}{\n-1} a_i^\dagger (\sum_k a_k^\dagger a_k) a_j$, and we can write all the terms in $H_{\text{fermi}}$ in the form $a_i^\dagger a_j^\dagger a_l a_k$.  

We can view $H_{\text{fermi}}$ as describing a system with an \textit{arbitrary} number of particles; $H_{\text{fermi}}$ simply specifies a 2-particle interaction, which acts on all pairs of particles.  (Note that, when written in second-quantized notation, $H_{\text{fermi}}$ has the same form irrespective of the number of particles.)  In particular, we can view $H_{\text{fermi}}$ as describing a system of 2 particles.  Now, suppose this system is in state $\rho$, and suppose that $\rho$ is $\n$-representable, that is, there exists an $\n$-particle state $\sigma$ such that $\Tr_{3,\ldots,\n} (\sigma) = \rho$.  Then, using the identity (\ref{ch3-2rdm-2ndquant}) for the matrix elements of the 2-RDM, we have that 
\[
\Tr(H_{\text{fermi}} \rho) = \frac{1}{\n(\n-1)} \Tr(H_{\text{fermi}} \sigma).  
\]
This says that the 2-particle state $\rho$ has the same energy as the $\n$-particle state $\sigma$, scaled by a factor of $1/\n(\n-1)$.  

We construct a convex program that finds a 2-fermion density matrix $\rho$ that is $\N$-representable, and that minimizes $\Tr(H_{\text{fermi}} \rho)$.  This tells us the ground state energy of $H_{\text{fermi}}$ (for the $\n$-particle system).  Note that this program has polynomially many variables, the set of $\N$-representable states is convex, and $\Tr(H_{\text{fermi}} \rho)$ is a linear function of $\rho$.  Assuming that we have an efficient algorithm for $\N$-representability, we claim that we can solve this convex program in polynomial time.  

\vskipline

One technical point concerns the geometry of the set $K$ of feasible solutions.  The set $K$ must be full-dimensional, i.e., $K$ cannot lie in a lower-dimensional subspace.  (We also need $K$ to have outer radius $R$ and inner radius $r$, such that $R/r$ is at most polynomially large; we will revisit this issue later.)  So we have to represent the 2-fermion state $\rho$ in such a way that there are no redundant variables.  To this end, let $\SSS$ be the complete set of 2-particle observables introduced in section \ref{ch3-observables}, and let $\ell = |\SSS|$; note that $\ell < \m^4$.  

We represent $\rho$ in terms of its expectation values $\alpha_S = \Tr(S\rho)$, for all observables $S \in \SSS$.  Let $\vec{\alpha} \in \RR^\ell$ denote the vector of these expectation values, $\vec{\alpha} = (\alpha_S)_{S \in \SSS}$.  Then we define $K$ to be the set of all vectors $\vec{\alpha} \in \RR^\ell$ such that the corresponding 2-fermion state $\rho$ is $\N$-representable.  Note that the $\N$-representability algorithm lets us test whether a given point $\vec{\alpha}$ is in $K$.  

We write our Hamiltonian in the form 
\[
H_{\text{fermi}} = \gamma_0 I + \sum_{S \in \SSS} \gamma_S S.  
\]
Since we are viewing $H_{\text{fermi}}$ as an operator on the space of 2-particle states, we have that the operators $S \in \SSS$ are orthogonal (see section \ref{ch3-observables}).  So the coefficients $\gamma_0$ and $\gamma_S$ are given by the formulas 
\[
\gamma_0 = \Tr(H_{\text{fermi}} Z_L)
\]
\[
\gamma_S = \frac{\Tr(H_{\text{fermi}} S) - \gamma_0 \Tr(S)}{\Tr(S^2)}.
\]

Define $\vec{\gamma} \in \RR^\ell$ to be the vector $\vec{\gamma} = (\gamma_S)_{S \in \SSS}$.  Then we can write 
\[
\Tr(H_{\text{fermi}} \rho) = \gamma_0 + \sum_{S \in \SSS} \gamma_S \alpha_S = \gamma_0 + \vec{\gamma} \cdot \vec{\alpha}.  
\]
So our convex program can be written as follows: find some $\vec{\alpha} \in K$ that minimizes the function $f(\vec{\alpha}) = \gamma_0 + \vec{\gamma} \cdot \vec{\alpha}$.

For future reference, let us bound the size of $\gamma_0$ and $\vec{\gamma}$.  Recall that, when restricted to the space of 2-particle states, the operators $S \in \SSS$ have rank 1 or 2, with eigenvalues 1 or $\pm 1$ (see section \ref{ch3-observables}).  So $|\gamma_0| \leq \norm{H_{\text{fermi}}}$, and $|\gamma_S| \leq |\Tr(H_{\text{fermi}} S)| + |\gamma_0 \Tr(S)| \leq 3\norm{H_{\text{fermi}}}$.  Also, recall that $H_{\text{fermi}}$ is defined by a 2-particle interaction $A$ where $\norm{A} \leq 1$.  Since we have only 2 particles, $H_{\text{fermi}} = A$, hence $\norm{H_{\text{fermi}}} \leq 1$.  So $|\gamma_0| \leq 1$ and $|\gamma_S| \leq 3$, and hence $\norm{\vec{\gamma}} \leq 3\sqrt{\ell}$.  

\vskipline

Given an algorithm for $\n$-representability, we can solve the above convex program (and thus Fermionic Local Hamiltonian) in polynomial time.  The logic is as follows:  Fermionic Local Hamiltonian reduces to the weak optimization problem $WOPT^*$ on the set $K$, which reduces to the weak membership problem $WMEM^*$ on the set $K$, which reduces to $\n$-representability.  Numerical precision is a concern here, because the algorithm for $N$-representability is allowed to make mistakes near the boundary of the set $K$.  We claim that, in order to solve Fermionic Local Hamiltonian with error $b-a$, we require an algorithm for $\n$-representability with error $\beta$, where $\beta \geq \poly((b-a), 1/\m)$.  Also, the overall reduction runs in time $\poly(\m, 1/(b-a))$.  

The first and last steps in the reduction are easy, using the definitions of $WOPT^*$ and $WMEM^*$.  Using the remarks above, we have that Fermionic Local Hamiltonian (with error $b-a$) reduces to $WOPT^*_\varepsilon$ with $\varepsilon \geq \frac{1}{\n(\n-1)} \cdot \frac{1}{3\sqrt{\ell}} \cdot \frac{b-a}{2}$.  And, using Lemma \ref{ch3-lem-observables}, $WMEM^*_\delta$ reduces to $\n$-representability with error $\beta \geq \delta/\sqrt{\ell}$.  

The middle step in the reduction makes use of Theorem \ref{ch3-thm-opt-mem}, and requires some further explanation.  This step requires a guarantee that $K$ is contained in a ball of radius $R$ centered at 0, and $K$ contains a ball of radius $r$ centered at some point $p$, such that $R/r$ is at most polynomially large.  In our case, we have the following bounds, which we prove in the next section.  
\begin{lem}
\label{ch3-geometry}
$K$ is contained in a ball of radius $R = \sqrt{\ell}$, and $K$ contains a ball of radius $r = 1/\ell^2\m^5$.  
\end{lem}
(Also recall that $\ell < \m^4$.)  Substituting into Theorem \ref{ch3-thm-opt-mem}, we get that $WOPT^*_\varepsilon$ reduces to $WMEM^*_\delta$ with $\delta \geq \poly(\varepsilon, 1/\m)$.  








Thus, given an efficient algorithm for $N$-representability, we get an efficient algorithm for Fermionic Local Hamiltonian.  This completes the proof that $\N$-represent\-ability is QMA-hard. $\square$


\subsection{Bounds on the Geometry of $K$} 

\noindent
Proof of Lemma \ref{ch3-geometry}:  We claim that $K$ is contained in a ball of radius $R = \sqrt{\ell}$, and $K$ contains a ball of radius $r = 1/\ell^2\m^5$.  

The first statement is easy to see, since for all $\vec{\alpha} \in K$, 
and for all $S \in \SSS$, we have $-1 \leq \alpha_S \leq 1$.  

The second statement is less trivial.  The obvious argument is as follows:  let $\sigma$ be the maximally mixed state on $\n$ particles, let $\rho$ be the corresponding reduced 2-particle state, and show that for any small perturbation of $\rho$, one can perturb $\sigma$ in a way that agrees with $\rho$.  But this argument runs into complications, because it is hard to perturb $\sigma$ in a way that affects just two modes; one usually ends up affecting $\n$ modes simultaneously.  

Instead, we use the following indirect argument.  (We first sketch the overall argument, then fill in the details.)  We consider $\N$-representability for different values of $\N$; let $K_\n$ denote the set of all vectors $\vec{\alpha}$ that are $\n$-representable.  We also define the ``particle-hole'' observables, where the roles of $a_i$ and $a_i^\dagger$ are reversed.  Let $\SSS'$ be the set of 2-hole observables, 
\begin{align*}
X'_{IJ} &= a_I a_J^\dagger + a_J a_I^\dagger, \text{ for all $I \prec J$}, \\
Y'_{IJ} &= -i a_I a_J^\dagger +i a_J a_I^\dagger, \text{ for all $I \prec J$}, \\
Z'_I &= a_I a_I^\dagger, \text{ for all $I$ except the last one}. 
\end{align*}
Let $\vec{\alpha}'$ denote a vector containing expectation values for these observables, and let $K'_\n$ be the set of all $\vec{\alpha}'$ that are $\n$-representable.  

It is easy to see that $K_2$ contains a ball of radius $1/\poly(\ell)$ (this is the trivial case).  Now, using the anti-commutation relations, we can write each 2-particle observable as a linear combination of 2-hole observables, and vice versa.  (This holds for states where the total number of particles is fixed.)  This implies an invertible linear transformation $A$ that maps $K_2$ to $K'_2$.  We show that this transformation does not shrink $K_2$ by more than a polynomial factor.

Next, note that $\tfrac{2}{(\m-2)(\m-3)} K'_2 = K_{\m-2}$, since a state with 2 holes can be viewed as a state with $\m-2$ particles.  (There is also a normalization factor, to account for the increase in the number of particles.)  Thus $K_{\m-2}$ contains a ball of radius $1/\poly(\ell)$.  Also, note that if a vector $\vec{\alpha}$ is $\n$-representable, then it is also $(\n-1)$-representable; so, for all $3 \leq \n \leq \m-2$, we have $K_\n \subseteq K_{\n-1}$.  Thus, for all $2 \leq \n \leq \m-2$, $K_\n$ contains a ball of radius $1/\poly(\ell)$.  This completes the argument; now we describe the details.

\vskipline

First, some remarks about the definition of the set $K_\n$.  We define 
\[
\begin{split}
K_\n = \set{ \vec{\alpha} \in \RR^\ell \;|\; & \text{there exists a 2-particle state $\rho$,} \\
 & \text{such that $\rho$ is $\n$-representable,} \\
 & \text{and for all observables $S \in \SSS$, $\alpha_S = \Tr(S\rho)$} }.  
\end{split}
\]
We can also describe $K_\n$ in terms of an $\n$-particle state $\sigma$, where $\Tr_{3,\ldots,\n} (\sigma) = \rho$.  However, some care is needed with the normalization factor for the expectation values $\Tr(S\sigma)$.  Recall that 
\[
\rho_{ijkl} = \tfrac{1}{\n(\n-1)} \Tr(a_k^\dagger a_l^\dagger a_j a_i \sigma)
 = \tfrac{1}{2} \Tr(a_k^\dagger a_l^\dagger a_j a_i \rho).  
\]
Thus $\Tr(S\sigma) = \tfrac{\n(\n-1)}{2} \Tr(S\rho)$.  So $K_\n$ is given by 
\[
\begin{split}
K_\n = \set{ \vec{\alpha} \in \RR^\ell \;|\; & \text{there exists an $\n$-particle state $\sigma$,} \\
 & \text{such that for all observables $S \in \SSS$, $\alpha_S = \tfrac{2}{\n(\n-1)} \Tr(S\sigma)$} }.  
\end{split}
\]

The definition of the set $K'_\n$ is exactly the same, but using the set of observables $\SSS'$ in place of $\SSS$.  

\vskipline

We claim that $K_2$ contains a ball of radius $1/\poly(\ell)$ (we will give a precise bound below).  Note that this is the trivial case of $\N$-representability; $K_2$ is the set of all vectors $\vec{\alpha}$ that correspond to 2-particle fermionic states.  Consider the vector $\vec{\alpha}$ that corresponds to the maximally mixed state on two fermions, $\sigma = I/\binom{\m}{2}$.  The components of the vector $\vec{\alpha}$ are given by 
\begin{align*}
\alpha_{(Z_I)} &= \Tr(Z_I\sigma) = 1/\tbinom{\m}{2}, \\
\alpha_{(X_{IJ})} &= \Tr(X_{IJ}\sigma) = 0, \\
\alpha_{(Y_{IJ})} &= \Tr(Y_{IJ}\sigma) = 0.  
\end{align*}

We claim that, for any perturbation $\vec{\alpha} + \vec{\eta}$, $\norm{\vec{\eta}} \leq 1/\poly(\ell)$, we can perturb $\sigma$ in such a way that it agrees with $\vec{\alpha} + \vec{\eta}$.  We construct this perturbation as follows.  Recall that when we defined the set of observables $\SSS$, we chose an ordering on all the pairs of modes.  Let $L$ denote the pair of modes that comes last in this ordering.  (Also recall that we excluded the observable $Z_L$ from the set $\SSS$.)  Now consider the following perturbation:  
\[
\sigma' = 
\sigma + \sum_{I \prec L} \eta_{(Z_I)} (Z_I-Z_L) 
 + \tfrac{1}{2} \sum_{I \prec J} \eta_{(X_{IJ})} X_{IJ} 
 + \tfrac{1}{2} \sum_{I \prec J} \eta_{(Y_{IJ})} Y_{IJ}.  
\]
Here we view $(Z_I-Z_L)$, $X_{IJ}$ and $Y_{IJ}$ as operators on the space of 2-particle states.  

This is a legal density matrix (positive semidefinite with trace 1), provided that $\vec{\eta} \leq 1/\ell\m^2$.  To see this, note that the operators $(Z_I-Z_L)$, $X_{IJ}$ and $Y_{IJ}$ have trace 0 and norm at most 1, and note that $\sigma = I/\binom{\m}{2}$.  

Also, we have that $\sigma'$ agrees with $\vec{\alpha} + \vec{\eta}$, that is, 
\begin{align*}
\Tr(Z_I\sigma') &= \alpha_{(Z_I)} + \eta_{(Z_I)}, \\
\Tr(X_{IJ}\sigma') &= \alpha_{(X_{IJ})} + \eta_{(X_{IJ})}, \\
\Tr(Y_{IJ}\sigma') &= \alpha_{(Y_{IJ})} + \eta_{(Y_{IJ})}.  
\end{align*}
This follows from the orthogonality properties of $Z_I$, $X_{IJ}$ and $Y_{IJ}$, shown in section \ref{ch3-observables}.  (We emphasize that we are viewing these as operators on 2-particle states.  $Z_I$ and $Z_{I'}$ are not orthogonal when we view them as operators on $\n$-particle states.)  

Thus we have shown that $K_2$ contains a ball of radius $1/\ell\m^2$.  

\vskipline

Next, we construct an invertible linear transformation $A$ that maps $K_2$ to $K'_2$.  We begin with the following identity, which comes from repeated application of the anticommutation relations:\footnote{Note that the subscript $a$ refers to one of the modes, while $a$ in regular type is an annihilation operator.}  
\[
a_a^\dagger a_b^\dagger a_d a_c = 
 \delta_{bd} \delta_{ac} - \delta_{ad} \delta_{bc} 
 + \delta_{ad} a_c a_b^\dagger + \delta_{bc} a_d a_a^\dagger 
 - \delta_{ac} a_d a_b^\dagger - \delta_{bd} a_c a_a^\dagger 
 + a_d a_c a_a^\dagger a_b^\dagger.  
\]
Thus if we write $I = \set{a,b}$ and $J = \set{c,d}$, we get the following expressions for $a_I^\dagger a_J$:  
\[
\begin{split}
a_I^\dagger a_J
 &= a_J a_I^\dagger, \text{ if $I \cap J = \emptyset$}, \\
 &= 1 - a_b a_b^\dagger - a_a a_a^\dagger + a_I a_I^\dagger, \text{ if $I=J$}, \\
 &= -a_c a_a^\dagger + a_J a_I^\dagger, \text{ if $a \neq c$ and $b=d$}, \\
 &\text{etc.}
\end{split}
\]
This shows that $a_I^\dagger a_J$, which is a 2-particle operator, can be written as a linear combination of 1-hole and 2-hole operators.  Now we restrict all operators to act on the space of states with exactly 2 particles (or equivalently, $\m-2$ holes).  Then we have the identity 
\[
a_f a_e^\dagger = (\tfrac{1}{\m-3} \sum_{g \notin \set{e,f}} a_g a_g^\dagger) a_f a_e^\dagger 
 = \tfrac{1}{\m-3} \sum_{g \notin \set{e,f}} a_g a_f a_e^\dagger a_g^\dagger.  
\]
So a 1-hole operator can be written in terms of 2-hole operators.  Substituting into the previous equation, we get that any 2-particle operator can be written as a linear combination of 2-hole operators.  

Furthermore, the 2-particle observables $Z_I$, $X_{IJ}$ and $Y_{IJ}$ can be written as linear combinations of the 2-hole observables $Z'_I$, $X'_{IJ}$ and $Y'_{IJ}$ (note that $X_{IJ}$ is constructed from $a_I^\dagger a_J$ and its adjoint; $Y_{IJ}$ is similar).  Thus the expectation values of the 2-particle observables are linear functions of the expectation values of the 2-hole observables.  So we have a linear transformation that maps $K'_2$ to $K_2$; this is $A^{-1}$.  

Similarly, any 2-hole operator $a_I a_J^\dagger$ can be written as a linear combination of 2-particle operators $a_{(J')}^\dagger a_{(I')}$.  The argument is almost the same as before:  first we use the anticommutation relations, then we use the identity 
\[
a_f^\dagger a_e = (\sum_{g \notin \set{e,f}} a_g^\dagger a_g) a_f^\dagger a_e 
 = \sum_{g \notin \set{e,f}} a_f^\dagger a_g^\dagger a_g a_e
\]
to replace 1-particle operators with 2-particle operators.  This allows us to construct the linear transformation $A$ that maps $K_2$ to $K'_2$.  

We now show that the linear transformation $A$ does not shrink $K_2$ by more than a polynomial factor.  Write the singular value decomposition $A = UDV$, where $U$ and $V$ are unitary, and $D$ is diagonal, with diagonal entries $D_{ii}>0$.  Let $B = A^{-1}$.  Looking at the matrix elements of $B$, we can see that 
\[
\Tr(B^\dagger B) = \sum_{i,j=1}^\ell |B_{ij}|^2 \leq \ell^2 \m^2.  
\]
At the same time, 
\[
\Tr(B^\dagger B) = \Tr(U D^{-1} V V^{-1} D^{-1} U^{-1}) = \Tr(D^{-2}) \geq D_{ii}^{-2}, 
\]
for all $i$.  So we have $D_{ii} \geq 1/\ell\m$, for all $i$.  That is, $A$ does not shrink by more than a $\ell\m$ factor in any direction.

This implies that $K'_2$ contains a ball of radius $1/\ell^2\m^3$.  

\vskipline

Next, we show that $\tfrac{2}{(\m-2)(\m-3)} K'_2 = K_{\m-2}$.  Consider what happens when we exchange the creation operator $a_i^\dagger$ with the annihilation operator $a_i$, for each mode $i$.  This transforms 2-hole observables into 2-particle observables, and vice versa.  In addition, this transforms 2-particle Slater basis states into $(\m-2)$-particle Slater basis states, and vice versa:  the 2-particle state with modes $i$ and $j$ occupied corresponds to the $(\m-2)$-particle state with modes $i$ and $j$ empty.  

So take any point $\vec{\alpha} \in K'_2$, which represents the expectation values of the 2-hole observables for some 2-particle state $\sigma$.  Use $\sigma$ to construct the corresponding $(\m-2)$-particle state $\tau$, as described above.  Then the expectation values of the 2-hole observables for $\sigma$ are exactly the expectation values of the 2-particle observables for $\tau$.  So $\tfrac{2}{(\m-2)(\m-3)} \vec{\alpha}$ is in $K_{\m-2}$.  (Note that we normalize $\vec{\alpha}$ to account for the increased number of particles---see the definition of $K_\n$.)  This shows that $\tfrac{2}{(\m-2)(\m-3)} K'_2 \subseteq K_{\m-2}$.  A similar argument shows that $K_{\m-2} \subseteq \tfrac{2}{(\m-2)(\m-3)} K'_2$.  This proves the claim.

Hence $K_{\m-2}$ contains a ball of radius $1/\ell^2\m^5$.  

\vskipline

Next, we show that $K_\n \subseteq K_{\n-1}$, for all $3 \leq \n \leq \m-2$.  Take any point $\vec{\alpha} \in K_\n$, which represents the expectation values of the observables $S \in \SSS$ for some 2-particle state $\rho$, where $\rho$ is $\n$-representable.  But if $\rho$ is $\n$-representable, then it is also $(\n-1)$-representable.  To see this, take some $\n$-particle state $\sigma$, such that $\Tr_{3,\ldots,\n} (\sigma) = \rho$; trace out the $\n$'th particle to get an $(\n-1)$-particle state $\sigma' = \Tr_\n (\sigma)$; and note that $\Tr_{3,\ldots,\n-1} (\sigma') = \rho$.  Thus $\vec{\alpha} \in K_{\n-1}$, which proves the claim.  

Hence $K_\n$ contains a ball of radius $1/\ell^2\m^5$, for all $3 \leq \n \leq \m-2$.  $\square$

\section{Fermionic Problems in QMA}

\begin{thm}
Fermionic Local Hamiltonian and $\N$-representability are in QMA.
\end{thm}

\noindent
Proof:  A problem is in QMA if there exists a poly-time quantum verifier $V$ that takes two inputs:  a description of the problem $x$, and a ``witness'' $\tau$ (which is a quantum state on polynomially many qubits).  $V$ should have the following property:  if $x$ is a ``YES'' instance, then there exists a state $\tau$ that causes $V$ to output ``true'' with probability $\geq p_1$; if $x$ is a ``NO'' instance, then for all possible states $\tau$, $V$ outputs ``true'' with probability $\leq p_0$; and $p_1 - p_0 \geq 1/\poly(\n)$.  

Suppose we have a ``YES'' instance of Fermionic Local Hamiltonian or $\N$-represent\-ability.  Intuitively, the witness should be an $\n$-fermion state $\sigma$ (i.e., the ground state of the fermionic Hamiltonian, or the $\n$-fermion state that agrees with the given 2-RDM).  Then the verifier works by measuring 2-fermion observables (we will discuss the measurement procedure later).  However, the standard model of quantum computation uses qubits, so we need to represent the fermionic state $\sigma$ using qubits, in such a way that the fermionic observables can be implemented efficiently.  

We represent the fermionic state $\sigma$ using $\m$ qubits, via the following mapping:
\[
(a_1^\dagger)^{i_1} \cdots (a_\m^\dagger)^{i_\m} \ket{\Omega}
\mapsto \ket{i_1} \otimes \cdots \otimes \ket{i_\m}.
\]
Call the resulting qubit state $\tilde{\sigma}$.  Note that, if $\sigma$ has exactly $\n$ fermions, then $\tilde{\sigma}$ lies in the subspace of states $\ket{i_1,\ldots,i_d}$ where $i_1+\cdots+i_d = \n$.  

We use the Jordan-Wigner transform to map the fermionic annihilation operators $a_i$ to qubit operators $A_i$:  
\[
a_i \mapsto A_i = -\Bigl( \bigotimes_{k=1}^{i-1} \sigma_k^z \Bigr) \otimes \ket{0}\bra{1}_i.  
\]
Likewise, 
\[
a_i^\dagger \mapsto A_i^\dagger = -\Bigl( \bigotimes_{k=1}^{i-1} \sigma_k^z \Bigr) \otimes \ket{1}\bra{0}_i.  
\]
One can check that the action of $A_i$ on the qubit states agrees with the action of $a_i$ on the fermionic states (recall the definition of $a_i$ in section \ref{ch3-2ndquant}).  

Thus, we can transform a fermionic observable $O = a_i^\dagger a_j^\dagger a_l a_k + a_k^\dagger a_l^\dagger a_j a_i$ into a qubit observable $\tilde{O} = A_i^\dagger A_j^\dagger A_l A_k + A_k^\dagger A_l^\dagger A_j A_i$.  This is a tensor product of many single-qubit observables and one four-qubit observable, so it can be measured efficiently.  Similar arguments apply for all of the 2-fermion observables in the set $\SSS$ (introduced in section \ref{ch3-observables}).  

We now describe the verifier $V$.  This is quite similar to the verifier for the Local Hamiltonian and Consistency problems on qubits (see chapter 2).  The witness $\tau$ consists of several (i.e., polynomially many) blocks, where each block has $\m$ qubits, supposedly representing one copy of the state $\tilde{\sigma}$.  The verifier $V$ acts as follows:  
\begin{verse}
On each block, $V$ first measures the observable $T = \sum_{k=1}^\m \ket{1} \bra{1}_k$, and if the outcome does not equal $\n$, $V$ outputs ``false.''  This projects each block onto the space of $\n$-fermion states.\\  
Next, in the case of Fermionic Local Hamiltonian, $V$ transforms the fermionic Hamiltonian $H$ into a qubit operator $\tilde{H}$ (note that $H$ is a linear combination of the 2-fermion observables $S \in \SSS$), then uses phase estimation to estimate the expectation value of $\tilde{H}$ for the state $\tilde{\sigma}$.  $V$ compares this with the energy threshold specified in the description of the problem, and outputs ``true'' or ``false'' accordingly.\\  
In the case of $\N$-representability, $V$ picks a fermionic observable $S \in \SSS$ at random, transforms it into a qubit observable $\tilde{S}$, and measures it on each block to estimate the expectation value for the state $\tilde{\sigma}$.  $V$ compares this with the expectation value for the state $\rho$ specified in the description of the problem, and outputs ``true'' or ``false'' accordingly.  
\end{verse}

The analysis of the verifier $V$ uses the same arguments as in chapter 2.  One technical difference is the use of the local fermionic observables $S \in \SSS$, rather than the local Pauli matrices; however, the observables $S \in \SSS$ can be used in a similar way to extract information from the witness $\sigma$ (see section \ref{ch3-observables}).  It is straightforward to see that, on a ``YES'' instance, given the correct witness $\tau = \tilde{\sigma}^{\tensor r}$, the verifier $V$ outputs ``true.''  On a ``NO'' instance, the situation is more complicated:  given an arbitrary state $\tau$, we want $V$ to output ``false.''  First, note that if the measurement of the observable $T$ returns a value different from $\n$ on some block, then $V$ automatically returns ``false.''  So without loss of generality, we can assume that $\tau$ lies in the simultaneous eigenspace of the observables $T$ (with eigenvalue $\n$) on all the blocks.  In other words, $\tau$ has exactly $\n$ fermions on each block.  However, $\tau$ might not be a tensor product state, i.e., the different blocks could be entangled.  But this does not fool the verifier, by the same Markov argument as in chapter 2 (originally due to \cite{AR}).  

Thus we have that Fermionic Local Hamiltonian and $\N$-representability are in QMA.  $\square$

\subsection{Pure-state $N$-representability is in QMA(2)}

The pure-state $\N$-representability problem is similar to the $\N$-representability problem, but with the extra constraint that the $\n$-particle state must be pure.  
\begin{quote}
In addition to $\rho$ and $\beta$, we are given a real number $\delta \geq 1/\poly(\n)$, specified with $\poly(\n)$ bits of precision.  We have to distinguish between these two cases:
\begin{itemize}
\item There exists an $\n$-fermion state $\sigma$ such that $\sigma$ is pure (hence $\Tr(\sigma^2) = 1$) and $\Tr_{3,\ldots,\n}(\sigma) = \rho$.  In this case, answer ``YES.''
\item For all $\n$-fermion states $\sigma$, either $\Tr(\sigma^2) \leq 1-\delta$ or $\lVert \Tr_{3,\ldots,\n}(\sigma) - \rho \rVert_1 \geq \beta$.  In this case, answer ``NO.''
\end{itemize}
\end{quote}
Note that we use $\Tr(\sigma^2)$ to measure the purity of the state $\sigma$, and we allow an error tolerance $\delta \geq 1/\poly(\n)$.

The class QMA(2) is similar to QMA, except that here the verifier $V$ receives two unentangled quantum witnesses, $\tau$ and $\eta$ (so the combined state is $\tau \tensor \eta$) \cite{Matsumoto}.  $V$ is required to have the following property:  if $x$ is a ``YES'' instance, then there exists a product state $\tau \tensor \eta$ that causes $V$ to output ``true'' with probability $\geq p_1$; if $x$ is a ``NO'' instance, then for all possible states of the form $\tau \tensor \eta$, $V$ outputs ``true'' with probability $\leq p_0$; and $p_1 - p_0 \geq 1/\poly(\n)$.  (Note that for a QMA(2) verifier, it is not known whether one can use parallel repetition to amplify the gap between the probabilities $p_1$ and $p_0$.)  

\begin{prop}
Pure-state $\N$-representability is in QMA(2).  
\end{prop}

\noindent
Proof:  First we describe the ``swap test.''  Given two unentangled states $\nu$ and $\eta$, on two quantum systems of equal dimension, the swap test allows us to estimate the quantity $\Tr(\nu\eta)$.  Let $Swap$ denote the operation of exchanging the two systems.  This is a unitary operation, but it is also Hermitian, and it can be viewed as an observable with eigenvalues 1 and $-1$.  Thus one can measure the $Swap$ observable, using the same procedure for measuring the Pauli matrices (see section \ref{ch2-in-qma}).  This procedure returns ``0'' with probability $\frac{1}{2} + \frac{1}{2} \Tr(Swap (\nu\tensor\eta))$, and ``1'' with probability $\frac{1}{2} - \frac{1}{2} \Tr(Swap (\nu\tensor\eta))$.  Then a straightforward calculation shows that 
\[
\begin{split}
\Tr(Swap (\nu\tensor\eta))
 &= \Tr((I\tensor\nu) Swap (I\tensor\eta)) \\
 &= \Tr(Swap (I\tensor(\eta\nu))) \\
 &= \Tr(\eta\nu) = \Tr(\nu\eta).  
\end{split}
\]

The swap test can be used to check the purity of the state $\nu$, as follows.  If $\nu$ is pure, and $\eta = \nu$, then $\Tr(\nu\eta) = \Tr(\nu^2) = 1$, so the test returns ``0'' with probability 1.  But if $\nu$ is not pure (and in particular $\Tr(\nu^2) \leq 1-\varepsilon$), then for all states $\eta$, 
\[
\Tr(\nu\eta) \leq \sqrt{\Tr(\nu^2)\Tr(\eta^2)} \leq \sqrt{1-\varepsilon} \leq 1-\varepsilon/2, 
\]
so the test returns ``0'' with probability $\leq 1-\varepsilon/4$.  (Intuitively, $\eta$ serves as a ``witness'' to the purity of the state $\nu$.  Note that it is essential that $\nu$ and $\eta$ are independent states.)

Now we describe the verifier for pure-state $\n$-representability.  The witness is a product state $\tau \tensor \eta$, where $\tau$ is the usual witness for $\n$-representability, consisting of polynomially many blocks, while $\eta$ consists of a single block, which is guaranteed to be unentangled with $\tau$.  (Each block consists of $d$ qubits, and supposedly represents a copy of the $\n$-fermion state $\sigma$ (or, to be precise, the corresponding qubit state $\tilde{\sigma}$).)  The verifier $V$ works as follows:  
\begin{verse}
First, $V$ measures the observable $T = \sum_{k=1}^\m \ket{1} \bra{1}_k$ on each block, and if the outcome does not equal $\n$, $V$ outputs ``false.''  This projects each block onto the space of $\n$-fermion states.\\  
Then $V$ flips a coin, and does one of two things with equal probability.\\  
If the coin comes up ``heads,'' $V$ discards the state $\eta$, and performs the usual verification procedure for $\n$-representability on the state $\tau$ (i.e., $V$ uses $\tau$ to estimate the expectation values of the 2-fermion observables).\\  
If the coin comes up ``tails,'' $V$ picks one block of $\tau$, uniformly at random, and discards the rest of $\tau$.  This produces the state $\tau^* = (1/r) \sum_{j=1}^r \tau^{(j)}$, where $r$ is the number of blocks, and $\tau^{(j)}$ is the reduced state of the $j$'th block.  $V$ now has the state $\tau^* \tensor \eta$, and $V$ checks the purity of $\tau^*$, using the swap test as described above.  
\end{verse}

On a ``YES'' instance, given the witness $\tau \tensor \eta$ where $\tau = \tilde{\sigma}^{\tensor r}$ and $\eta = \tilde{\sigma}$, the verifier $V$ returns ``true'' with probability close to 1.  On a ``NO'' instance, for any witness of the form $\tau \tensor \eta$, we claim that $V$ returns ``true'' with probability bounded away from 1.  Without loss of generality, we can assume that $\tau$ and $\eta$ lie in the subspace of states with exactly $\n$ fermions per block.  However, $\tau$ might be an arbitrary entangled state (not an $r$-fold product state).  Nonetheless, we consider the state $\tau^* = (1/r) \sum_{j=1}^r \tau^{(j)}$ (defined above) on a single block.  Both the purity test and the $\n$-representability test act on this state.  Since this is a ``NO'' instance, we know that either $\Tr((\tau^*)^2) \leq 1-\delta$ or $\lVert \Tr_{3,\ldots,\n}(\tau^*) - \rho \rVert_1 \geq \beta$ (note that we are abusing notation, using $\tau^*$ to denote both the $\n$-fermion state and its representation as a qubit state).  Hence either the purity test or the $\n$-representability test will fail with significant probability, so $V$ will return ``false.''  $\square$

\section{Discussion}

\subsection{Related Work in Quantum Information}

It is remarkable that checking consistency of 2-body reduced density
operators is so hard, while checking consistency
of 1-body reduced density operators is simple \cite{Coleman}. This
can be understood from the previous discussion: 
intuitively, 1-body density operators $\langle
a_i^\dagger a_j\rangle$ correspond to Hamiltonians
only containing bilinear terms in $a_i^\dagger$ and $a_j$; such
Hamiltonians can easily be diagonalized as they represent systems of
free fermions. As shown in \cite{Coleman}, consistency can be decided in
that case based solely on the eigenvalues of the
reduced density operators. A number of related problems have been
investigated recently~\cite{reduced,reduced2,reduced3,reduced4,reduced5}; in particular, see \cite{Klyachko05}.

These results have to be contrasted with our problem of deciding $N$-represent\-ability for 2-body density operators,
where the eigenvalues alone are not enough to decide consistency but also the eigenvectors are relevant. Actually,
let us consider the simpler problem where only the diagonal elements of the 2-body density operators, 
$D_{ij}=\langle a_i^\dagger a_j^\dagger a_j a_i\rangle$, are specified. 
Using the mapping from spins to fermions discussed above, one easily finds that these $D_{ij}$ 
correspond to local spin Hamiltonians which only contain commuting $\sigma^z$ operators. These
are spin-glasses, and so the problem of deciding $\N$-representability of $\{D_{ij}\}$ is NP-hard
\cite{Barahona}. It was indeed pointed out a long time ago that $\N$-representability restricted to the diagonal
elements is equivalent to a combinatorial problem \cite{Kuhn} that was later shown to be equivalent to the NP-hard
problem of deciding membership in the boolean quadric polytope \cite{Deza}.

\subsection{Applications to Quantum Chemistry}

There are various methods for calculating the 2-RDM corresponding to the ground state of a molecular system \cite{book1,book,Mazziotti}.  These methods necessarily involve solving some instances of the $N$-representability problem.  Typically, one imposes a set of constraints, called $N$-representability conditions, which can be efficiently computed, but only give an approximation of the true set of $N$-representable 2-RDM's.  For example, one can impose positivity constraints on the $p$-particle reduced states, where $p$ is a small constant, say 2 or 3; these are called $p$-positivity conditions.  One can then perform a variational minimization, or use a more sophisticated method such as the contracted Schrodinger equation (CSE).  In the CSE method, one first integrates the $N$-particle Schrodinger equation to get an equation that relates the 2-, 3- and 4-RDM's.  The 3- and 4-RDM's can then be approximated in terms of the 2-RDM, and one can solve for the 2-RDM using an iterative procedure.  Here, the $N$-representability conditions are expressed in the approximate reconstruction of the 3- and 4-RDM's from the 2-RDM, and in the iterative procedure.

We have shown that finding ground state energies by means of the $\N$-represent\-ability problem is intractable in the worst case.  
This leaves open the possibility of finding efficient algorithms that give accurate results for \textit{particular} physical systems (though they must break down in the general case).  
The hope is that some physical systems may have special features that make the problem easier. 
One example is one-dimensional translational invariant
spin systems, where the density matrix renormalization group allows for a systematic approximation of the convex set of allowed reduced density operators from within \cite{VC06}. 
Also, for some molecular systems, variational minimization using 3-positivity conditions gives promising results \cite{Mazzz}; this gives an approximation of the convex set from the outside.  
The non-variational CSE method looks promising as well, and is especially intriguing, as it combines $p$-positivity ideas with a particular ansatz for the $N$-particle wave function; see \cite{mazziotti-acse} for a recent development in this area.  

It would be very interesting to investigate the conditions under which these approximations are justified.  While there is empirical evidence that these methods work well, it seems that certain questions---especially concerning the accuracy of these methods on larger molecules---can only be answered through a better theoretical understanding.  Most of the previous work has focused on applying these methods to small molecules or simple ``toy models,'' and measuring the accuracy of the results against those obtained from brute-force calculations (full configuration interaction) or exact analytic solutions.  However, based on this evidence it is hard to predict how well these methods will scale to larger, more complex molecules.  In particular, does the accuracy decrease when we move to larger molecules?  Ideally, one would wish to have some guarantee of the accuracy of the result, in cases where the true ground state energy is not already known.  

It may be that, on larger molecules, there is a tradeoff between the speed and accuracy of these numerical methods.  (For instance, one can always improve the accuracy by using $p$-positivity conditions with larger $p$, but the complexity grows exponentially with $p$; and indeed, in practice, 3-positivity conditions are much more computationally intensive than 2-positivity conditions.)  Although it is very hard to answer these questions completely, theoretical investigations may shed some light.

Finally, we remark that there are proposals for finding ground state energies of molecular systems by using a quantum computer \cite{Aspuru-Guzik-et-al-05,Abrams-Lloyd-99}.  These methods offer an exponential speedup, in that the quantum computer can actually represent the full $N$-particle state, and measure its energy via phase estimation.  However, to prepare an approximate ground state on the quantum computer, one must use heuristic methods, such as adiabatic evolution starting from the Hartree-Fock ground state.  These heuristic methods are not expected to work in all cases, which is consistent with our result that Fermionic Local Hamiltonian is QMA-hard.  

\vskipline

In conclusion, we investigated the problem of $\N$-representability, and characterized its computational
complexity by showing that it is QMA-complete. Obviously, the theory of quantum computing was a prerequisite to
understanding the complexity of this classic problem.

\vskipline
\vskipline
\noindent \textit{Acknowledgements:} Y.K.L. and M.C. thank the Institute for Quantum Information for its
hospitality. Y.K.L. is supported by an ARO/DTO QuaCGR Fellowship. M.C. acknowledges
an EPSRC Postdoctoral and a Nevile Research Fellowship which he holds at Magdalene College Cambridge, and is
supported by the EU under the FP6-FET Integrated Project SCALA, CT-015714. F.V. is supported by the Gordon and
Betty Moore Foundation through Caltech's Center for the Physics of Information, and by the NSF under Grant No. PHY-0456720.

\vskipline
\noindent A shorter version of this paper appeared as \cite{Liu-N-rep}:  Y.-K. Liu, M. Christandl and F. Verstraete, ``$N$-representability is QMA-complete,'' \textit{Phys. Rev. Lett.} 98, 110503 (2007).  That version is copyrighted by the American Physical Society.  This use is permitted under the copyright agreement.

\chapter{The Consistency Problem for 1-D and Stoquastic Systems}

\ifthenelse{\boolean{ucsdformat}}{\thispagestyle{chappage}}{}



(This chapter contains preliminary results.  It has been superseded by the paper:  Y.-K. Liu, ``The Local Consistency Problem for Stoquastic and 1-D Quantum Systems,'' ArXiv:0712.1388 [quant-ph], 2007.)

\section{Introduction}

Previously we showed that Consistency is QMA-complete, which implies that the Consistency and Local Hamiltonian problems have the same complexity (up to poly-time oracle reductions).  In this chapter we will prove similar statements about some special cases of these problems, which are not known to be QMA-hard, and in fact seem to be strictly easier than QMA.  We consider the Local Hamiltonian problem for certain 1-dimensional spin chains, and also for so-called ``stoquastic'' systems; these cases are not known to be QMA-hard.  We show that there are corresponding special cases of the Consistency problem that have the same complexity, up to poly-time oracle reductions.  

One direction is easy:  Local Hamiltonian reduces to Consistency, using the same techniques as in the previous chapters.  But the opposite direction, reducing Consistency to Local Hamiltonian, is nontrivial.  In the general case, we could get such a reduction using the QMA-hardness of Local Hamiltonian; but we want a reduction to a special case of Local Hamiltonian that is not QMA-hard.  Here we devise a different reduction from Consistency to Local Hamiltonian, that works in these special cases.  This reduction uses convex optimization with a membership oracle, combined with a new trick:  a connection between Local Hamiltonian and Consistency, via Lagrange duality.  (This is section 4.2.)

This duality idea is similar to recent work by Hall \cite{Hall} on the ``subsystem compatibility problem.''  This problem is very much like Consistency, except that the input consists of density matrices describing all subsets of size $n-1$ (for a system of $n$ qubits), rather than subsets of size $k$ for some constant $k$.  Thus the input is exponentially large in $n$, and the problem can be solved in time polynomial in the length of the input.  In contrast, for the Consistency problem, the input is polynomially large in $n$, and we show a poly-time reduction to Local Hamiltonian.

\vskipline

Then we apply these ideas to the special case of one-dimensional spin chains.  Specifically, we have $n$ qudits (a qudit is a $d$-dimensional particle), arranged in a line with nearest-neighbor interactions (that is, interactions between particles $i$ and $i+1$, for $i = 1,\ldots,n-1$).  Many simple models studied in condensed-matter physics are of this form, and moreover there are heuristic methods, such as the density-matrix renormalization group (DMRG), which solve these models efficiently in practice \cite{schollwock}.  Although the performance of these heuristics is not fully understood, this experience suggested that 1-D systems are computationally tractable, in contrast to systems in 2 or more dimensions.  (One rigorous result along these lines is given by \cite{osborne}.)  So it was a surprise when Aharonov, Gottesman and Kempe showed that Local Hamiltonian on a 1-D chain of qudits (with $d=12$) is QMA-hard \cite{qma1d-1,qma1d-2}.  It is still an open problem whether the problem is QMA-hard for smaller values of $d$, and for qubits in particular.  

We define the Consistency problem on a 1-D chain of qudits, where we are given density matrices describing each pair of adjacent qudits.  We show that, for a 1-D chain of qubits ($d=2$), Consistency and Local Hamiltonian have the same complexity (up to poly-time oracle reductions).  We also sketch how this result can be generalized to a 1-D chain of qudits ($d>2$).  (This is section 4.3.)  

We remark that the complex behavior of 1-D quantum systems is a sharp contrast to what happens in the classical world.  For instance, Max-2-SAT, which is the classical analogue of Local Hamiltonian, is poly-time solvable when restricted to a 1-dimensional chain \cite{qma1d-1}.  Also, inference in graphical models can be solved exactly in poly-time when the underlying graph is a tree.  This has an intuitive explanation.  Consider the Gibbs distribution associated with a (classical) tree-structured graphical model.  Deleting any single node $i$ breaks the tree into two or more disconnected components; moreover, variables in different components are independent conditioned on the variable at node $i$.  Thus the correlations among variables have a simple structure (they are a Markov random field).  However, this is no longer true when one considers the Gibbs state of a quantum Hamiltonian, even when interactions are restricted to lie on a tree.

\vskipline

Finally, we consider the class of ``stoquastic'' quantum systems, introduced in \cite{stoq-1,stoq-2}.  A Hamiltonian is called ``stoquastic'' if all of its off-diagonal matrix elements (relative to the standard basis) are less than or equal to 0.  By the Perron-Frobenius theorem \cite{Bellman}, this implies that the ground state can be chosen to have the form $\ket{\psi} = \sum_z c_z \ket{z}$, where $\ket{z}$ are the standard basis states and the coefficients $c_z$ are all real and nonnegative.  Since the coefficients $c_z$ all have the same complex phase, they can be viewed as analogous to a probability density, with $\sum_z c_z^2 = 1$.  

Stoquastic Hamiltonians appear in many natural physical systems, as well as some versions of the adiabatic algorithm for combinatorial optimization \cite{FGGS}.  However, there is some evidence that the Local Hamiltonian problem in this case is not QMA-hard.  Bravyi et al \cite{stoq-1} showed that Stoquastic Local Hamiltonian is in AM, a class which is believed to lie ``just above'' NP in the polynomial hierarchy.  If Stoquastic Local Hamiltonian were QMA-hard, this would imply that QMA is in AM, which is possible but perhaps a little unlikely.

On the other hand, Bravyi et al \cite{stoq-1} also showed that Stoquastic Local Hamiltonian is MA-hard, so it cannot be very much easier than general Local Hamiltonian.  Indeed, it could be that Stoquastic Local Hamiltonian is QMA-hard, and we are simply ignorant.  (However, such ignorance may be long-lived.  It is still an open problem to show that BQP is in the polynomial hierarchy, a much weaker result that would follow trivially if QMA were in AM.)

We propose a stoquastic version of the Consistency problem.  We believe this problem ie equivalent to Stoquastic Local Hamiltonian (up to poly-time oracle reductions), and we give a heuristic argument, modulo some technical details, for why this should be true.  (This is section 4.4.)  


\section{Reductions from Consistency to Local Hamiltonian}
\label{ch4-consistency-localham}

Consider the standard versions of the Consistency and Local Hamiltonian problems, as defined in Chapter 2.  Previously we gave reductions from Local Hamiltonian to Consistency (thus showing that Consistency is QMA-hard); now let us consider reductions in the opposite direction.  One way is to use the QMA-hardness of Local Hamiltonian \cite{KSV,KKR}:  since Consistency is in QMA, one can ``encode'' an instance of Consistency into an instance of Local Hamiltonian.  Here we will give a different reduction, based on Lagrange duality, which does not involve QMA-hardness.  This reduction illustrates a simple and quite transparent relationship between the two problems, which is interesting in its own right.  It will also be useful in dealing with special cases of these problems which are not QMA-hard.

The idea comes from a theorem of ``strong alternatives'' in semidefinite programming \cite{BV}.  Let $F_1,\ldots,F_d$ be complex Hermitian matrices of dimension $N$.  Consider the following matrix inequality:
\begin{equation}
\label{ch4-eqn-a}
\sum_{i=1}^d x_i F_i + I \prec 0, 
\end{equation}
where $x \in \RR^d$ is a variable.  (Notation:  $M \prec 0$ means $M$ is strictly negative definite, $M \succeq 0$ means $M$ is positive semidefinite, etc.)  Also consider the following system of inequalities:
\begin{equation}
\label{ch4-eqn-b}
Z \succeq 0, \quad Z \neq 0, \quad \Tr(F_i Z) = 0 \; (\forall i = 1,\ldots,d), 
\end{equation}
where $Z$, a complex Hermitian matrix of dimension $N$, is a variable.  The theorem states that exactly one of the two inequalities (\ref{ch4-eqn-a}) and (\ref{ch4-eqn-b}) is feasible.  In other words, if (\ref{ch4-eqn-b}) is feasible, then (\ref{ch4-eqn-a}) is not; and if (\ref{ch4-eqn-b}) is not feasible, then (\ref{ch4-eqn-a}) is.  (When this property holds, we say that (\ref{ch4-eqn-a}) and (\ref{ch4-eqn-b}) are strong alternatives.)

Observe that inequality (\ref{ch4-eqn-b}) can be used to express the Consistency problem:  $Z$ is a global density matrix (unnormalized, but note that all the constraints remain the same if we divide across by $\Tr(Z)$), and we can choose the constraints $\Tr(F_i Z) = 0$ to ensure that $Z$ agrees with the desired local density matrices (note that the matrices $F_i$ will then be local observables).  But now the expression $\sum_{i=1}^d x_i F_i + I$ in inequality (\ref{ch4-eqn-a}) is simply a local Hamiltonian, and estimating its largest eigenvalue is precisely the Local Hamiltonian problem (modulo a sign flip).  So a Local Hamiltonian oracle allows us to test membership in the convex set defined by inequality (\ref{ch4-eqn-a}); and, using the methods of convex optimization described in Chapter 2, we can then decide the feasibility of (\ref{ch4-eqn-a}).  Since (\ref{ch4-eqn-a}) and (\ref{ch4-eqn-b}) are strong alternatives, this lets us solve the Consistency problem.

This is the intuition, but some further work is needed to make it rigorous.  We have to allow for the inverse-polynomial precision in the Consistency and Local Hamiltonian problems.  Also, in order to do convex optimization with a membership oracle, the set of feasible solutions $K$ must satisfy certain geometric properties.  So we have to formulate inequality (\ref{ch4-eqn-a}) in a different way.  We will show a finite-precision, ``algorithmic'' version of the theorem of strong alternatives.

\begin{thm}
\label{ch4-thm-consistency-localham}
There is a poly-time oracle reduction from Consistency to Local Hamiltonian.
\end{thm}

\noindent
Proof:  First, recall the statement of the Consistency problem:  
\begin{quote}
We have a system of $n$ qubits, and we are given local density matrices $\rho_1,\ldots,\rho_m$, where $\rho_i$ describes the subset of qubits $C_i \subseteq \set{1,\ldots,n}$.  (We assume $|C_i| \leq k$ for some constant $k$.)  

In addition, we are given a string ``$1^s$'' and a real number $\beta \geq 1/s$.  (All numbers are specified with $\poly(n)$ bits of precision.)  The problem is to distinguish between the following two cases:  
\begin{itemize}
\item There exists an $n$-qubit state $\sigma$ such that, for all $i$, $\Tr_{\set{1,\ldots,n}-C_i}(\sigma) = \rho_i$.  In this case, answer ``YES.''
\item For all $n$-qubit states $\sigma$, there exists an $i$ such that $\norm{\Tr_{\set{1,\ldots,n}-C_i}(\sigma) - \rho_i}_1 \geq \beta$.  In this case, answer ``NO.''
\end{itemize}
\end{quote}

As before, we consider the $n$-qubit Pauli matrices $P = \Tensor_{i=1}^n P_i$, where $P_i \in \set{I,X,Y,Z}$.  We say that $P$ is supported inside a subset $C \subseteq \set{1,\ldots,n}$ if, for all $i \notin C$, $P_i = I$.  Then we define $\SSS$ to be the set of ``local'' Pauli matrices, excluding the identity matrix, 
\[
\SSS = \bigcup_{i=1}^m \set{P \;|\; \text{$P$ is supported inside $C_i$}} - \set{I}.
\]
We also let $d = |\SSS|$, and note that $d \leq 4^k m - 1$.  These are the local observables, and knowing their expectation values is equivalent to knowing the local reduced density matrices $\rho_1,\ldots,\rho_m$.  

Suppose we have an instance of the Consistency problem.  For each observable $P \in \SSS$, we define $\alpha_P$ to be the desired expectation value, which we compute as follows:  pick some subset $C_i$ such that $P$ is supported in $C_i$, then set $\alpha_P = \Tr(P\rho_i)$.  

Let us make a couple of observations.  Clearly, if this is a ``YES'' instance of Consistency, then there exists an $n$-qubit state $\sigma$ such that, for all $P \in \SSS$, $\Tr(P\sigma) = \alpha_P$.  

We claim that, if this is a ``NO'' instance of Consistency, then for all $n$-qubit states $\sigma$, $\sum_{P \in \SSS} |\Tr(P\sigma) - \alpha_P| \geq \beta$.  This can be seen as follows.  Note that, for any $\sigma$, there is some subset $C_i$ such that $\norm{\tilde{\sigma}-\rho_i}_1 \geq \beta$, where $\tilde{\sigma} = \Tr_{\set{1,\ldots,n}-C_i}(\sigma)$.  Using the matrix Cauchy-Schwarz inequality \cite{Bhatia}, $\norm{\tilde{\sigma}-\rho_i}_1 \leq \norm{\tilde{\sigma}-\rho_i}_2 \sqrt{2^k}$.  By Fourier analysis, 
\[
\begin{split}
\norm{\tilde{\sigma}-\rho_i}_2
 &= \frac{1}{\sqrt{2^k}} \Bigl( \sum_\text{$P$ supp. on $C_i$} \Tr(P(\tilde{\sigma}-\rho_i))^2 \Bigr)^{1/2} \\
 &\leq \frac{1}{\sqrt{2^k}} \sum_{P \in \SSS} |\Tr(P(\tilde{\sigma}-\rho_i))|
 = \frac{1}{\sqrt{2^k}} \sum_{P \in \SSS} |\Tr(P\sigma) - \alpha_P|.  
\end{split}
\]
The claim follows by combining these inequalities.

\vskipline

Next we write down a convex program and its dual.  For each local observable $P \in \SSS$, we define a new observable 
\[
F_P = P - \alpha_P I, 
\]
which is shifted so that the desired expectation value now equals 0.  We also define $F(x)$ to be a linear combination of these observables, 
\[
F(x) = \sum_{P \in \SSS} x_P F_P + I, \quad \text{for $x \in \RR^d$}.  
\]

Now consider the following convex program:  
\begin{verse}
Find some $x \in [-1,1]^d$ and $s \in [1-2d,1+2d]$ that \\
minimize $s$ such that $F(x) \preceq sI$.
\end{verse}
To see that this is a convex program, recall that the largest eigenvalue of $F(x)$ is a convex function of $x$, since it can be written as the pointwise minimum over a family of affine functions of $x$.  The variable $s$ is redundant here, but it will play a role later when we apply algorithms to solve this program.  We will refer to this as the primal program; let $p^*$ denote the optimal value of the objective function $s$.

The dual program is as follows:
\begin{verse}
Find some $2^n \times 2^n$ complex matrix $Z$ that \\
maximizes $g(Z)$ such that $Z \succeq 0$ and $\Tr(Z) = 1$,
\end{verse}
where the dual function $g(Z)$ is given by 
\[
\begin{split}
g(Z) &= \inf_{\substack{x \in [-1,1]^d \\ s \in [1-2d,1+2d]}} s + \Tr(Z(F(x)-sI)) \\
     &= \inf_{x \in [-1,1]^d} \Tr(ZF(x)) \\
     &= \inf_{x \in [-1,1]^d} \sum_{P \in \SSS} x_P \Tr(ZF_P) + 1.
\end{split}
\]
Let $d^*$ denote the optimal value of the objective function $g(Z)$.  Strong duality holds because the primal problem is convex and satisfies a generalized Slater condition \cite{BV} (to see this, note that the point $(x,s) = (0,2)$ is strictly feasible, i.e., it lies in the relative interior of the domain, and it satisfies $F(x) \prec sI$).  Strong duality implies that $p^* = d^*$, i.e., the optimal values of the primal and dual programs are equal.

\vskipline

We now give a poly-time oracle reduction from Consistency to Local Hamiltonian.  We show that Consistency reduces to the weak optimization problem $WOPT^*$, which reduces to the weak membership problem $WMEM^*$, which reduces to Local Hamiltonian.  

First, suppose we have a ``YES'' instance of Consistency.  Then there exists an $n$-qubit state $\sigma$ such that, for all $P \in \SSS$, $\Tr(P\sigma) = \alpha_P$.  So in the dual program there exists some $Z \succeq 0$, $\Tr(Z) = 1$, such that for all $P \in \SSS$, $\Tr(ZF_P) = 0$.  This implies $g(Z) = 1$, hence the dual program has optimal value $d^* \geq 1$.  By strong duality, the primal program has optimal value $p^* \geq 1$.  

On the other hand, suppose we have a ``NO'' instance of Consistency.  Then for all $n$-qubit states $\sigma$, $\sum_{P \in \SSS} |\Tr(P\sigma) - \alpha_P| \geq \beta$.  So, in the dual program, for all $Z$ such that $Z \succeq 0$ and $\Tr(Z) = 1$, we have that $\sum_{P \in \SSS} |\Tr(ZF_P)| \geq \beta$, which implies $g(Z) \leq 1-\beta$.  Thus the dual program has optimal value $d^* \leq 1-\beta$.  By strong duality, the primal program has optimal value $p^* \leq 1-\beta$.  

So we have reduced Consistency to the problem of distinguishing between the two cases $p^* \geq 1$ and $p^* \leq 1-\beta$ for the primal program.  This is an instance of the weak optimization problem $WOPT^*_{\beta/2}$ over the convex set 
\[
K = \set{(x,s) \in [-1,1]^d \times [1-2d,1+2d] \;|\; F(x)-sI \preceq 0}.
\]

\vskipline

Now we will reduce $WOPT^*$ to $WMEM^*$.  First we need some bounds on the geometry of $K$.  It is easy to see that $K$ is contained within a ball of radius $R = \sqrt{d+(1+2d)^2} \leq O(d)$.  In addition, we claim that $K$ contains a ball around the point $(0,\ldots,0,2)$ of radius $r = \frac{1}{4(d+1)}$.  To see this, consider an arbitrary point $(y,t+2)$ where $y \in \RR^d$, $t \in \RR$ and $\sqrt{\norm{y}^2+t^2} \leq \frac{1}{4(d+1)}$.  The operator 
\[
F(y) - (t+2)I = \sum_{P \in \SSS} y_P F_P - tI - I
\]
has all of its eigenvalues bounded above by $\sum_{P \in \SSS} \frac{1}{4(d+1)} \norm{F_P} + \frac{1}{4(d+1)} - 1 \leq -\frac{1}{2}$ (using the fact that $\norm{F_P} \leq 2$).  Thus $(y,t+2)$ is in $K$.  This proves the claim.  

So we have $R/r \leq O(d^2)$.  By theorem \ref{ch3-thm-opt-mem}, $WOPT^*_{\beta/2}$ reduces to $WMEM^*_\delta$ where $\delta \geq \poly(\beta, 1/d)$, with running time $\poly(d,1/\beta)$.  

\vskipline

Finally, we reduce $WMEM^*$ to the Local Hamiltonian problem.  Observe that, since the $F_P$ are local operators, $F(x)$ is a local Hamiltonian.  Given an oracle that solves the Local Hamiltonian problem, we can estimate the largest eigenvalue of $F(x)$ (i.e., the smallest eigenvalue of $-F(x)$), and thus decide whether $(x,s)$ is in the set $K$.  

Suppose we have a ``YES'' instance of $WMEM^*_\delta$.  Then $(x,s) \in K$, so $F(x) \preceq sI$, i.e., all eigenvalues of $-F(x)$ are $\geq -s$.  So this is a ``NO'' instance of Local Hamiltonian.  

Now suppose we have a ``NO'' instance of $WMEM^*_\delta$.  Then $(x,s) \notin S(K,\delta)$, and in particular, $(x,s+\delta) \notin K$.  So $F(x) \npreceq (s+\delta)I$, i.e., $-F(x)$ has an eigenvalue that is $\leq -s-\delta$.  So this is a ``YES'' instance of Local Hamiltonian.  

Note that $\norm{F(x)} \leq \sum_{P \in \SSS} \norm{F_P} + 1 \leq 2d+1$.  Thus, $WMEM^*_\delta$ reduces to Local Hamiltonian with precision $\delta/(2d+1)$.  

Thus we conclude that Consistency (with precision $\beta$) reduces to Local Hamiltonian (with precision $\poly(\beta, 1/d)$), and the running time is $\poly(d, 1/\beta)$.  Note that $d < 4^km$ is polynomial in the size of the input.  $\square$

\section{Consistency for 1-D Systems}

Let us consider a 1-dimensional chain of $n$ qudits (a qudit is a $d$-dimensional particle), with nearest-neighbor interactions (i.e., interactions between particle $i$ and particle $i+1$, for $i = 1,\ldots,n-1$).  

First consider the case of qubits ($d=2$).  The reduction from Local Hamiltonian to Consistency shown in chapter 2 (theorem \ref{thm-QMA-hard}), and the reverse reduction shown above (theorem \ref{ch4-thm-consistency-localham}), both preserve the neighborhood structure of the problems---that is, each local term in the Hamiltonian corresponds to a local density matrix, and vice versa.  Thus we have:  
\begin{thm}
On a 1-D chain of qubits ($d=2$), Local Hamiltonian and Consistency are equivalent (with respect to poly-time oracle reductions).  
\end{thm}

We will now sketch one way of extending these results to the case of qudits ($d>2$).  The first step is to define a set of observables for a single qudit, with nice properties similar to the Pauli matrices.  Let $\ket{i}$, $i = 0,\ldots,d-1$ denote the standard basis states for a single qudit.  Also, let $\text{i}$ (in plain, not italic type) denote the square root of $-1$.
\begin{align*}
X_{ij} &= \ket{j}\bra{i} + \ket{i}\bra{j}, \qquad 0 \leq i<j \leq d-1 \\
Y_{ij} &= \text{i}\ket{j}\bra{i} - \text{i}\ket{i}\bra{j}, \qquad 0 \leq i<j \leq d-1 \\
Z_i &= \Bigl( \frac{1}{i+1} \sum_{a=0}^i \ket{a}\bra{a} \Bigr) - \ket{i+1}\bra{i+1}, 
\qquad 0 \leq i \leq d-2
\end{align*}
Note that $Z_i$ is the diagonal matrix whose diagonal consists of $\frac{1}{i+1}$ in the first $i+1$ positions, followed by $-1$, followed by $0$ in all the remaining positions.  We have a total of $2\binom{d}{2} + (d-1) = d(d-1) + (d-1) = d^2 - 1$ observables.

These observables satisfy the following orthogonality relations:
\begin{center}
\begin{tabular}{|c|c|l|}
	\hline
	$A$ & $B$ & $\Tr(AB)$ \\
	\hline
	$I$ & $I$ & $d$ \\
	$I$ & $X_{kl}$ & 0 \\
	$I$ & $Y_{kl}$ & 0 \\
	$I$ & $Z_k$ & 0 \\
	\hline
	$X_{ij}$ & $X_{kl}$ & 2 if $(i,j)=(k,l)$; 0 otherwise \\
	$X_{ij}$ & $Y_{kl}$ & 0 \\
	$X_{ij}$ & $Z_k$ & 0 \\
	\hline
	$Y_{ij}$ & $Y_{kl}$ & 2 if $(i,j)=(k,l)$; 0 otherwise \\
	$Y_{ij}$ & $Z_k$ & 0 \\
	\hline
	$Z_i$ & $Z_k$ & $1+\frac{1}{i+1}$ if $i=k$; 0 otherwise \\
	\hline
\end{tabular}
\end{center}
In addition, note that $\norm{X_{ij}} = \norm{Y_{ij}} = \norm{Z_i} = 1$.

We can now use these qudit observables in the same way that we used the Pauli matrices for qubits.  We construct $n$-qudit observables by taking tensor products of single-qudit observables:  
\[
P = \Tensor_{a=1}^n P_a, \qquad P_a \in \set{I,X_{ij},Y_{ij},Z_i}.  
\]
Note that for any $n$-qudit observables $P$ and $Q$, $\Tr(PQ) = \prod_{a=1}^n \Tr(P_a Q_a)$.  Any $n$-qudit density matrix $\rho$ can be written in the form 
\[
\rho = \sum_P \frac{\alpha_P}{\Tr(P^2)} P, \qquad \alpha_P = \Tr(P\rho).
\]

We say that $P$ is supported inside a subset $C \subseteq \set{1,\ldots,n}$ if for all $i \notin C$, $P_i = I$.  If this is the case, we define $P|_C = \Tensor_{i \in C} P_i$, which we call the ``restriction'' of $P$ to the subset $C$.  We can write the reduced density matrix for the subset $C$ in the form 
\[
\begin{split}
\rho^{[C]} &= \Tr_{\set{1,\ldots,n}-C}(\rho) \\
&= \sum_{\text{$P$ supported in $C$}} \frac{\alpha_P}{\Tr(P^2)} \Tr_{\set{1,\ldots,n}-C}(P) \\
&= \sum_{\text{$P$ supported in $C$}} \frac{\alpha_P}{\Tr((P|_C)^2)} P|_C.
\end{split}
\]

Now we can use essentially the same reductions as before, from Local Hamiltonian to Consistency and vice versa, for systems of qudits.  (Details omitted.)  

\section{Consistency for Stoquastic Systems}

We say that a Hamiltonian is ``stoquastic'' if, when written in the standard basis, all of its off-diagonal matrix elements are less than or equal to 0.  (Note that the diagonal elements can be made to be $\leq 0$ by adding a multiple of the identity to the Hamiltonian; this shifts the eigenvalues but does not change the eigenvectors.)  This implies that the ground state can be chosen to have the form $\ket{\psi} = \sum_z c_z \ket{z}$ where $\ket{z}$ are the standard basis vectors and $c_z \geq 0$.  The Stoquastic Local Hamiltonian problem is simply the Local Hamiltonian problem with the additional promise that the local terms that make up the Hamiltonian are stoquastic.  As discussed previously, this makes the problem potentially easier.  

In this section we propose a ``stoquastic'' version of the Consistency problem, that has the same complexity as Stoquastic Local Hamiltonian (up to poly-time reductions).  We will describe a few different versions of the problem, all of which are at least as hard as Stoquastic Local Hamiltonian.  However, one version of the problem is especially interesting, because we believe it is no harder than Stoquastic Local Hamiltonian.  We provide a heuristic argument, though not a formal proof.  

First, let us say that a density matrix is ``stoquastic'' if, when written in the standard basis, all of its off-diagonal matrix elements are greater than or equal to 0.  (Its diagonal elements must be $\geq 0$ since the matrix is positive semidefinite.)  Note that the set of stoquastic density matrices is convex.  Now consider an obvious way of defining the stoquastic Consistency problem:
\begin{quote}
Given local density matrices $\rho_1,\ldots,\rho_m$, does there exist a global density matrix $\rho$ that is stoquastic and agrees with $\rho_1,\ldots,\rho_m$?
\end{quote}
Stoquastic Local Hamiltonian reduces to this problem, using an argument like the one in chapter 2.  But this problem does not seem to be in QMA, since the verifier does not have a way to test whether $\rho$ is indeed stoquastic.  

Another way of defining the stoquastic Consistency problem is as follows:
\begin{quote}
Given local density matrices $\rho_1,\ldots,\rho_m$ which are stoquastic, does there exist a global density matrix $\rho$ that agrees with $\rho_1,\ldots,\rho_m$?
\end{quote}
Again, Stoquastic Local Hamiltonian reduces to this problem.  Unlike our previous attempt, this problem is in QMA.  However, it is not clear whether this problem reduces to Stoquastic Local Hamiltonian; when we apply the technique from section \ref{ch4-consistency-localham}, we instead get a reduction from this problem to standard Local Hamiltonian.

It turns out that the most interesting way to define the stoquastic Consistency problem is as follows:
\begin{quote}
Given local density matrices $\rho_1,\ldots,\rho_m$, does there exist a global density matrix $\rho$ such that, for all $i=1,\ldots,m$, $\Tr_{\set{1,\ldots,n}-C_i}(\rho) \geq_e \rho_i$?  (Here $C_i$ is the subset of qubits described by $\rho_i$, and $\geq_e$ denotes element-wise inequality between two matrices written in the standard basis; we assume all matrices are real.)
\end{quote}
This definition is a little unusual, but we believe that it has the following interesting properties.  First, Stoquastic Local Hamiltonian reduces to this problem.  Second, this problem is in QMA.  Finally, this problem reduces to Stoquastic Local Hamiltonian.  In the following sections we explain these statements, though we do not present a formal proof.  

%
%

\subsection{Reducing from Stoquastic Local Hamiltonian to Stoquastic Consistency}

First we show a reduction from Stoquastic Local Hamiltonian to Stoquastic Consistency.  The basic idea is as follows.  We are given a local Hamiltonian $H = \sum_{i=1}^m H_i$, where the $H_i$ are real and stoquastic.  Without loss of generality, we can assume $H_i \leq_e 0$ (we simply add a multiple of the identity to $H_i$).  Now consider local density matrices $\rho_1,\ldots,\rho_m$, where $\rho_i$ acts on the same subset of qubits as $H_i$.  We want to find $\rho_1,\ldots,\rho_m$ that correspond to the ground state of $H$.  Now consider the following convex program:
\begin{verse}
Find $\rho_1,\ldots,\rho_m$ that minimize $\sum_{i=1}^m \Tr(H_i\rho_i)$, 
subject to two constraints: \\
(1) For all $i$, $\rho_i \succeq 0$ and $\Tr(\rho_i) = 1$. \\
(2) There exists $\sigma$ s.t. $\sigma \succeq 0$, $\Tr(\sigma) = 1$, 
and for all $i$, $\Tr_{\set{1,\ldots,n}-C_i}(\sigma) \geq_e \rho_i$. \\
Here, $\rho_i$ is a $2^{|C_i|} \times 2^{|C_i|}$ real matrix, 
and $\sigma$ is a $2^n \times 2^n$ real matrix.
\end{verse}

We claim that this convex program is equivalent to the Stoquastic Local Hamiltonian problem.  If $H$ has an eigenstate $\ket{\varphi}$ with eigenvalue $\leq \lambda$, then the convex program has optimal value $\leq \lambda$; to see this, set $\rho_i = \Tr_{\set{1,\ldots,n}-C_i} \ket{\varphi}\bra{\varphi}$.  On the other hand, if all the eigenvalues of $H$ are $\geq \lambda + \delta$, then the convex program has optimal value $\geq \lambda + \delta$; to see this, observe that for any feasible $\rho_1,\ldots,\rho_m$, we have $\sum_{i=1}^m \Tr(H_i\rho_i) \geq \sum_{i=1}^m \Tr(H_i\sigma) = \Tr(H\sigma)$, using constraint (2) and the fact that $H_i \leq_e 0$.  

Finally, the task of solving this convex program reduces to Stoquastic Consistency.  If we have an oracle for Stoquastic Consistency, we can use it to check whether constraint (2) is satisfied.  This provides a membership oracle for the set $K$ of feasible solutions, which then allows us to solve the convex program (theorem \ref{ch3-thm-opt-mem}).  The main technical detail is to formulate the problem so that the set $K$ is full-dimensional, with inner and outer radii that satisfy $R/r \leq \poly(n)$.  This can be done using a subset of the local Pauli observables, where we account for the constraint that the density matrices must be real; we omit the details.

\subsection{Reducing from Stoquastic Consistency to Stoquastic Local Hamiltonian}

Next we show a reduction from Stoquastic Consistency to Stoquastic Local Hamiltonian.  The reduction uses strong duality, as in Section \ref{ch4-consistency-localham}.  

The first step is to represent $\rho_1,\ldots,\rho_m$ as the expectation values of certain observables.  However, we use a different set of observables, instead of the Pauli matrices, so that we can deal with inequalities involving the matrix elements of $\rho_i$.  For each $i$, define the following observables acting on the subset $C_i$:  
\[
X^{(i)}_{st} = \frac{1}{2}(\ket{s}\bra{t} + \ket{t}\bra{s}), 
  \qquad s,t \in \set{0,1}^{|C_i|}, \; s \preceq t, 
\]
where $s \preceq t$ denotes lexicographic order.  We can think of these observables as acting on the full $n$-qubit system (we tensor them with the identity matrix).  For any real $n$-qubit state $\sigma$, the matrix elements of $\Tr_{\set{1,\ldots,n}-C_i}(\sigma)$ are given by the expectation values of these observables:  
\[
\Tr(X^{(i)}_{st}\sigma) = \bra{s}\Tr_{\set{1,\ldots,n}-C_i}(\sigma)\ket{t}.  
\]
Then the conditions for a ``YES'' instance of Stoquastic Consistency can be written as:  
\[
\Tr(X^{(i)}_{st}\sigma) \geq \bra{s}\rho_i\ket{t}.  
\]

We let $\SSS$ be the set of all these observables $X^{(i)}_{st}$, for all of the subsets $C_i$, $i=1,\ldots,m$.  We also let $d = |\SSS|$.  Note that $\norm{X^{(i)}_{st}} \leq 1$.  These observables do not have any nice orthogonality properties, but the reduction from Consistency to Local Hamiltonian does not require that.  

Next, we formulate a convex program, together with its dual.  Define new observables 
\[
F^{(i)}_{st} = X^{(i)}_{st} - \bra{s}\rho_i\ket{t} I, 
\]
which are shifted so that our goal is to satisfy the inequalities $\Tr(F^{(i)}_{st}\sigma) \geq 0$.  For notational convenience, let us refer to these observables as $F_p$, for $p = 1,\ldots,d$.  Define $F(x)$ to be a linear combination of these observables, 
\[
F(x) = \sum_{p=1}^d x_p F_p + I, \quad \text{for $x \in \RR^d$}.  
\]

We construct a convex program which is similar to the one in Section \ref{ch4-consistency-localham}, except that we restrict $x$ to lie in the domain $[0,1]^d$ instead of $[-1,1]^d$.  
\begin{verse}
Find some $x \in [0,1]^d$ and $s \in [1-2d,1+2d]$ that \\
minimize $s$ such that $F(x) \preceq sI$.
\end{verse}
This is the primal program; let $p^*$ denote the optimal value of the objective function $s$.

The dual program is as follows:
\begin{verse}
Find some $2^n \times 2^n$ real matrix $Z$ that \\
maximizes $g(Z)$ such that $Z \succeq 0$ and $\Tr(Z) = 1$,
\end{verse}
where the dual function $g(Z)$ is given by 
\[
g(Z) = \inf_{x \in [0,1]^d} \Tr(ZF(x)) = \inf_{x \in [0,1]^d} \sum_{p=1}^d x_p \Tr(ZF_p) + 1.
\]
Let $d^*$ denote the optimal value of the objective function $g(Z)$.  Strong duality holds because the primal problem is convex and satisfies a generalized Slater condition \cite{BV} (to see this, note that the point $(x,s) = ((1/3d)\vec{1},2)$ is strictly feasible).  Strong duality implies that $p^* = d^*$.

Now, suppose we have a ``YES'' instance of Stoquastic Consistency.  Then in the dual program there exists some $Z \succeq 0$, $\Tr(Z) = 1$, such that for all $p$, $\Tr(ZF_p) \geq 0$.  This implies $g(Z) = 1$, hence the dual program has optimal value $d^* \geq 1$.  By strong duality, the primal program has optimal value $p^* \geq 1$.  

On the other hand, suppose we have a ``NO'' instance of Stoquastic Consistency.  Then for all $Z$ such that $Z \succeq 0$ and $\Tr(Z) = 1$, there is some $p$ such that $\Tr(ZF_p) \leq -\beta$, which implies $g(Z) \leq 1-\beta$.  Thus the dual program has optimal value $d^* \leq 1-\beta$.  By strong duality, the primal program has optimal value $p^* \leq 1-\beta$.  

Thus it suffices to solve the primal problem.  We claim that we can do this, given an oracle for Stoquastic Local Hamiltonian.  Observe that the $F_p$ are local operators, whose off-diagonal elements are all $\geq 0$.  Thus $-F(x)$ is a stoquastic local Hamiltonian, and we can use the oracle to estimate its ground state energy.  This is equivalent to estimating the largest eigenvalue of $F(x)$, which allows us to test whether the constraint $F(x) \preceq sI$ is satisfied.  Thus we have a membership oracle for the set $K$ of feasible solutions.  Using a similar analysis to section \ref{ch4-consistency-localham}, we can show that $K$ has inner and outer radii that satisfy $R/r \leq \poly(n)$.  Then, by theorem \ref{ch3-thm-opt-mem}, this allows us to solve the primal problem.  



\vskipline

\noindent
\textit{Acknowledgements:}  Thanks to Frank Verstraete and Daniel Nagaj for useful discussions.

\chapter{Gibbs States and the Consistency of Local Density Matrices}

\ifthenelse{\boolean{ucsdformat}}{\thispagestyle{chappage}}{}

Suppose we have an $n$-qubit system, and we are given a collection of local density matrices $\rho_1,\ldots,\rho_m$, where each $\rho_i$ describes some subset of the qubits.  We say that $\rho_1,\ldots,\rho_m$ are ``consistent'' if there exists a global state $\sigma$ (on all $n$ qubits) whose reduced density matrices match $\rho_1,\ldots,\rho_m$.  

We prove the following result:  if $\rho_1,\ldots,\rho_m$ are consistent with some state $\sigma \succ 0$, then they are also consistent with a state $\sigma'$ of the form $\sigma' = (1/Z) \exp(M_1+\cdots+M_m)$, where each $M_i$ is a Hermitian matrix acting on the same qubits as $\rho_i$, and $Z$ is a normalizing factor.  (This is known as a Gibbs state.)  Actually, we show a more general result, on the consistency of a set of expectation values $\expect{T_1},\ldots,\expect{T_r}$, where the observables $T_1,\ldots,T_r$ need not commute.  This result was previously proved by Jaynes (1957) in the context of the maximum-entropy principle; here we provide a somewhat different proof, using properties of the partition function.  


\section{Introduction}

Many-body systems have an intriguing property:  under the right circumstances, local interactions can conspire to produce long-range or global effects.  This behavior leads to phase transitions in statistical mechanics, and it also appears in combinatorial problems such as 3-SAT.  If we consider quantum systems, the situation is more complicated, due to non-commuting measurements and the possibility of entanglement.  This leads to new kinds of quantum phase transitions \cite{Sachdev}, and new examples such as the Local Hamiltonian problem \cite{AN}.  

A basic question in all of these examples is:  if we know local information about various parts of a system, what can we say about the system as a whole?  This paper gives one answer to this question, for quantum systems.  

Suppose we have an $n$-qubit system, and we are given a collection of local density matrices $\rho_1,\ldots,\rho_m$, where each $\rho_i$ describes a subset $C_i \subseteq \set{1,\ldots,n}$ of the qubits.  We say that $\rho_1,\ldots,\rho_m$ are ``consistent'' if there exists a global state $\sigma$ (on all $n$ qubits) whose reduced density matrices match $\rho_1,\ldots,\rho_m$; in other words, for all $i = 1,\ldots,m$, $\Tr_{\set{1,\ldots,n}-C_i}(\sigma) = \rho_i$.  

Clearly, if $\rho_1,\ldots,\rho_m$ are consistent, then whenever two density matrices $\rho_i$ and $\rho_j$ describe overlapping subsets of qubits ($C_i \intersect C_j \neq \emptyset$), they must agree on the intersection $C_i \intersect C_j$; that is, $\Tr_{C_i-(C_i \intersect C_j)}(\rho_i) = \Tr_{C_j-(C_i \intersect C_j)}(\rho_j)$.  This gives a necessary condition for consistency.  

However, the above condition is not sufficient to guarantee consistency.  To see this, consider the following example:  we have three qubits, and we are told that qubits 1 and 2 are in the Bell state $\ket{\Phi^+} = (\ket{00}+\ket{11})/\sqrt{2}$, and qubits 2 and 3 are also in the same state $\ket{\Phi^+}$.  More formally, let $\rho_A = \ket{\Phi^+}\bra{\Phi^+}$, $A = \set{1,2}$, and let $\rho_B = \ket{\Phi^+}\bra{\Phi^+}$, $B = \set{2,3}$.  In this case, $\rho_A$ and $\rho_B$ both agree on qubit 2, since $\Tr_1(\rho_A) = I/2 = \Tr_3(\rho_B)$.  But there is no state $\sigma$ on all three qubits such that $\Tr_3(\sigma) = \rho_A$ and $\Tr_1(\sigma) = \rho_B$; one way to see this is to apply the strong subadditivity inequality, $S(1,2,3) + S(2) \leq S(1,2) + S(2,3)$.  

Thus the consistency of $\rho_1,\ldots,\rho_m$ would seem to be a more subtle question.  We prove the following result:  
\begin{thm}\label{consistent-rhos}
If $\rho_1,\ldots,\rho_m$ are consistent with some state $\sigma \succ 0$, then they are also consistent with a state $\sigma'$ of the form $\sigma' = (1/Z) \exp(M_1+\cdots+M_m)$, where each $M_i$ is a Hermitian matrix acting on the qubits in $C_i$, and $Z = \Tr(\exp(M_1+\cdots+M_m))$.  
\end{thm}
Here, $\sigma \succ 0$ means that $\sigma$ is a positive definite matrix.  The state $\sigma'$ is known as a Gibbs state.  

Essentially, this result says that a Gibbs state $\sigma'$ can simulate an arbitrary state $\sigma \succ 0$, with respect to an observer who can only access subsets $C_1,\ldots,C_m$ of the qubits.  For example, consider a physical system with local interactions, described by a Hamiltonian $H$.  It is easy to see that the ground state of $H$ can be approximated by $\eta = (1/Z) \exp(-\beta H)$, for $\beta$ large; and since $H$ is a sum of local terms, $\eta$ is a Gibbs state.  Our result extends this simple observation to a much more general setting.  

Actually, we prove the following more general result:  Consider a finite quantum system, and let $T_1,\ldots,T_r$ be observables (Hermitian matrices).  Without loss of generality, assume that the collection of matrices $I,T_1,\ldots,T_r$ is linearly independent (over $\RR$).  We say that a state $\rho$ has expectation values $t_1,\ldots,t_r$ if $\Tr(T_i \rho) = t_i$ for all $i = 1,\ldots,r$.  
\begin{thm}\label{consistent-ts}
If there exists some state $\rho \succ 0$ which has expectation values $t_1,\ldots,t_r$, then there exists a state $\rho'$ which has the same expectation values $t_1,\ldots,t_r$, and is of the form $\rho' = (1/Z) \exp(\theta_1 T_1 + \cdots + \theta_r T_r)$, where $\theta_1,\ldots,\theta_r \in \RR$.  
\end{thm}
This statement holds even when the observables $T_1,\ldots,T_r$ do not commute.  

This result was previously proved by Jaynes, as part of the maximum entropy principle in statistical mechanics \cite{Jaynes-2,Jaynes-Brandeis}.  Jaynes showed that the Gibbs state $\rho'$ is the state which maximizes the entropy $S(\rho) = -\Tr(\rho \log \rho)$ subject to the constraints $\expect{T_i} = t_i$; implicitly, he also showed that the Gibbs state $\rho'$ is always feasible, in the sense that it can produce the same expectation values $\expect{T_i}$ as an arbitrary state $\rho \succ 0$.  

However, Jaynes' motivation was somewhat different from ours.  Jaynes was interested in statistical mechanics, which deals with large systems with many degrees of freedom and only a few constraints.  Feasibility is not usually a concern in such cases, while the maximum-entropy property is crucial in making plausible inferences about the ``true'' state of the system.  

In this paper, we focus on finite quantum systems, with many non-commuting constraints; we are interested in the relationship between local constraints and the global state of the system.  For us, feasibility of the Gibbs state is important, since it is possible for the system to become overdetermined.  Statistical inference is less important, because the systems we study are small enough that their state can be completely determined (at least in principle).  Rather than viewing this as an inference problem, we can speak directly about what states are allowed under a given set of constraints.  

Finally, we prove our result using a technique which is different from Jaynes.  Jaynes used the Lagrange dual of the entropy-maximization problem, while we use some analytic properties of the partition function.  Our analysis bears some resemblance to classical results on exponential families in statistics \cite{Brown}---although the technical details are quite different.  Our proof also contains some geometric intuition which may be of interest.


\section{Proofs of our results}

First, we will review some useful facts about the partition function for a Gibbs state.  Then we will prove theorem \ref{consistent-ts}, and obtain theorem \ref{consistent-rhos} as a special case.  

\subsection{The partition function}

Recall the situation described in theorem \ref{consistent-ts}:  we have a finite quantum system, and observables $T_1,\ldots,T_r$, such that $I,T_1,\ldots,T_r$ are linearly independent (over $\RR$).  We are interested in states of the form 
\[
\rho(\theta) = \exp(\theta_1 T_1 + \cdots + \theta_r T_r) / Z(\theta), 
\quad \theta \in \RR^r, 
\]
where $Z(\theta) = \Tr(\exp(\theta_1 T_1 + \cdots + \theta_r T_r))$.  $Z(\theta)$ is called the partition function, and we also define the log partition function $\psi(\theta) = \log Z(\theta)$.  

Note that, in the above definition, we can translate each observable $T_i$ by a multiple of the identity, without changing the state $\rho(\theta)$.  More precisely, if we define new observables $P_i = T_i + \lambda_i I$, with $\lambda_i \in \RR$, we have that:  
\[
\frac{\exp(\theta_1 P_1 + \cdots + \theta_r P_r)}
     {\Tr(\exp(\theta_1 P_1 + \cdots + \theta_r P_r))}
 = \frac{\exp(\theta_1 T_1 + \cdots + \theta_r T_r)}
        {\Tr(\exp(\theta_1 T_1 + \cdots + \theta_r T_r))}.  
\]
Using subscripts $T$ and $P$ to denote the two sets of observables, we arrive at the same state, $\rho_P(\theta) = \rho_T(\theta)$, although the partition functions are different, $Z_P(\theta) \neq Z_T(\theta)$.  

The log partition function $\psi$ has some nice analytic properties:  it is convex, and its derivatives encode the expectation values of the observables $T_i$.  We briefly sketch these results, which can be found in quantum statistical mechanics \cite{Jaynes-Brandeis}, as well as quantum information geometry \cite{Ingarden}.  

\begin{prop}\label{convexity}
$\psi$ is convex on $\RR^r$.  
\end{prop}
Proof sketch:  This follows from some facts in matrix analysis \cite{Bhatia}.  First, the Golden-Thompson inequality:  If $A$ and $B$ are Hermitian matrices, then 
\[
\Tr(\exp(A+B)) \leq \Tr(\exp(A)\exp(B)).  
\]
Next, a matrix version of H\"older's inequality:  For any matrix $A$, define the Frobenius or Hilbert-Schmidt norm to be $\norm{A}_2 = (\Tr(A^\dagger A))^{1/2}$.  Also, let $|A|$ denote the unique positive semidefinite square root of $A^\dagger A$.  Then we have that, for all square matrices $A$ and $B$, 
\[
\norm{AB}_2 \leq \norm{|A|^p}_2^{1/p} \norm{|B|^q}_2^{1/q}, 
\]
for $\tfrac{1}{p} + \tfrac{1}{q} = 1$, $p > 1$.  $\square$

\begin{prop}\label{derivatives}
$\psi$ is differentiable on $\RR^r$, and 
\[
\frac{\partial\psi}{\partial\theta_i} = \Tr(T_i \rho(\theta)) = \expect{T_i}.  
\]
\end{prop}
Proof sketch:  Use ``parameter differentiation'' \cite{Wilcox}:  If $H$ is a Hermitian matrix which depends on a parameter $\lambda$, and $\partial H / \partial \lambda$ and $\partial^2 H / \partial \lambda^2$ exist and are continuous, then $\partial(\exp(H)) / \partial \lambda$ exists and is equal to 
\[
\frac{\partial}{\partial\lambda} \exp(H)
 = \int_0^1 \exp((1-u)H) \frac{\partial H}{\partial\lambda} \exp(uH) du.  \quad \square
\]


\subsection{Proof of theorem 2}

Proof:  We are given expectation values $t_1,\ldots,t_r$, and we want to find a state 
\[
\rho'(\theta) = \exp(\theta_1 T_1 + \cdots + \theta_r T_r) / Z'(\theta) 
\]
that has these expectation values.  (Here, $Z'(\theta)$ is the partition function, and $\psi'(\theta) = \log Z'(\theta)$ is the log partition function.)  By translating the observables $T_i$, we can assume that $t_i = 0$, for all $i = 1,\ldots,r$.  We can now restate the problem in terms of the log partition function:  we are looking for some $\theta \in \RR^r$ such that $\nabla \psi'(\theta) = 0$.  

We know there exists a state $\rho \succ 0$ which has the desired expectation values $t_1,\ldots,t_r$.  Now choose some observables $U_1,\ldots,U_s$, such that the set $\set{I,T_1,\ldots,T_r,$ $U_1,\ldots,U_s}$ is complete and linearly independent (in other words, any $2^n$-dimensional Hermitian matrix can be written uniquely as a real linear combination of the matrices in this set).  Let $u_1,\ldots,u_s$ be the expectation values of $\rho$ for the observables $U_1,\ldots,U_s$; that is, $u_i = \Tr(U_i \rho)$.  By translating the $U_i$, we can assume that $u_i = 0$, for all $i = 1,\ldots,s$.  

We will consider states of the form 
\[
\begin{split}
\rho(\theta,\phi)
 = \exp\bigl( & \theta_1 T_1 + \cdots + \theta_r T_r + \\
              & \phi_1 U_1 + \cdots + \phi_s U_s \bigr) / Z(\theta,\phi).  
\end{split}
\]
(Here, $Z(\theta,\phi)$ is the partition function, and $\psi(\theta,\phi) = \log Z(\theta,\phi)$ is the log partition function.)  Completeness of the $T_i$ and the $U_i$ implies that we can write $\rho$ in the form $\rho = \rho(\theta,\phi)$ for some $(\theta,\phi) \in \RR^{r+s}$.  This implies that $\nabla \psi(\theta,\phi) = 0$ for some $(\theta,\phi) \in \RR^{r+s}$.  

Furthermore, we claim that there is a unique point $(\theta,\phi)$ such that $\rho(\theta,\phi)$ has the expectation values $t_i$ and $u_i$.  This is because the expectation values $t_i$ and $u_i$ uniquely determine the state $\rho$, and setting $\rho = \rho(\theta,\phi)$ uniquely determines the values of $\theta$ and $\phi$.  This in turn follows from the completeness and linear independence of the $T_i$ and the $U_i$.  So we conclude that $\nabla \psi(\theta,\phi) = 0$ at exactly one point $(\theta,\phi)$.  

\begin{figure}
\begin{center}
\includegraphics{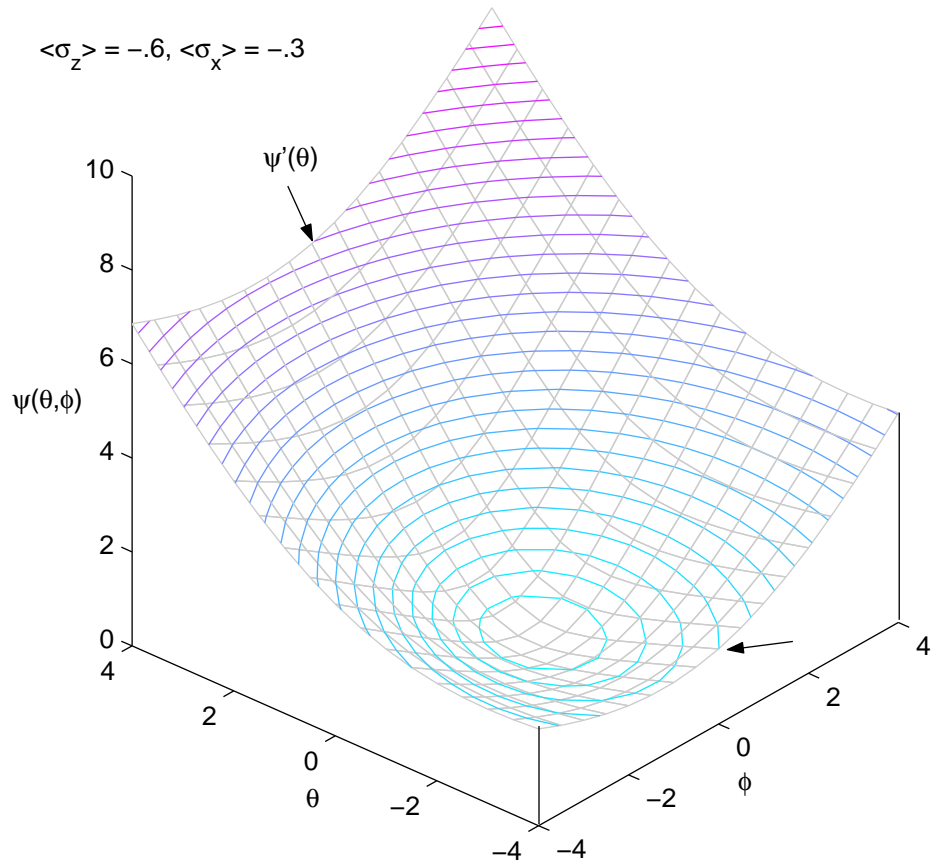}
\caption{A single-qubit example.  We want to find a Gibbs state $\rho'$ that satisfies $\expect{\sigma_z} = -0.6$; we have one observable $T = \sigma_z + 0.6$.  We know there exists some state $\rho \succ 0$ that satisfies $\expect{\sigma_z} = -0.6$; in this case, $\rho$ also satisfies $\expect{\sigma_x} = -0.3$, and we let $U = \sigma_x + 0.3$ play the role of the ``extra'' observables.  As the graph shows, $\nabla \psi(\theta,\phi)$ vanishes at exactly one point; $\psi'(\theta) = \psi(\theta,0)$; and $\nabla \psi'(\theta)$ vanishes for some $\theta$.}
\label{fig-geometry}
\addcontentsline{toc}{chapter}{\protect\numberline{}Figure 5.1:  A single-qubit example}
\end{center}
\end{figure}

To complete the proof, we will carry out the following plan:  we will show that $\psi(\theta,\phi) \rightarrow \infty$ as $\norm{\theta,\phi} \rightarrow \infty$, where $\norm{\theta,\phi}$ denotes the norm of the vector $(\theta,\phi)$.  This implies that $\psi'(\theta) \rightarrow \infty$ as $\norm{\theta} \rightarrow \infty$; and hence $\nabla \psi'(\theta) = 0$ for some $\theta \in \RR^r$.  (See figure \ref{fig-geometry} for a simple example that shows the geometric intuition for the proof.)  

Let $(\theta_0,\phi_0)$ be the unique point where $\nabla \psi$ vanishes.  We claim that $(\theta_0,\phi_0)$ is the unique global minimum of $\psi$.  [Since $\psi$ is convex (proposition \ref{convexity}), it follows that $\psi$ is bounded below, and $(\theta_0,\phi_0)$ is a global minimum.  Also, $\psi$ is differentiable everywhere on the domain $\RR^{r+s}$, which has no boundaries (proposition \ref{derivatives}); so any extremum $(\theta,\phi)$ must satisfy $\nabla \psi(\theta,\phi) = 0$.  But this happens only at $(\theta_0,\phi_0)$, and so $(\theta_0,\phi_0)$ is the unique global minimum.]  

Let $S$ be the set of all unit vectors in $\RR^{r+s}$.  Define the function $f(\nu,z) = \psi((\theta_0,\phi_0) + z\nu)$, for $\nu \in S$, and $z \in \RR$.  Say we fix $z = 1$.  We claim that there exists some $b > 0$ such that, for all $\nu$, $f(\nu,1) \geq \psi(\theta_0,\phi_0) + b$.  [Since $(\theta_0,\phi_0)$ is the unique global minimum, we have that $f(\nu,1) > \psi(\theta_0,\phi_0)$, for all $\nu$.  Moreover, $f(\nu,1)$ is a continuous function of $\nu$, and $S$ is a compact set, hence its image $f(S,1)$ is compact.  Hence $f(\nu,1)$ must be bounded away from $\psi(\theta_0,\phi_0)$, for all $\nu$.]  

Next we claim that, for all $\nu$, and for all $z \geq 1$, $(\partial f / \partial z)(\nu,z) \geq b$.  [Fix any $\nu$.  $f(\nu,z)$ is a differentiable function of $z$, so by the mean value theorem, there exists some $z \in (0,1)$ such that $(\partial f / \partial z)(\nu,z) = f(\nu,1) - f(\nu,0) \geq b$.  In addition, since $\psi$ is convex, $(\partial f / \partial z)(\nu,z)$ is nondecreasing in $z$.  This proves the claim.]  

Now, say we are given some $(\theta,\phi)$, and assume that $\norm{(\theta,\phi)-(\theta_0,\phi_0)} \geq 1$.  We can write $(\theta,\phi)$ in the form 
\[
(\theta,\phi) = (\theta_0,\phi_0) + \norm{(\theta,\phi)-(\theta_0,\phi_0)} \nu, 
\]
for some unit vector $\nu \in S$.  Then we have:  
\[
\begin{split}
\psi(\theta,\phi)
 &= f(\nu,\norm{(\theta,\phi)-(\theta_0,\phi_0)})\\
 &= f(\nu,1) + \int_1^{\norm{(\theta,\phi)-(\theta_0,\phi_0)}} 
               (\partial f / \partial z)(\nu,z) dz\\
 &\geq \psi(\theta_0,\phi_0) + b + b (\norm{(\theta,\phi)-(\theta_0,\phi_0)} - 1)\\
 &= \psi(\theta_0,\phi_0) + b \norm{(\theta,\phi)-(\theta_0,\phi_0)}.  
\end{split}
\]
From this, we conclude that $\psi(\theta,\phi) \rightarrow \infty$ as $\norm{\theta,\phi} \rightarrow \infty$.  

Notice that the partition functions for $\rho'(\theta)$ and $\rho(\theta,\phi)$ are related:  
\[
\psi'(\theta) = \psi(\theta,0).  
\]
Hence, $\psi'(\theta) \rightarrow \infty$ as $\norm{\theta} \rightarrow \infty$.  

We will use the following fact:  if $f:\; \RR^n \rightarrow \RR$ is continuous, and $f(x) \rightarrow \infty$ as $\norm{x} \rightarrow \infty$, then $f$ is bounded below, and $f$ attains its minimum at some point $x_* \in \RR^n$.  [To see this, let $S = \set{x \in \RR^n \;|\; f(x) \leq \alpha}$, choosing $\alpha$ large enough that $S \neq \emptyset$.  Note that $S$ is bounded; otherwise, there would exist a sequence $\set{x_i}$ such that $\norm{x_i} \rightarrow \infty$ and $f(x_i) \leq \alpha$, a contradiction.  Also, note that $S$ is closed; this is because the interval $(-\infty,\alpha]$ is closed, and $f$ is continuous.  So we have that $S$ is compact.  This implies that $f(S)$ is compact.  Hence $f(S)$ is closed and bounded; also note that $f(S) \neq \emptyset$.  This implies that $f$ is bounded below, and attains its minimum.]  

From this, we conclude that $\psi'$ attains its minimum at some point $\theta_* \in \RR^r$.  $\RR^r$ has no boundaries, and $\psi'$ is differentiable everywhere on $\RR^r$, so it follows that $\nabla \psi'(\theta_*) = 0$.  This completes the proof.  $\square$


\subsection{Proof of theorem \ref{consistent-rhos}}

Proof:  We will obtain theorem \ref{consistent-rhos} as a special case of theorem \ref{consistent-ts}.  The basic idea is that specifying the local density matrices $\rho_1,\ldots,\rho_m$ is equivalent to specifying the expectation values of all Pauli matrices on the subsets $C_1,\ldots,C_m$.  

Let $X$, $Y$ and $Z$ denote the Pauli matrices for a single qubit, and define $\PP = \set{I,X,Y,Z}$.  We can construct $n$-qubit Pauli matrices by taking tensor products $P = P_1 \tensor \cdots \tensor P_n \in \PP^{\tensor n}$.  Any $2^n$-dimensional Hermitian matrix can be written as a real linear combination of $n$-qubit Pauli matrices.  Furthermore, the $n$-qubit Pauli matrices are orthogonal with respect to the Hilbert-Schmidt inner product:  $\Tr(P^\dagger Q) = 2^n$ if $P = Q$, and 0 otherwise.  

We make the following claim:  Let $\sigma$ be a density matrix on $n$ qubits, and let $\rho$ be a density matrix on a subset of the qubits $C \subseteq \set{1,\ldots,n}$, with $|C| = k$.  We claim that $\Tr_{\set{1,\ldots,n}-C}(\sigma) = \rho$, if and only if, for all Pauli matrices $P$ on the subset $C$, $\Tr((P \tensor I) \sigma) = \Tr(P \rho)$.  (Notation:  we write $n$-qubit Pauli matrices in the form $P \tensor Q$, where $P$ acts on the subset $C$, and $Q$ acts on the rest of the qubits.)  

The ($\Rightarrow$) direction is obvious, but we need to show ($\Leftarrow$).  We write $\sigma$ and $\rho$ as linear combinations of Pauli matrices, with real coefficients $\beta_{(P \tensor Q)}$ and $\alpha_P$:  
\begin{align*}
\sigma &= \sum_{(P \tensor Q) \in \PP^{\tensor n}} \beta_{(P \tensor Q)} P \tensor Q \\
\rho &= \sum_{P \in \PP^{\tensor k}} \alpha_P P.  
\end{align*}
We know that, for all Pauli matrices $P$ on the subset $C$, $\Tr((P \tensor I) \sigma) = 2^n \beta_{(P \tensor I)} = \Tr(P \rho) = 2^k \alpha_P$.  But this implies:  
\[
\begin{split}
\Tr_{\set{1,\ldots,n}-C}(\sigma)
 &= \sum_{P \in \PP^{\tensor k}} 2^{n-k} \beta_{(P \tensor I)} P \\
 &= \sum_{P \in \PP^{\tensor k}} \alpha_P P = \rho, 
\end{split}
\]
which proves the claim.  

Thus, theorem \ref{consistent-rhos} is a special case of theorem \ref{consistent-ts}, where the observables $T_1,\ldots,T_r$ consist of all the Pauli matrices on the subsets $C_1,\ldots,C_m$.  $\square$


\vskipline

\noindent
\textit{Acknowledgements:}  I am grateful to Dorit Aharonov, Chris Fuchs and David Meyer for helpful discussions about this work.  Funded by an ARO/NSA Quantum Computing Graduate Research Fellowship.



\chapter{Conclusions}

\ifthenelse{\boolean{ucsdformat}}{\thispagestyle{chappage}}{}

In this dissertation we have studied the complexity of the Consistency and $N$-represent\-ability problems.  We showed that these problems are QMA-complete, using reductions based on convex optimization with a membership oracle.  In addition, we showed that certain special cases of Consistency and Local Hamiltonian have the same complexity (even though they are not known to be QMA-hard).  

A number of interesting open problems remain.  Are there better reductions from convex optimization to membership?  (In particular, are there reductions that have a less stringent precision requirement for the membership oracle?)  Can one give a mapping reduction from Local Hamiltonian to Consistency, rather than an oracle reduction?  Can one show that \textit{approximately} solving Local Hamiltonian is QMA-hard?  (This would be a quantum analogue of the celebrated PCP theorem \cite{vv,al,dinur}.)  

We are also starting to understand the complexity of special classes of quantum systems.  In chapter 3 we remarked that translationally-invariant systems seem to be an easy special case.  However, a recent result \cite{kay} shows that this is no longer true if one allows interactions involving $\log(n)$ particles, or particles that have $n$ states---in these cases, Local Hamiltonian is once more QMA-complete.

On the positive side, recent work suggests that there is a polynomial-time approximation scheme (PTAS) for Local Hamiltonian on planar graphs \cite{bbt}, even though solving the problem exactly is QMA-hard.  Also, it seems likely that one can get a PTAS for Local Hamiltonian on a 1-D chain, by reducing to the Consistency problem and applying $p$-positivity conditions \cite{liu-ppos}.


\bibliographystyle{plain}
\bibliography{thesis}

\begin{thebibliography}{10}

\bibitem{Aaronson}
S.~Aaronson.
\newblock Quantum computing, postselection, and probabilistic polynomial-time.
\newblock {\em Proc. Royal Society A}, 461(2063):3473--3482, 2005.

\bibitem{Abrams-Lloyd-97}
D.S. Abrams and S.~Lloyd.
\newblock Simulation of many-body fermi systems on a universal quantum
  computer.
\newblock {\em Phys. Rev. Lett.}, 79(13):2586--2589, 1997.
\newblock Arxiv: quant-ph/9703054.

\bibitem{Abrams-Lloyd-99}
D.S. Abrams and S.~Lloyd.
\newblock Quantum algorithm providing exponential speed increase for finding
  eigenvalues and eigenvectors.
\newblock {\em Phys. Rev. Lett.}, 83(24):5162--5165, 1999.

\bibitem{ADH97}
L.~Adleman, J.~DeMarrais, and M.~Huang.
\newblock Quantum computability.
\newblock {\em SIAM J. Comput.}, 26(5):1524--1540, 1997.

\bibitem{A}
D.~Aharonov.
\newblock Private communication, 2004.

\bibitem{qma1d-1}
D.~Aharonov, D.~Gottesman, and J.~Kempe.
\newblock The power of quantum systems on a line.
\newblock Arxiv:0705.4077v1 [quant-ph], 2007.

\bibitem{AN}
D.~Aharonov and T.~Naveh.
\newblock Quantum {NP} - a survey.
\newblock Arxiv: quant-ph/0210077, 2002.

\bibitem{AR}
D.~Aharonov and O.~Regev.
\newblock A lattice problem in quantum {NP}.
\newblock In {\em IEEE Foundations of Computer Science (FOCS '03)}, page 210,
  2003.
\newblock Arxiv: quant-ph/0307220.

\bibitem{ADKLLR}
D.~Aharonov, W.~van Dam, J.~Kempe, Z.~Landau, S.~Lloyd, and O.~Regev.
\newblock Adiabatic quantum computation is equivalent to standard quantum
  computation.
\newblock In {\em IEEE Foundations of Computer Science (FOCS'04)}, pages
  42--51, 2004.
\newblock Arxiv: quant-ph/0405098.

\bibitem{al}
S.~Arora and C.~Lund.
\newblock Hardness of approximations.
\newblock In D.S. Hochbaum, editor, {\em Approximation Algorithms for NP-Hard
  Problems}. PWS Publishing, Boston, 1997.

\bibitem{Aspuru-Guzik-et-al-05}
A.~Aspuru-Guzik, A.D. Dutoi, P.J. Love, and M.~Head-Gordon.
\newblock Simulated quantum computation of molecular energies.
\newblock {\em Science}, 309:1704--1707, 2005.

\bibitem{bbt}
N.~Bansal, S.~Bravyi, and B.M. Terhal.
\newblock A classical approximation scheme for the ground-state energy of
  {I}sing spin {H}amiltonians on planar graphs.
\newblock Arxiv preprint: 0705.1115, 2007.

\bibitem{Barahona}
F.~Barahona.
\newblock On the computational complexity of {I}sing spin glass models.
\newblock {\em J. Phys. A: Math. Gen.}, 15(10):3241--3253, 1982.

\bibitem{Bellman}
R.~Bellman.
\newblock {\em Introduction to Matrix Analysis}.
\newblock McGraw-Hill, New York, 1970.

\bibitem{BBBV97}
C.H. Bennett, E.~Bernstein, G.~Brassard, and U.~Vazirani.
\newblock Strengths and weaknesses of quantum computing.
\newblock {\em SIAM J. Comput.}, 26(5):1510--1523, 1997.

\bibitem{BV97}
E.~Bernstein and U.~Vazirani.
\newblock Quantum complexity theory.
\newblock {\em SIAM J. Comput.}, 26(5):1411--1473, 1997.

\bibitem{BV}
D.~Bertsimas and S.~Vempala.
\newblock Solving convex programs by random walks.
\newblock {\em J. ACM}, 51(4):540--556, 2004.

\bibitem{Bhatia}
R.~Bhatia.
\newblock {\em Matrix analysis}.
\newblock Springer-Verlag, New York, 1997.

\bibitem{reduced2}
S.~Bravyi.
\newblock Compatibility between local and multipartite states.
\newblock {\em Quant. Info. and Comput.}, 4(1):12--26, 2004.

\bibitem{Bravyi}
S.~Bravyi.
\newblock Efficient algorithm for a quantum analogue of 2-{SAT}.
\newblock Arxiv: quant-ph/0602108, 2006.

\bibitem{stoq-2}
S.~Bravyi, A.J. Bessen, and B.M. Terhal.
\newblock {M}erlin-{A}rthur games and stoquastic complexity.
\newblock Arxiv: quant-ph/0611021, 2006.

\bibitem{stoq-1}
S.~Bravyi, D.P. DiVincenzo, R.I. Oliveira, and B.M. Terhal.
\newblock The complexity of stoquastic local hamiltonian problems.
\newblock Arxiv: quant-ph/0606140, 2006.

\bibitem{Bravyi-Kitaev-00}
S.~Bravyi and A.~Kitaev.
\newblock Fermionic quantum computation.
\newblock Arxiv: quant-ph/0003137, 2000.

\bibitem{BravyiVyalyi}
S.~Bravyi and M.~Vyalyi.
\newblock Commutative version of the local {H}amiltonian problem and common
  eigenspace problem.
\newblock {\em Quantum Info. and Comput.}, 5(3):187--215, 2005.

\bibitem{Brown}
L.D. Brown.
\newblock {\em Fundamentals of statistical exponential families with
  applications in statistical decision theory}.
\newblock IMS Lecture Notes---Monograph Series, 9. Institute of Mathematical
  Statistics, Hayward, CA, 1986.

\bibitem{reduced3}
M.~Christandl and G.~Mitchison.
\newblock The spectra of quantum states and the {K}ronecker coefficients of the
  symmetric group.
\newblock {\em Commun. Math. Phys.}, 261(3):789--797, 2006.

\bibitem{book}
J.~Cioslowski, editor.
\newblock {\em Many-Electron Densities and Reduced Density Matrices}.
\newblock Kluwer Academic, New York, 2000.

\bibitem{Coleman}
A.~J. Coleman.
\newblock Structure of fermion density matrices.
\newblock {\em Rev. Mod. Phys.}, 35(3):668--686, Jul 1963.

\bibitem{book1}
A.J. Coleman and V.I. Yukalov.
\newblock {\em Reduced Density Matrices: Coulson's Challenge}.
\newblock Springer-Verlag, Berlin, 2000.

\bibitem{Coulson}
C.~A. Coulson.
\newblock Present state of molecular structure calculations.
\newblock {\em Rev. Mod. Phys.}, 32(2):170--177, Apr 1960.

\bibitem{reduced5}
S.~Daftuar and P.~Hayden.
\newblock Quantum state transformations and the {S}chubert calculus.
\newblock {\em Ann. Phys.}, 315(1):80--122, 2005.

\bibitem{DHHMNO}
C.M. Dawson, H.L. Haselgrove, A.P. Hines, D.~Mortimer, M.A. Nielsen, and T.J.
  Osborne.
\newblock Quantum computing and polynomial equations over the finite field
  {Z2}.
\newblock Arxiv: quant-ph/0408129, 2004.

\bibitem{Deza}
M.~Deza and M.~Laurent.
\newblock Applications of cut polyhedra — {II}.
\newblock {\em J. Comput. Appl. Math.}, 55(2):217--247, 1994.

\bibitem{dinur}
I.~Dinur.
\newblock The {PCP} theorem by gap amplification.
\newblock {\em J. ACM}, 54(3), 2007.

\bibitem{FGGS}
E.~Farhi, J.~Goldstone, S.~Gutmann, and M.~Sipser.
\newblock Quantum computation by adiabatic evolution.
\newblock Arxiv: quant-ph/0001106, 2000.

\bibitem{FR98}
L.~Fortnow and J.D. Rogers.
\newblock Complexity limitations on quantum computation.
\newblock In {\em Proc. IEEE Complexity'98}, pages 202--209, 1998.

\bibitem{GLS-1981}
M.~Grotschel, L.~Lovasz, and A.~Schrijver.
\newblock The ellipsoid method and its consequences in combinatorial
  optimization.
\newblock {\em Combinatorica}, 1(2):169--197, 1981.

\bibitem{GLS}
M.~Gr{\"o}tschel, L.~Lov{\'a}sz, and A.~Schrijver.
\newblock {\em Geometric algorithms and combinatorial optimization}.
\newblock Springer-Verlag, Berlin, 1988.

\bibitem{Grover}
L.K. Grover.
\newblock A fast quantum mechanical algorithm for database search.
\newblock In {\em ACM Symp. on Theory of Computing (STOC '96)}, pages 212--219,
  1996.

\bibitem{Gurvits}
L.~Gurvits.
\newblock Classical deterministic complexity of {E}dmonds' problem and quantum
  entanglement.
\newblock In {\em ACM Symp. on Theory of Computing (STOC '03)}, pages 10--19,
  2003.

\bibitem{Hall}
W.~Hall.
\newblock Compatibility of subsystem states and convex geometry.
\newblock Arxiv: quant-ph/0610031, 2006.

\bibitem{reduced}
A.~Higuchi, A.~Sudbery, and J.~Szulc.
\newblock One-qubit reduced states of a pure many-qubit state: Polygon
  inequalities.
\newblock {\em Phys. Rev. Lett.}, 90(10):107902, Mar 2003.

\bibitem{Horodecki}
M.~Horodecki, P.~Horodecki, and R.~Horodecki.
\newblock Mixed-state entanglement and quantum communication.
\newblock In {\em Quantum Information: An Introduction to Basic Theoretical
  Concepts and Experiments}. Springer, 2001.
\newblock Arxiv: quant-ph/0109124.

\bibitem{Ingarden}
R.S. Ingarden, H.~Janyszek, A.~Kossakowski, and T.~Kawaguchi.
\newblock Information geometry of quantum statistical systems.
\newblock {\em Tensor (N.S.)}, 37:105--111, 1982.

\bibitem{qma1d-2}
S.~Irani.
\newblock The complexity of quantum systems on a one-dimensional chain.
\newblock Arxiv:0705.4067v1 [quant-ph], 2007.

\bibitem{JWB}
D.~Janzing, P.~Wocjan, and T.~Beth.
\newblock Identity check is {QMA}-complete.
\newblock Arxiv: quant-ph/0305050, 2003.

\bibitem{Jaynes-2}
E.~T. Jaynes.
\newblock Information theory and statistical mechanics. {II}.
\newblock {\em Phys. Rev.}, 108(2):171--190, Oct 1957.

\bibitem{Jaynes-Brandeis}
E.T. Jaynes.
\newblock Information theory and statistical mechanics (lectures at brandeis).
\newblock Reprinted in \textit{E.T. Jaynes: Papers on Probability, Statistics
  and Statistical Physics}, R.D. Rosenkrantz (ed.), D. Reidel Publishing
  Company, 1983, 1962.

\bibitem{KV}
A.T. Kalai and S.~Vempala.
\newblock {Simulated Annealing for Convex Optimization}.
\newblock {\em Mathematics of Operations Research}, 31(2):253--266, 2006.

\bibitem{kay}
A.~Kay.
\newblock A {QMA}-complete translationally invariant {H}amiltonian problem and
  the complexity of finding ground state energies in physical systems.
\newblock Arxiv preprint: 0704.3142, 2007.

\bibitem{KKR}
J.~Kempe, A.~Kitaev, and O.~Regev.
\newblock The complexity of the local {H}amiltonian problem.
\newblock {\em SIAM J. Comput.}, 35(5):1070--1097, 2006.

\bibitem{KR}
J.~Kempe and O.~Regev.
\newblock 3-{L}ocal {H}amiltonian is {QMA}-complete.
\newblock {\em Quantum Info. and Comput.}, 3(3):258--264, 2003.

\bibitem{KSV}
A.~Yu. Kitaev, A.~H. Shen, and M.~N. Vyalyi.
\newblock {\em Classical and quantum computation}.
\newblock American Mathematical Society, Providence, RI, 2002.

\bibitem{KW}
A.Yu. Kitaev and J.~Watrous.
\newblock {QMA} is contained in {PP}.
\newblock Unpublished, 2001?

\bibitem{reduced4}
A.~Klyachko.
\newblock Quantum marginal problem and representations of the symmetric group.
\newblock Arxiv: quant-ph/0409113, 2004.

\bibitem{Klyachko05}
A.~Klyachko.
\newblock Quantum marginal problem and {N}-representability.
\newblock {\em J. of Physics: Conference Series}, 36:72--86, 2006.

\bibitem{Knuth-vol2}
D.E. Knuth.
\newblock {\em The Art of Computer Programming: Volume 2, Seminumerical
  Algorithms (3rd ed.)}.
\newblock Addison Wesley, 1998.

\bibitem{Matsumoto}
H.~Kobayashi, K.~Matsumoto, and T.~Yamakami.
\newblock Quantum certificate verification: Single versus multiple quantum
  certificates.
\newblock Arxiv: quant-ph/0110006, 2001.

\bibitem{Liu-ZK}
Y.-K. Liu.
\newblock Computational zero-knowledge proofs for the consistency of local
  quantum states.
\newblock Unpublished, 2006.

\bibitem{Liu-consistency-qma}
Y.-K. Liu.
\newblock Consistency of local density matrices is {QMA}-complete.
\newblock In {\em Approximation, Randomization and Combinatorial Optimization
  (APPROX + RANDOM '06)}, pages 438--449, 2006.
\newblock LNCS 4110, Springer.

\bibitem{liu-ppos}
Y.-K. Liu.
\newblock In preparation, 2007.

\bibitem{Liu-N-rep}
Y.-K. Liu, M.~Christandl, and F.~Verstraete.
\newblock Quantum computational complexity of the {N}-representability problem:
  {QMA} complete.
\newblock {\em Phys. Rev. Lett.}, 98(11):110503, 2007.

\bibitem{LS93}
L.~Lovasz and M.~Simonovits.
\newblock Random walks in a convex body and an improved volume algorithm.
\newblock {\em Random Structures and Algorithms}, 4(4), 1993.

\bibitem{Mazziotti}
D.A. Mazziotti.
\newblock Quantum chemistry without wave functions: Two-electron reduced
  density matrices.
\newblock {\em Acc. Chem. Res.}, 39:207--215, 2006.

\bibitem{Mazzz}
D.A. Mazziotti.
\newblock Variational reduced-density-matrix method using three-particle
  {N}-representability conditions with application to many-electron molecules.
\newblock {\em Phys. Rev. A}, 74(3):032501, 2006.

\bibitem{mazziotti-acse}
D.A. Mazziotti.
\newblock Anti-{H}ermitian part of the contracted {S}chrodinger equation for
  the direct calculation of two-electron reduced density matrices.
\newblock {\em Phys. Rev. A}, 75(2):022505, 2007.

\bibitem{Motwani-Raghavan}
R.~Motwani and P.~Raghavan.
\newblock {\em Randomized Algorithms}.
\newblock Cambridge University Press, 1995.

\bibitem{NC}
M.A. Nielsen and I.L. Chuang.
\newblock {\em Quantum computation and quantum information}.
\newblock Cambridge University Press, Cambridge, UK, 2000.

\bibitem{OT}
R.~Oliveira and B.M. Terhal.
\newblock The complexity of quantum spin systems on a two-dimensional square
  lattice.
\newblock Arxiv: quant-ph/0504050, 2005.

\bibitem{Ortiz-et-al-00}
G.~Ortiz, J.E. Gubernatis, E.~Knill, and R.~Laflamme.
\newblock Quantum algorithms for fermionic simulations.
\newblock {\em Phys. Rev. A}, 64(2):022319, 2001.
\newblock Arxiv: cond-mat/0012334.

\bibitem{osborne}
T.J. Osborne.
\newblock Efficient approximation of the dynamics of one-dimensional quantum
  spin systems.
\newblock {\em Phys. Rev. Lett.}, 97(15):157202, 2006.

\bibitem{Papadimitriou}
C.H. Papadimitriou.
\newblock {\em Computational complexity}.
\newblock Addison Wesley Longman, 1994.

\bibitem{Sachdev}
S.~Sachdev.
\newblock {\em Quantum Phase Transitions}.
\newblock Cambridge University Press, 2000.

\bibitem{schollwock}
U.~Schollw\"{o}ck.
\newblock The density-matrix renormalization group.
\newblock {\em Rev. Mod. Phys.}, 77(1):259, 2005.

\bibitem{Shor2}
P.W. Shor.
\newblock Fault-tolerant quantum computation.
\newblock In {\em IEEE Foundations of Computer Science (FOCS'96)}, pages
  56--65, 1996.

\bibitem{Shor}
P.W. Shor.
\newblock Polynomial-time algorithms for prime factorization and discrete
  logarithms on a quantum computer.
\newblock {\em SIAM J. Comput.}, 26(5):1484--1509, 1997.

\bibitem{Ostlund-Szabo}
A.~Szabo and N.S. Ostlund.
\newblock {\em Modern quantum chemistry: introduction to advanced electronic
  structure theory}.
\newblock Macmillan, New York, 1982.

\bibitem{Tredgold}
R.~H. Tredgold.
\newblock Density matrix and the many-body problem.
\newblock {\em Phys. Rev.}, 105(5):1421--1423, Mar 1957.

\bibitem{vv}
V.~Vazirani.
\newblock {\em Approximation Algorithms}.
\newblock Springer, 2001.

\bibitem{Vsurvey}
S.~Vempala.
\newblock Geometric random walks: A survey.
\newblock In J.E. Goodman, J.~Pach, and E.~Welzl, editors, {\em Combinatorial
  and Computational Geometry}, volume~52 of {\em MSRI publications}. Cambridge
  University Press, New York, 2005.

\bibitem{VC06}
F.~Verstraete and J.~I. Cirac.
\newblock Matrix product states represent ground states faithfully.
\newblock {\em Phys. Rev. B}, 73(9):094423, 2006.

\bibitem{VC}
F.~Verstraete and J.I. Cirac.
\newblock Mapping local {H}amiltonians of fermions to local {H}amiltonians of
  spins.
\newblock {\em J. Stat. Mech.}, 2005(09):P09012, 2005.

\bibitem{Vyal}
M.N. Vyalyi.
\newblock {QMA} = {PP} implies that {PP} contains {PH}.
\newblock ECCC report no. 21, 2003.

\bibitem{Wat00}
J.~Watrous.
\newblock Succinct quantum proofs for properties of finite groups.
\newblock In {\em Proc. IEEE FOCS'2000}, pages 537--546, 2000.

\bibitem{Wat}
J.~Watrous.
\newblock Zero-knowledge against quantum attacks.
\newblock In {\em ACM Symp. on Theory of Computing (STOC '06)}, pages 296--305,
  2006.
\newblock Arxiv: quant-ph/0511020.

\bibitem{Wilcox}
R.~M. Wilcox.
\newblock Exponential operators and parameter differentiation in quantum
  physics.
\newblock {\em J. Math. Phys.}, 8(4):962--982, 1967.

\bibitem{Kuhn}
M.L. Yoseloff and H.W. Kuhn.
\newblock Combinatorial approach to the {N}-representability of {P}-density
  matrices.
\newblock {\em J. Math. Phys.}, 10(4):703--706, 1969.

\bibitem{YN}
D.B. Yudin and A.S. Nemirovskii.
\newblock Informational complexity and efficient methods for the solution of
  convex extremal problems.
\newblock {\em Ekonomika i Matematicheskie Metody}, 12:357--369, 1976.
\newblock English translation: Matekon 13 (3) pp.25-45 (1977).

\end{thebibliography}

\end{document}